\documentclass[aps,prfluids, superscriptaddress,nofootinbib]{revtex4-2}

\usepackage{amsmath,amssymb,bm,mathtools}
\usepackage{graphicx}
\usepackage{xcolor}
\usepackage{geometry}
\usepackage{hyperref}
\usepackage{physics}
\usepackage{enumitem}
\hypersetup{colorlinks=true,linkcolor=blue,citecolor=blue,urlcolor=blue}

\usepackage{CJKutf8}

\definecolor{orange}{RGB}{230,120,0}
\makeatletter
\def\HLTXT_#1#{\csname HLTXT@#1\endcsname}
\def\HLTXT@BLUE#1{{\color{black}#1}}
\def\HLTXT@ORANGE#1{{\color{black}#1}}
\makeatother

\newcommand{\ResFigNoKsix}[1]{\detokenize{./#1}}
\newcommand{\ResFigWithKsix}[1]{\detokenize{./#1}}
\newcommand{\ResFig}[1]{\ResFigWithKsix{#1}}
\renewcommand{\dd}{\mathrm{d}}
\newcommand{\p}{\partial}
\newcommand{\D}{\mathrm{D}}
\renewcommand{\grad}{\bm{\nabla}}
\newcommand{\Div}{\bm{\nabla}\!\cdot}
\newcommand{\Curl}{\bm{\nabla}\!\times}

\newcommand{\avg}[1]{\left\langle #1 \right\rangle}

\newcommand{\vort}{\bm{\zeta}}
\newcommand{\vortmode}{\zeta}

\begin{document}

\title{Distinct transverse-response signatures of retained-spin, eliminated-spin, and polynomial Burnett-type surrogate closures}
\author{Satori Tsuzuki (\begin{CJK}{UTF8}{ipxm}都築怜理\end{CJK})}
\affiliation{Research Center for Advanced Science and Technology, The University of Tokyo}

\begin{abstract}
High-curvature observables in incompressible flows, including $k^4$-weighted spectra, can arise from explicit internal rotation, elimination of a fast spin variable, or polynomial higher-gradient closure. Building on a retained-spin micropolar closure derived separately from the Boltzmann--Curtiss equation, we show that these mechanisms are dynamically distinguishable in transverse linear response. In a fast-spin regime the retained-spin theory reduces to a one-field model with a rational $k$-dependent kernel whose low-$k$ expansion generates $k^4$ and $k^6$ terms, while preserving the large-$k$ roll-off of the eliminated degree of freedom. We compare four closures: incompressible Navier--Stokes, a polynomial Burnett-type surrogate, the explicit-spin micropolar theory, and the eliminated-spin rational-kernel theory. The explicit-spin theory has two poles, the eliminated-spin theory retains only the slow pole, and finite polynomial truncations fail qualitatively: a strict $k^4$ truncation becomes over-damped, while a matched $k^6$ truncation develops near-critical amplification and finite-$k$ instability. Many-particle event-driven simulations of perfectly rough spheres show that these observables are measurable and, in targeted campaigns, discriminating at the microscopic level: fixed-$k$ and multi-$k$ harmonic forcing resolve a finite spin-to-vorticity phase lag that strongly favors retained-spin dynamics over instantaneous adiabatic elimination, while the stronger-drive multi-$k$ vorticity response rejects a pure $k^2$ closure and favors the rational eliminated-spin kernel over a polynomial surrogate. Transverse response thus provides a practical diagnostic for separating retained rotational microphysics, eliminated-spin effective dynamics, and ordinary polynomial higher-gradient closures.
\end{abstract}

\date{\today}

\maketitle

\section{Introduction}
\label{sec:intro}
Many continuum descriptions of complex fluids extend the classical Navier--Stokes
(NS) framework by promoting internal microstructural degrees of freedom to
continuum fields. A prominent example is \emph{micropolar} (or \emph{spin})
hydrodynamics, in which an intrinsic micro-rotation $\bm{\omega}_0$ and the
associated couple stress $\bm{m}$ enter alongside the velocity $\bm{u}$ and
Cauchy stress $\bm{\sigma}$ \cite{CondiffDahler1964,Eringen1966,Eringen1999,Eringen2001}.
Such theories are natural when local internal rotation is not instantaneously
slaved to the macroscopic vorticity, and when antisymmetric stresses or spin
diffusion modify the large-scale response. What is much less developed is the
\emph{inverse} use of this framework: given macroscopic decay or frequency-response
data, can one tell whether an internal rotational degree of freedom is actually
dynamical?

This question is nontrivial because short-wavelength observables are often
mechanistically ambiguous. Consider, for example, quantities that weight
curvature by $k^{4}$, such as $k^{4}E(k,t)$ or spectra associated with
$|\nabla\times\vort|^{2}$, where $\vort=\Curl\bm{u}$. Large values of such
observables can arise from at least three distinct mechanisms:
(i) explicit internal-spin dynamics;
(ii) a one-field effective theory obtained by eliminating a fast spin variable;
or
(iii) a conventional higher-gradient closure of Burnett type constructed
without retaining an internal rotational field from the outset
\cite{Burnett1935}. At the level of a low-order derivative expansion, cases
(ii) and (iii) may look deceptively similar. Dynamically, however, they need not
be equivalent: eliminating a fast internal variable produces a \emph{rational}
$k$-dependent kernel, whereas a Burnett truncation is polynomial in $k$ to the
order retained.

The present paper should therefore be read as the response-theory companion to
Ref.~\cite{Tsuzuki2026CompanionRetainedSpin}. That companion paper derives the
retained-spin micropolar closure directly from the Boltzmann--Curtiss equation,
including the exact balance laws, the extended Chapman--Enskog bookkeeping in
which the local spin is treated as a quasi-slow variable, and dilute-gas
rough-sphere estimates of $\eta_r$ and $\beta+\gamma$. Here we take that
retained-spin closure as the starting point and ask a different question: once
such a closure is available, which dynamical observables distinguish explicit
retained spin, adiabatic elimination of a fast spin variable, and a finite
polynomial higher-gradient surrogate?

This shift of emphasis is important. The main novelty claimed here is \emph{not}
the existence of a first-order retained-spin closure by itself. Such closures
have clear antecedents in kinetic theory and microcontinuum mechanics, and the
detailed line-by-line derivation is now available in the companion paper
\cite{Tsuzuki2026CompanionRetainedSpin}. The new element of the present paper is
the use of that closure as a controlled starting point for a response-theoretic
comparison between explicit-spin dynamics, fast-spin elimination, and
polynomial higher-gradient truncation. The polynomial model used below is a
Burnett-type \emph{surrogate}: it is introduced to isolate the consequences of
replacing the rational eliminated-spin kernel by a finite polynomial in $k$.
It is not presented as a coefficient-complete Burnett derivation for the
microscopic collision model used later in the EDMD section.

Within that scope, the paper makes three concrete contributions. First, we
recast the retained-spin closure into a transverse-response language and show
that fast elimination of spin yields a one-field model with a rational
$k$-kernel whose low-$k$ series reproduces Burnett-type $k^{4}$ and $k^{6}$
terms while preserving a distinct large-$k$ structure and, away from the
quasi-steady limit, an exponential memory kernel. Second, we compare four
response classes---incompressible Navier--Stokes, a polynomial Burnett-type
surrogate, the explicit-spin micropolar theory, and the eliminated-spin
rational-kernel theory---by means of reduced single-mode benchmarks and
response maps. Third, we test the observability of the proposed diagnostics in
many-particle event-driven molecular dynamics (EDMD) of perfectly rough spheres
\cite{DahlerSather1963,MonchickYunMason1963,CondiffLuDahler1965,McCoySandlerDahler1966},
with emphasis on measurable late-time decay, coherent vorticity response, and
phase-locked spin response.

The reduced-model and EDMD calculations therefore play different roles. The
single-mode benchmarks are controlled consistency tests of the response classes
themselves. The EDMD section asks a narrower but experimentally relevant
question: which of the proposed observables can actually be extracted from noisy
microscopic data? In this sense the two companion papers are complementary: the
companion derivation paper \cite{Tsuzuki2026CompanionRetainedSpin} is
coefficient- and closure-oriented, whereas the present manuscript is explicitly
response- and diagnostic-oriented.

\HLTXT_BLUE{The discrimination criteria developed here are deliberately linear-response
criteria. They do not establish nonlinear closure equivalence or nonlinear
stability in the full hydrodynamic system; rather, they identify which
signatures survive in transverse decay and harmonic response once the dynamics
is linearized about a homogeneous state. Nonlinear mode coupling,
finite-amplitude instabilities, and turbulent transfer remain outside the
scope of the present analysis.}

The remainder of the paper is organized as follows.
Section~\ref{sec:samchfoundation} recalls the retained-spin closure used in the
present study and derives the fast-spin reduction needed for the response
comparison.
Section~\ref{sec:translineardiagno} develops the transverse linear-response
comparison.
Section~\ref{sec:analysis} presents reduced-model single-mode benchmarks, and
Sec.~\ref{sec:edmd_benchmarks} turns to microscopic EDMD observability
benchmarks.
Section~\ref{sec:discussion} discusses the diagnostic implications and the
limitations of the present comparison, and Sec.~\ref{sec:conclusion}
summarizes the main conclusions.

\section{Retained-spin closure and fast-spin reduction}
\label{sec:samchfoundation}

The detailed Boltzmann--Curtiss derivation of the retained-spin micropolar
closure adopted here is presented in the companion paper
\cite{Tsuzuki2026CompanionRetainedSpin}. In particular, that work derives the
exact mass, momentum, and intrinsic-angular-momentum balances, formulates the
first-order Chapman--Enskog construction in which the local spin is treated as a
quasi-slow extended variable, and records dilute-gas rough-sphere estimates of
$\eta_r$ and $\beta+\gamma$. For the response problem studied here, it suffices
to recall the closed equations and the fast-spin reduction that links the
explicit-spin and eliminated-spin descriptions.

\subsection{Starting retained-spin equations}
\label{sec:notationandtarget}
We work in three spatial dimensions. Bold symbols denote vectors or
second-order tensors, and repeated indices are summed over.
The material derivative is defined by
\[
\frac{\D}{\D t}(\cdot)=\p_t(\cdot)+\bm{u}\cdot\grad(\cdot).
\]
The macroscopic vorticity is $\vort=\Curl \bm{u}$; this notation is used
throughout to avoid confusion with the particle angular velocity $\bm{\omega}$
of the kinetic description underlying the companion derivation.
The macroscopic fields are the mass density $\rho$, velocity $\bm{u}$,
pressure $P$, and intrinsic spin (gyration) rate $\bm{\omega}_0$.
For isotropic microinertia we write $J=I/m$, the moment of inertia per unit mass.

The retained-spin Navier--Stokes-order closure used throughout is
\begin{align}
\p_t\rho+\Div(\rho\bm{u})&=0,
\label{eq:target_mass}\\[2mm]
\rho\frac{\D \bm{u}}{\D t}
&=
-\grad P
+(\eta+\eta_r)\grad^2\bm{u}
+\Bigl(\frac{\eta}{3}+\xi-\eta_r\Bigr)\grad(\Div\bm{u})
+2\eta_r\,\Curl\bm{\omega}_0
+\rho\,\bm{F},
\label{eq:target_momentum}\\[2mm]
\rho J\frac{\D \bm{\omega}_0}{\D t}
&=
(\beta+\gamma)\grad^2\bm{\omega}_0
+(\alpha+\beta-\gamma)\grad(\Div\bm{\omega}_0)
+2\eta_r(\vort-2\bm{\omega}_0)
+\rho\,\bm{G}.
\label{eq:target_spin}
\end{align}
Here $\eta$ denotes the shear viscosity, $\xi$ the bulk viscosity,
$\eta_r$ the rotational viscosity, and $(\alpha,\beta,\gamma)$ the
couple-stress viscosities. For the present response problem, the key structural
point is that $\eta_r$ couples vorticity to intrinsic spin through the
antisymmetric stress channel, while $(\alpha,\beta,\gamma)$ determine spin
diffusion. The coefficient-level derivation and the corresponding dilute-gas
rough-sphere estimates are deferred to Ref.~\cite{Tsuzuki2026CompanionRetainedSpin}.

\subsection{Fast-spin elimination and effective one-field closure}
\label{sec:fast_spin_elimination}
To compare explicit retained spin with eliminated-spin and polynomial
higher-gradient models, it is useful to eliminate $\bm{\omega}_0$ from
Eq.~(\ref{eq:target_spin}) in a fast-spin regime. Define the spin-diffusion
operator
\begin{equation}
\mathcal{L}_s := (\beta+\gamma)\grad^2 + (\alpha+\beta-\gamma)\grad\Div.
\label{eq:spin_diff_operator}
\end{equation}
Then Eq.~(\ref{eq:target_spin}) may be written as
\begin{equation}
\left(\rho J\frac{\D}{\D t} - \mathcal{L}_s + 4\eta_r\right)\bm{\omega}_0
=
2\eta_r\vort + \rho\bm{G}.
\label{eq:spin_operator_rearranged}
\end{equation}
Introducing the characteristic spin-relaxation time and spin-diffusion lengths,
\begin{equation}
\tau_s := \frac{\rho J}{4\eta_r},
\qquad
\ell_1^2 := \frac{\beta+\gamma}{4\eta_r},
\qquad
\ell_2^2 := \frac{\alpha+\beta-\gamma}{4\eta_r},
\label{eq:spin_time_length_scales}
\end{equation}
this becomes
\begin{equation}
\left(1 + \tau_s\frac{\D}{\D t} - \ell_1^2\grad^2 - \ell_2^2\grad\Div\right)\bm{\omega}_0
=
\frac12\vort + \frac{\rho}{4\eta_r}\bm{G}.
\label{eq:spin_scaled_operator}
\end{equation}
If spin relaxes rapidly on the macroscopic time scale,
\begin{equation}
\tau_s\left\|\frac{\D\bm{\omega}_0}{\D t}\right\| \ll \|\bm{\omega}_0\|,
\label{eq:fast_spin_assumption}
\end{equation}
then the quasi-steady approximation gives
\begin{equation}
\left(1 - \ell_1^2\grad^2 - \ell_2^2\grad\Div\right)\bm{\omega}_0
=
\frac12\vort + \frac{\rho}{4\eta_r}\bm{G}.
\label{eq:spin_quasisteady_general}
\end{equation}
In the incompressible transverse sector,
\begin{equation}
\Div\bm{u}=0,
\qquad
\bm{G}=\bm{0},
\qquad
\Div\vort=0,
\qquad
\Div\bm{\omega}_0=0,
\label{eq:incompressible_conditions_section10}
\end{equation}
so Eq.~(\ref{eq:spin_quasisteady_general}) reduces to
\begin{equation}
\left(1-\ell_s^2\grad^2\right)\bm{\omega}_0 = \frac12\vort,
\qquad
\ell_s^2 := \frac{\beta+\gamma}{4\eta_r}.
\label{eq:spin_quasisteady_incompressible}
\end{equation}
In Fourier space,
\begin{equation}
\widehat{\bm{\omega}}_0(\bm{k})
=
\frac{1}{2\left(1+\ell_s^2 k^2\right)}\widehat{\vort}(\bm{k}),
\qquad
k:=|\bm{k}|.
\label{eq:spin_fourier_incompressible}
\end{equation}
Thus the eliminated-spin closure is governed by a \emph{rational} $k$-kernel,
not by a finite polynomial. Its low-$k$ expansion generates a Burnett-type
series,
\begin{equation}
\bm{\omega}_0
=
\frac12\left(\vort + \ell_s^2\grad^2\vort + \ell_s^4\grad^4\vort + \cdots\right),
\label{eq:spin_series_incompressible}
\end{equation}
which, when substituted into Eq.~(\ref{eq:target_momentum}), yields the effective
one-field incompressible closure
\begin{equation}
\rho\frac{\D\bm{u}}{\D t}
=
-\grad P
+\eta\grad^2\bm{u}
-\eta_r\ell_s^2\grad^4\bm{u}
-\eta_r\ell_s^4\grad^6\bm{u}
-\cdots
+\rho\bm{F}.
\label{eq:effective_burnett_series_u}
\end{equation}
The leading higher-gradient term is therefore
$-(\beta+\gamma)\grad^4\bm{u}/4$. This is the sense in which adiabatic spin
elimination produces a Burnett-type correction without ever introducing a
finite polynomial closure as a microscopic starting point. If the inertial term
in Eq.~(\ref{eq:spin_scaled_operator}) is retained instead of neglected, exact
elimination yields a temporally nonlocal memory kernel; the quasi-steady limit
singled out here is the regime in which comparison with Burnett-type
higher-gradient surrogates is most direct.

\section{Transverse linear response and diagnostic models} \label{sec:translineardiagno}
\HLTXT_BLUE{All response functions in this section should be understood as
linear transverse-sector objects. The subsequent pole-counting and
kernel-shape diagnostics are therefore statements about the linearized response
around a homogeneous rest state, not about the full nonlinear closure problem.}

Consider now an observable that strongly weights high curvature. A
canonical example is
\begin{equation}
Q(k,t) \propto k^4 E(k,t),
\label{eq:Q_k4E}
\end{equation}
which, in the incompressible isotropic setting, is associated with
$|\Curl\vort|^2$~\cite{Tsuzuki2026PRFluids, Tsuzuki2026arXiv}. 

If such an observable exhibits an early buildup
or a pronounced peak at relatively large $k$, then at least three
distinct mechanisms must be considered: genuine two-field dynamics in
which $(\bm{u},\bm{\omega}_0)$ both evolve explicitly; a one-field
Burnett-type dynamics obtained by adiabatically eliminating the spin
field from the underlying micropolar theory; and a Burnett closure
derived directly from a spinless kinetic theory. Observable
Eq.~(\ref{eq:Q_k4E}) alone cannot distinguish these possibilities, because
all three enhance the role of large wavenumbers. The distinction must
instead be made dynamically, for example through response functions~\cite{Kubo1957,Forster1975} or dispersion relations.

\subsection{Linear response in the transverse sector}
To expose the discriminating structure most cleanly, linearize about a
homogeneous rest state,
\begin{equation}
\bm{u}=\bm{0},
\qquad
\bm{\omega}_0=\bm{0},
\qquad
\rho=\rho_0=\text{const.},
\label{eq:linearization_rest_state}
\end{equation}
ignore body forces and body couples, and consider transverse plane-wave
disturbances. For the response analysis we select one transverse polarization,
so the Fourier amplitudes are scalars. We therefore write
\begin{equation}
\zeta(\bm{x},t)=\hat{\zeta}\,e^{st+i\bm{k}\cdot\bm{x}},
\qquad
\omega(\bm{x},t)=\hat{\omega}\,e^{st+i\bm{k}\cdot\bm{x}},
\qquad
s\in\mathbb{C},
\label{eq:plane_wave_ansatz}
\end{equation}
with forcing amplitudes $\hat{f}$ and $\hat{g}$ in the vorticity and spin
channels. For harmonic forcing one sets $s=-i\Omega$. This scalar notation is
equivalent to fixing one transverse component of the vector response. It is
also the notation used in the single-mode benchmarks of Sec.~\ref{sec:analysis}.

In the transverse incompressible sector, Eqs.~(\ref{eq:target_momentum}) and
(\ref{eq:target_spin}) reduce, after taking the curl of the momentum equation,
to
\begin{align}
\rho\,\p_t\zeta
&=
(\eta+\eta_r)\grad^2\zeta - 2\eta_r\grad^2\omega,
\label{eq:lin_vorticity_micropolar}
\\
\rho J\,\p_t\omega
&=
(\beta+\gamma)\grad^2\omega + 2\eta_r(\zeta-2\omega),
\label{eq:lin_spin_micropolar}
\end{align}
where the longitudinal spin-diffusion term drops out. Substitution of
Eq.~(\ref{eq:plane_wave_ansatz}) then gives
\begin{equation}
\begin{pmatrix}
\rho s + (\eta+\eta_r)k^2 & -2\eta_r k^2 \\
-2\eta_r & \rho J s + (\beta+\gamma)k^2 + 4\eta_r
\end{pmatrix}
\begin{pmatrix}
\hat{\zeta} \\
\hat{\omega}
\end{pmatrix}
=
\begin{pmatrix}
\hat{f} \\
\hat{g}
\end{pmatrix}.
\label{eq:response_matrix_micropolar}
\end{equation}
The analytic cases derived below map onto the numerical models in
Table~\ref{tab:transverse-models-summary} as follows:
Model~C = explicit retained spin,
Model~D = adiabatic elimination,
Model~B = polynomial Burnett-type surrogate,
and Model~A = the Navier--Stokes limit.

\subsection{Three diagnostic relations}
\paragraph*{Model C: explicit retained spin.}
For the explicit-spin theory, the response matrix is the inverse of the operator
in Eq.~(\ref{eq:response_matrix_micropolar}), whose determinant is
\begin{equation}
\Delta_{\mathrm{MP}}(s,k)
=
\left[\rho s + (\eta+\eta_r)k^2\right]
\left[\rho J s + (\beta+\gamma)k^2 + 4\eta_r\right]
-4\eta_r^2 k^2.
\label{eq:delta_micropolar}
\end{equation}
The corresponding dispersion relation is
\begin{equation}
\Delta_{\mathrm{MP}}(s,k)=0.
\label{eq:dispersion_micropolar}
\end{equation}
Because Eq.~(\ref{eq:delta_micropolar}) is quadratic in $s$, Model~C has two
branches: a hydrodynamic shear/vorticity mode and a fast spin-relaxation mode.
The vorticity-to-vorticity response function is
\begin{equation}
\chi_{\zeta\zeta}^{\mathrm{C}}(s,k)
=
\frac{\rho J s + (\beta+\gamma)k^2 + 4\eta_r}
{\Delta_{\mathrm{MP}}(s,k)}.
\label{eq:chi_case1}
\end{equation}
The additional pole is the clearest dynamical signature of an explicit internal
degree of freedom.

The same two-field structure also shows what is lost when the spin variable is
eliminated exactly rather than adiabatically. Solving Eq.~(\ref{eq:lin_spin_micropolar})
at fixed $k$ gives
\begin{equation}
\omega(t;k)
=
e^{-\lambda_k t}\,\omega(0;k)
+\frac{1}{\rho J}\int_0^t e^{-\lambda_k(t-t')}
\left[2\eta_r\,\zeta(t';k)+g(t';k)\right]\dd t',
\qquad
\lambda_k:=\frac{4\eta_r+(\beta+\gamma)k^2}{\rho J}.
\label{eq:omega_memory_solution}
\end{equation}
Substitution into the vorticity equation yields the exact one-field memory form
\begin{align}
\rho\,\partial_t\zeta(t;k)
+(\eta+\eta_r)k^2\zeta(t;k)
-\frac{4\eta_r^2k^2}{\rho J}\int_0^t
e^{-\lambda_k(t-t')}\zeta(t';k)\,\dd t'
&=
f(t;k)
+2\eta_r k^2 e^{-\lambda_k t}\omega(0;k)
\nonumber\\
&\quad
+\frac{2\eta_r k^2}{\rho J}\int_0^t
e^{-\lambda_k(t-t')}g(t';k)\,\dd t'.
\label{eq:zeta_memory_equation}
\end{align}
Thus exact elimination of spin is temporally nonlocal. The quasi-steady
Model~D derived next is the low-frequency limit of this memory equation.

\paragraph*{Model D: fast-spin elimination.}
In the quasi-steady fast-spin limit, Eq.~(\ref{eq:lin_spin_micropolar}) gives
\begin{equation}
\hat{\omega}
=
\frac{2\eta_r}{4\eta_r + (\beta+\gamma)k^2}\,\hat{\zeta}.
\label{eq:omega0_hat_eliminated}
\end{equation}
Substituting this relation into Eq.~(\ref{eq:lin_vorticity_micropolar}) yields
the closed one-field response
\begin{equation}
\left[
\rho s + \eta k^2 + \frac{\eta_r(\beta+\gamma)k^4}{4\eta_r + (\beta+\gamma)k^2}
\right]\hat{\zeta}
=
\hat{f},
\label{eq:case2_closed_response}
\end{equation}
and therefore
\begin{equation}
\chi_{\zeta\zeta}^{\mathrm{D}}(s,k)
=
\left[
\rho s + \eta k^2 + \frac{\eta_r(\beta+\gamma)k^4}{4\eta_r + (\beta+\gamma)k^2}
\right]^{-1}.
\label{eq:chi_case2}
\end{equation}
Model~D has only one pole in $s$, but its $k$-dependence is rational rather
than polynomial.

The controlled relation between Models~C and D is obtained by rewriting
Eq.~(\ref{eq:chi_case1}) as
\begin{equation}
\left[\chi_{\zeta\zeta}^{\mathrm{C}}(s,k)\right]^{-1}
=
\rho s + (\eta+\eta_r)k^2
-\frac{4\eta_r^2 k^2}{\rho J s + 4\eta_r + (\beta+\gamma)k^2}.
\label{eq:chi_case1_inverse}
\end{equation}
If
\begin{equation}
\varepsilon_{\mathrm{ad}}(s,k)
:=
\frac{|\rho J s|}{4\eta_r + (\beta+\gamma)k^2}
\ll 1,
\label{eq:adiabatic_small_parameter}
\end{equation}
then expanding the last denominator yields
\begin{align}
\left[\chi_{\zeta\zeta}^{\mathrm{C}}(s,k)\right]^{-1}
&=
\left[\chi_{\zeta\zeta}^{\mathrm{D}}(s,k)\right]^{-1}
+\frac{4\eta_r^2 k^2\,\rho J s}
{\left[4\eta_r+(\beta+\gamma)k^2\right]^2}
\nonumber\\
&\qquad
+O\!\left(
\frac{4\eta_r^2 k^2(\rho J s)^2}
{\left[4\eta_r+(\beta+\gamma)k^2\right]^3}
\right).
\label{eq:chi_case1_to_case2}
\end{align}
At $k=0$ this reduces to the familiar condition $|s|\tau_s\ll 1$, with
$\tau_s=\rho J/(4\eta_r)$. Model~D therefore reproduces the slow branch of
Model~C whenever $\varepsilon_{\mathrm{ad}}(s,k)\ll 1$, while the omitted fast
pole remains on the timescale
$s\sim-[4\eta_r+(\beta+\gamma)k^2]/(\rho J)$.
For free decay this reduction describes the late-time dynamics after the
initial transient; for harmonic forcing it applies whenever
$\rho J|\Omega|\ll 4\eta_r+(\beta+\gamma)k^2$.

The small-$k$ expansion of Model~D is
\begin{equation}
\left[\chi_{\zeta\zeta}^{\mathrm{D}}(s,k)\right]^{-1}
=
\rho s + \eta k^2 + \frac{\beta+\gamma}{4}k^4
-\frac{(\beta+\gamma)^2}{16\eta_r}k^6 + O(k^8),
\label{eq:case2_smallk_expansion}
\end{equation}
so the eliminated-spin theory reproduces a Burnett-type $k^4$ correction at
low $k$ while retaining a nonpolynomial exact kernel.

\paragraph*{Model B: polynomial Burnett-type surrogate.}
For a one-field closure written directly as a finite gradient expansion, the
transverse response has the schematic form
\begin{equation}
\left[\rho s + \eta k^2 + B_1 k^4 + B_2 k^6 + \cdots\right]\hat{\zeta}
=
\hat{f}.
\label{eq:case3_response_general}
\end{equation}
At strict Burnett order,
\begin{equation}
\chi_{\zeta\zeta}^{\mathrm{B}}(s,k)
=
\left[\rho s + \eta k^2 + B_1 k^4\right]^{-1}.
\label{eq:chi_case3}
\end{equation}
Model~B again has only one pole, but now the $k$-dependence is polynomial to
the truncation order retained. Its role here is diagnostic: it isolates what is
lost when the rational eliminated-spin kernel is replaced by a finite
polynomial.

\subsection{Diagnostic summary}
The comparison therefore depends on two independent structural features:
pole count and kernel shape. Model~C has two poles. Models~A, B, and D have
one. Among the one-pole models, Model~D is rational in $k$ and asymptotically
returns to an effective $k^2$ scaling, whereas Model~B is polynomial and
inherits the asymptote of its highest retained power. Consequently, the mere
appearance of a high-curvature observable such as Eq.~(\ref{eq:Q_k4E}) does not
identify the underlying mechanism. What can distinguish the possibilities is the
measured or computed transverse response function and the associated dispersion
relation.

% -------------------------------------------------------------------------
% Summary table of transverse-sector linear models (A--D)
% -------------------------------------------------------------------------
\begin{table*}[t]
\centering
\caption{
Summary of the four transverse-sector models used in the response comparison.
Here $\zeta(t;k)$ denotes the Fourier amplitude of the macroscopic vorticity
mode at wavenumber $k$, and $\omega_0(t;k)$ the internal spin
(micro-rotation) mode. Model~B is a polynomial Burnett-type surrogate used to
contrast finite polynomial truncation with the rational kernel of Model~D.}
\label{tab:transverse-models-summary}
\renewcommand{\arraystretch}{1.35}
\begin{tabular}{p{0.15\textwidth} p{0.5\textwidth} p{0.33\textwidth}}
\hline\hline
Model & Time evolution (single transverse mode at fixed $k$) & Response function $\chi_{\zeta\zeta}(s,k)$ \\
\hline
A &
$\displaystyle
\rho\,\partial_t\vortmode = -\eta k^2\,\vortmode + f
$
&
$\displaystyle
[\rho s+\eta k^2]^{-1}
$
\\
\hline
B &
$\displaystyle
\rho\,\partial_t\vortmode =
-\bigl(\eta k^2 + B_1 k^4 + B_2 k^6\bigr)\vortmode + f
$
&
$\displaystyle
[\rho s+\eta k^2 + B_1 k^4 + B_2 k^6]^{-1}
$
\\
\hline
C &
$\displaystyle
\begin{aligned}
\rho\,\partial_t\vortmode
&=
-(\eta+\eta_r)k^2\,\vortmode
+2\eta_r k^2\,\omega_0
+f,\\
\rho J\,\partial_t\omega_0
&=
2\eta_r\,\vortmode
-\bigl[(\beta+\gamma)k^2+4\eta_r\bigr]\omega_0
+g
\end{aligned}
$
&
$\displaystyle
\frac{\rho J s+(\beta+\gamma)k^2+4\eta_r}
{\Delta_{\mathrm{MP}}(s,k)}
$
\\
\hline
D &
$\displaystyle
\rho\,\partial_t\vortmode
=
-\left[
\eta k^2
+
\frac{\eta_r(\beta+\gamma)k^4}
{4\eta_r+(\beta+\gamma)k^2}
\right]\vortmode
+ f
$
&
$\displaystyle
\left[
\rho s+\eta k^2
+
\frac{\eta_r(\beta+\gamma)k^4}
{4\eta_r+(\beta+\gamma)k^2}
\right]^{-1}
$
\\
\hline\hline
\end{tabular}
\end{table*}

Table~\ref{tab:transverse-models-summary} summarizes four useful
transverse-sector diagnostic models. Model A is the incompressible
Navier--Stokes limit and has a single diffusive pole. Model B is a
polynomial Burnett-type truncation used as a diagnostic surrogate; it has
polynomial $k$-dependence and again has a single pole. Model C is the
full micropolar system with explicit spin and therefore has two poles,
corresponding to two relaxation branches. Model D is obtained from
Model C by fast-spin elimination and has a single pole with a rational
$k$-dependent kernel. In Laplace space the response is defined by 
$\zeta(s;k)=\chi_{\zeta\zeta}(s,k)f(s;k)$ when $g=0$.

For Model C, the two-field response is governed by the determinant
\begin{equation}
\Delta_{\mathrm{MP}}(s,k)
=
\bigl[\rho s+(\eta+\eta_r)k^2\bigr]
\bigl[\rho J s+(\beta+\gamma)k^2+4\eta_r\bigr]
-4\eta_r^2 k^2,
\label{eq:DeltaMP_snippet}
\end{equation}
so that $\chi_{\zeta\zeta}(s,k)$ has two poles given by the roots of
$\Delta_{\mathrm{MP}}(s,k)=0$. By contrast, Models A, B, and D reduce to
single-field closures for $\zeta(t;k)$ and therefore exhibit only one
pole. For Model D, the rational kernel admits the low-$k$ expansion
\begin{align}
\eta k^2+\frac{\eta_r(\beta+\gamma)k^4}{4\eta_r+(\beta+\gamma)k^2}
&=
\eta k^2
+\frac{\beta+\gamma}{4}\,k^4
-\frac{(\beta+\gamma)^2}{16\eta_r}\,k^6
+\mathcal{O}(k^8),
\label{eq:adiabatic_lowk_expansion_snippet}
\end{align}
which matches a polynomial Burnett-type expansion through the displayed
order at sufficiently small $k$, while remaining nonpolynomial at finite $k$.
This distinction is central for diagnosis via measured
$\chi_{\zeta\zeta}(k,\Omega)=\chi_{\zeta\zeta}(s=-i\Omega,k)$ and
the associated transverse dispersion relations.
%================================================================

\section{Reduced-model single-mode benchmarks}\label{sec:analysis}
In this section we benchmark, through minimal numerical experiments, the
transverse-sector dispersion relations and response functions developed in
Sec.~\ref{sec:translineardiagno} using the fast-spin reduction of
Sec.~\ref{sec:fast_spin_elimination}. We work directly with the single-mode
time-evolution equations summarized in
Table~\ref{tab:transverse-models-summary}: for each model we follow one
transverse Fourier mode (the scalar amplitude $\zeta(t;k)$, and $\omega_0(t;k)$
when present). These calculations are controlled checks of the reduced closures
themselves rather than independent DNS or experimental validation. Because the
single-mode systems in Table~\ref{tab:transverse-models-summary} are linear,
both the free-decay dynamics (Setting~1) and the harmonic steady response
(Setting~2) are determined in closed form by the corresponding response
kernels.

\paragraph*{Parameter set and two polynomial surrogate variants.}
All plots shown below use a nondimensional parameter set
\begin{equation}
\rho=1,\qquad \eta=1,\qquad \eta_r=0.30,\qquad \beta+\gamma=0.50,\qquad J=0.05.
\label{eq:analysis_params}
\end{equation}
\HLTXT_ORANGE{The values in Eq.~(\ref{eq:analysis_params}) are not intended as a
quantitative fit to a specific laboratory fluid or to a particular
rough-sphere gas state. They form a nondimensional illustrative benchmark set
chosen to place the reduced system in a clear fast-spin regime: the
explicit-spin model and its adiabatically eliminated limit are close, while
the qualitative distinction between the rational eliminated-spin kernel and
finite polynomial surrogates remains visible. Coefficient-level estimates for
dilute rough-sphere gases are discussed in the companion transport-oriented
paper; here the parameters are used only for reduced response-class
benchmarking.}

For the polynomial Model~B we consider two numerical instantiations:
the strict $k^4$ truncation $B^{(4)}$, in which only the leading Burnett-type
term is retained, and the matched $k^6$ truncation $B^{(6)}$, in which the
low-$k$ expansion of the eliminated-spin kernel
[Eq.~(\ref{eq:case2_smallk_expansion})] is matched through $O(k^6)$,
\begin{align}
B^{(4)}:&\qquad B_1=\frac{\beta+\gamma}{4}=0.125,\qquad B_2=0,
\label{eq:analysis_B4}\\
B^{(6)}:&\qquad B_1=\frac{\beta+\gamma}{4}=0.125,\qquad
B_2=-\frac{(\beta+\gamma)^2}{16\eta_r}\simeq -5.21\times 10^{-2}.
\label{eq:analysis_B6}
\end{align}
These are diagnostic surrogate choices, not claims about the unique
coefficients of a separate coefficient-complete Burnett derivation. Their role
is to expose how finite polynomial approximants differ from the rational kernel
of Model~D. Within that polynomial class, the two truncations fail in
qualitatively different ways: $B^{(4)}$ remains linearly stable but becomes
increasingly over-damped at large $k$, whereas $B^{(6)}$ develops near-critical
cancellation and eventual finite-$k$ instability.

\begin{table*}[t]
\centering
\caption{
Notation crosswalk for the benchmark and inference models used below.
The reduced-model surrogates $B^{(4)}$ and $B^{(6)}$ are fixed polynomial closures used in Sec.~\ref{sec:analysis},
whereas $B^{(\mathrm{free})}_{4}$ and $B^{(\mathrm{constr})}_{4}$ are one-pole fit families used only in the multi-$k$
EDMD response analysis of Sec.~\ref{sec:edmd_harmonic_spin} and Appendix~\ref{app:model_selection}.
}
\label{tab:model-notation-crosswalk}
\renewcommand{\arraystretch}{1.22}
\setlength{\tabcolsep}{4pt}
\small
\begin{tabular}{p{0.08\textwidth} p{0.13\textwidth} c p{0.17\textwidth} p{0.18\textwidth} p{0.21\textwidth}}
\hline\hline
Symbol & Status / where used & Poles in $\chi_{\zeta\zeta}$ & Kernel form & Large-$k$ / stability property & Primary role \\
\hline
$C$
& theory-constrained model; Secs.~\ref{sec:translineardiagno} and \ref{sec:edmd_harmonic_spin}
& 2
& explicit-spin two-field response [Eq.~(\ref{eq:delta_micropolar})]
& slow hydrodynamic pole plus fast spin-relaxation pole
& retained-spin reference; identified most cleanly through the spin-to-vorticity ratio $R_{\omega\zeta}$ \\
\hline
$D$
& theory-constrained model; Secs.~\ref{sec:translineardiagno} and \ref{sec:analysis}
& 1
& rational eliminated-spin kernel [Eq.~(\ref{eq:chi_case2})]
& returns to an effective $k^2$ scaling at large $k$
& eliminated-spin target in the reduced benchmarks and the one-field response comparison \\
\hline
$B^{(4)}$
& fixed benchmark surrogate; Sec.~\ref{sec:analysis}
& 1
& strict $k^4$ polynomial [Eq.~(\ref{eq:analysis_B4})]
& linearly stable, but increasingly over-damped at large $k$
& benchmark example of a stable finite polynomial truncation \\
\hline
$B^{(6)}$
& fixed benchmark surrogate; Sec.~\ref{sec:analysis}
& 1
& matched $k^4{+}k^6$ polynomial [Eq.~(\ref{eq:analysis_B6})]
& near-critical cancellation and eventual finite-$k$ instability
& benchmark example of a low-$k$-matched polynomial truncation that fails qualitatively at finite $k$ \\
\hline
$B^{(\mathrm{free})}_{4}$
& fit family; Fig.~\ref{fig:section_VC_multik_response_evidence} and Appendix~\ref{app:model_selection}
& 1
& one-pole polynomial fit with free $B_1$ in $\nu k^2+B_1k^4$
& polynomial tail set by the fitted sign and magnitude of $B_1$
& flexible polynomial competitor to $D$ in the multi-$k$ vorticity-response evidence \\
\hline
$B^{(\mathrm{constr})}_{4}$
& fit family; Fig.~\ref{fig:section_VC_multik_response_evidence} and Appendix~\ref{app:model_selection}
& 1
& same as $B^{(\mathrm{free})}_{4}$, but with $B_1\ge 0$
& explicitly excludes the sign choice that would generate negative large-$k$ damping
& constrained polynomial competitor in the multi-$k$ vorticity-response evidence \\
\hline
$D^{(\mathrm{free})}_{\mathrm{rational}}$
& fit family; Fig.~\ref{fig:section_VC_multik_response_evidence} and Appendix~\ref{app:model_selection}
& 1
& one-pole reference with free damping $K_k$ at each sampled $k$
& no theory-enforced continuum asymptote across the sampled wavenumbers
& information-criterion baseline for the one-field EDMD response fits \\
\hline\hline
\end{tabular}
\end{table*}

\paragraph*{Remark on notation.}
Table~\ref{tab:model-notation-crosswalk} separates two uses of the polynomial notation that play different roles in the paper.
The labels $B^{(4)}$ and $B^{(6)}$ refer to the fixed reduced-model surrogates of Sec.~\ref{sec:analysis}, introduced from the low-$k$ expansion of the eliminated-spin kernel to expose two distinct benchmark failure modes of finite polynomial truncation: stable large-$k$ over-damping and near-critical finite-$k$ instability.
By contrast, $B^{(\mathrm{free})}_{4}$ and $B^{(\mathrm{constr})}_{4}$ are not the same objects: they are one-pole fitting families used later in the multi-$k$ EDMD vorticity-response analysis, where only the $k^4$ polynomial structure is retained and the coefficient is inferred from data rather than fixed by Eq.~(\ref{eq:analysis_B4}) or Eq.~(\ref{eq:analysis_B6}).
Likewise, the symbol $D$ denotes the theory-constrained eliminated-spin kernel used in the benchmark calculations, whereas $D^{(\mathrm{free})}_{\mathrm{rational}}$ is a flexible one-pole reference used only to define the information-criterion baseline in Fig.~\ref{fig:section_VC_multik_response_evidence}.
Stating this separation explicitly prevents the later model-selection labels from being read as alternative names for the reduced-benchmark surrogates.

\subsection{Dispersion relations and Burnett stability}
\begin{figure*}[t]
\centering
\begin{minipage}[t]{0.49\textwidth}
\centering
\includegraphics[width=\linewidth]{\ResFigNoKsix{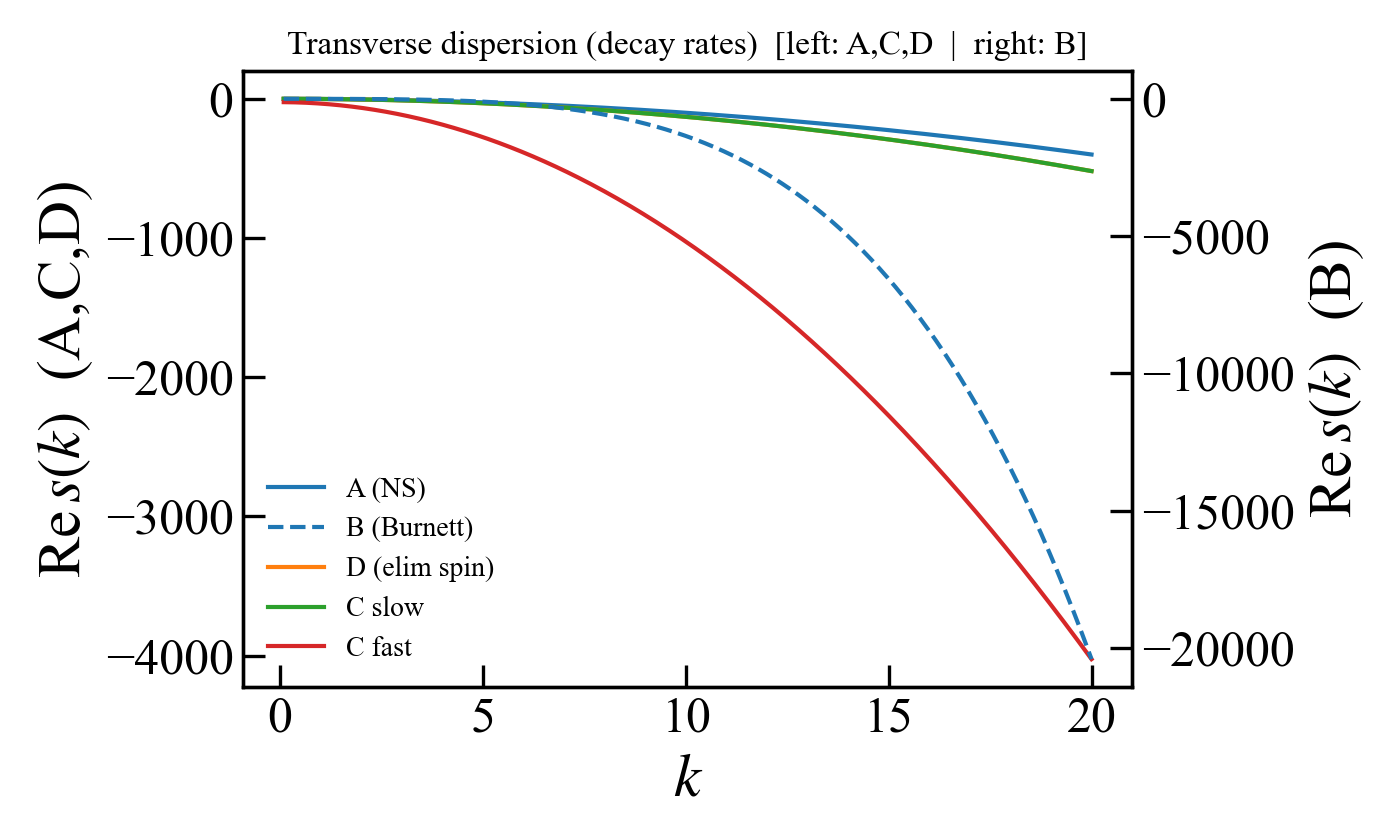}}
\end{minipage}\hfill
\begin{minipage}[t]{0.49\textwidth}
\centering
\includegraphics[width=\linewidth]{\ResFigWithKsix{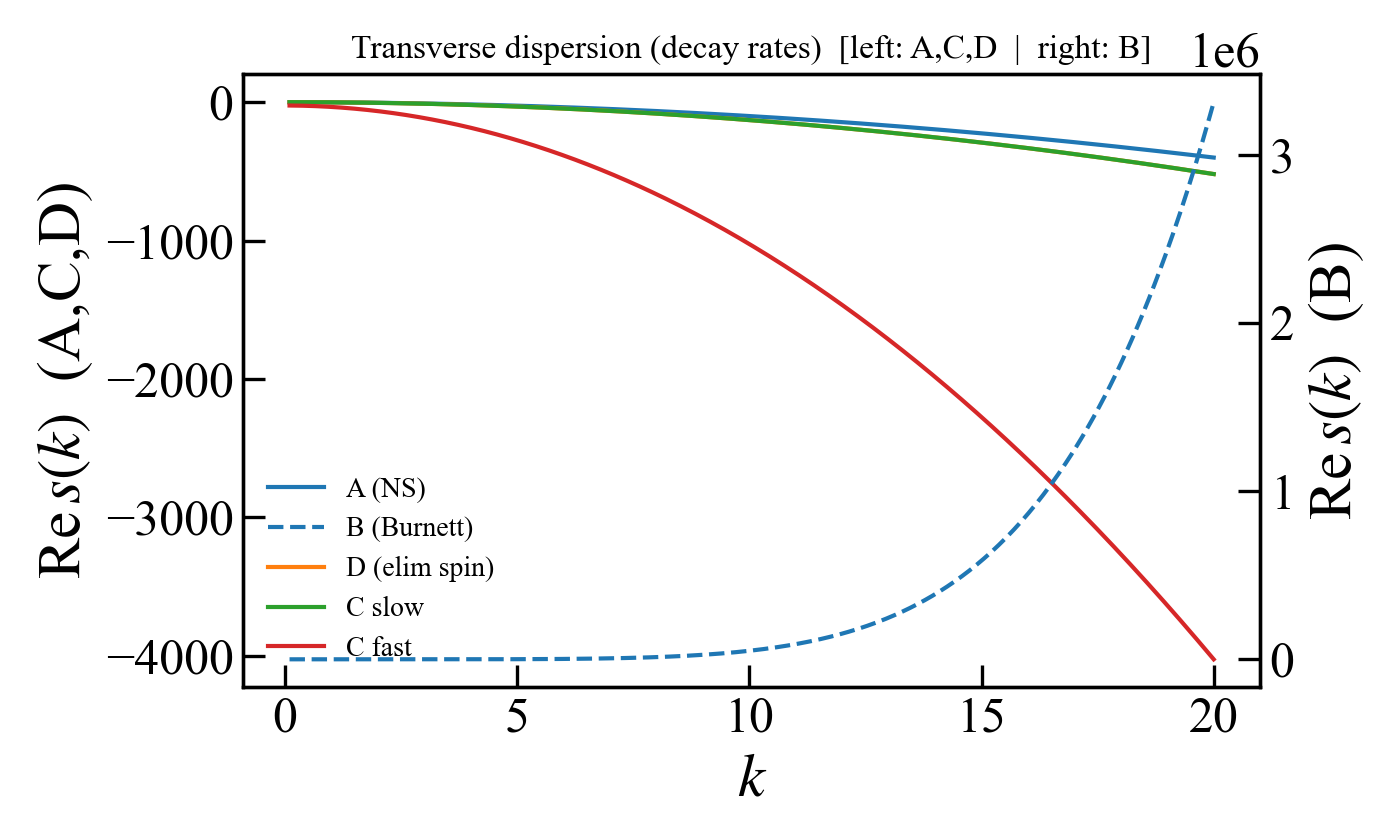}}
\end{minipage}
\caption{
Transverse dispersion relations $s(k)$ (decay/growth rates) for two polynomial surrogate choices.
Left: strict $k^4$ polynomial truncation $B^{(4)}$ (without the $k^6$ term), which remains stable but becomes increasingly
over-damped at large $k$.
Right: matched polynomial truncation $B^{(6)}$ (with the $k^6$ term), for which the damping coefficient changes sign and
Model~B becomes unstable for $k>k_{\mathrm{crit}}$.
In both panels the left axis shows $\mathrm{Re}\,s(k)$ for Models A, C (slow/fast), and D, while Model~B is plotted on a
separate right axis.
}
\label{fig:dispersion}
\end{figure*}

Figure~\ref{fig:dispersion} plots the transverse dispersion branches $s(k)$ (growth/decay rates)
for all models. Model~C (explicit spin) has two branches because the determinant
$\Delta_{\mathrm{MP}}(s,k)$ equation (\ref{eq:delta_micropolar}) is quadratic in $s$.
For the parameter set of Eq.~(\ref{eq:analysis_params}), the fast branch is strongly damped already at $k\!=\!0$,
with $s_{\mathrm{fast}}(0)=-4\eta_r/(\rho J)=-24$, and remains well separated from the hydrodynamic branch.
The slow branch is the hydrodynamic vorticity/shear mode.

The side-by-side comparison in Fig.~\ref{fig:dispersion} shows that the behavior of Model~B depends qualitatively on
whether the $k^6$ term is retained. In the strict truncation $B^{(4)}$, the polynomial damping coefficient
$\eta k^2+B_1k^4$ stays positive, so the model remains linearly stable over the plotted range, but its decay rate
steepens much more rapidly than that of the rational eliminated-spin kernel (D) at large $k$.
In the matched truncation $B^{(6)}$, the negative coefficient $B_2<0$ in Eq.~(\ref{eq:analysis_B6}) causes the damping
coefficient $\eta k^2+B_1k^4+B_2k^6$ to change sign at a finite wavenumber $k_{\mathrm{crit}}$, beyond which
Model~B exhibits a spurious linear instability (positive $\mathrm{Re}\,s$).
In contrast, Model~D remains linearly stable for all $k$ because its kernel is rational and
asymptotically returns to an effective $k^2$-scaling rather than a divergent polynomial truncation.

\begin{figure*}[t]
\centering
\begin{minipage}[t]{0.49\textwidth}
\centering
\includegraphics[width=\linewidth]{\ResFigNoKsix{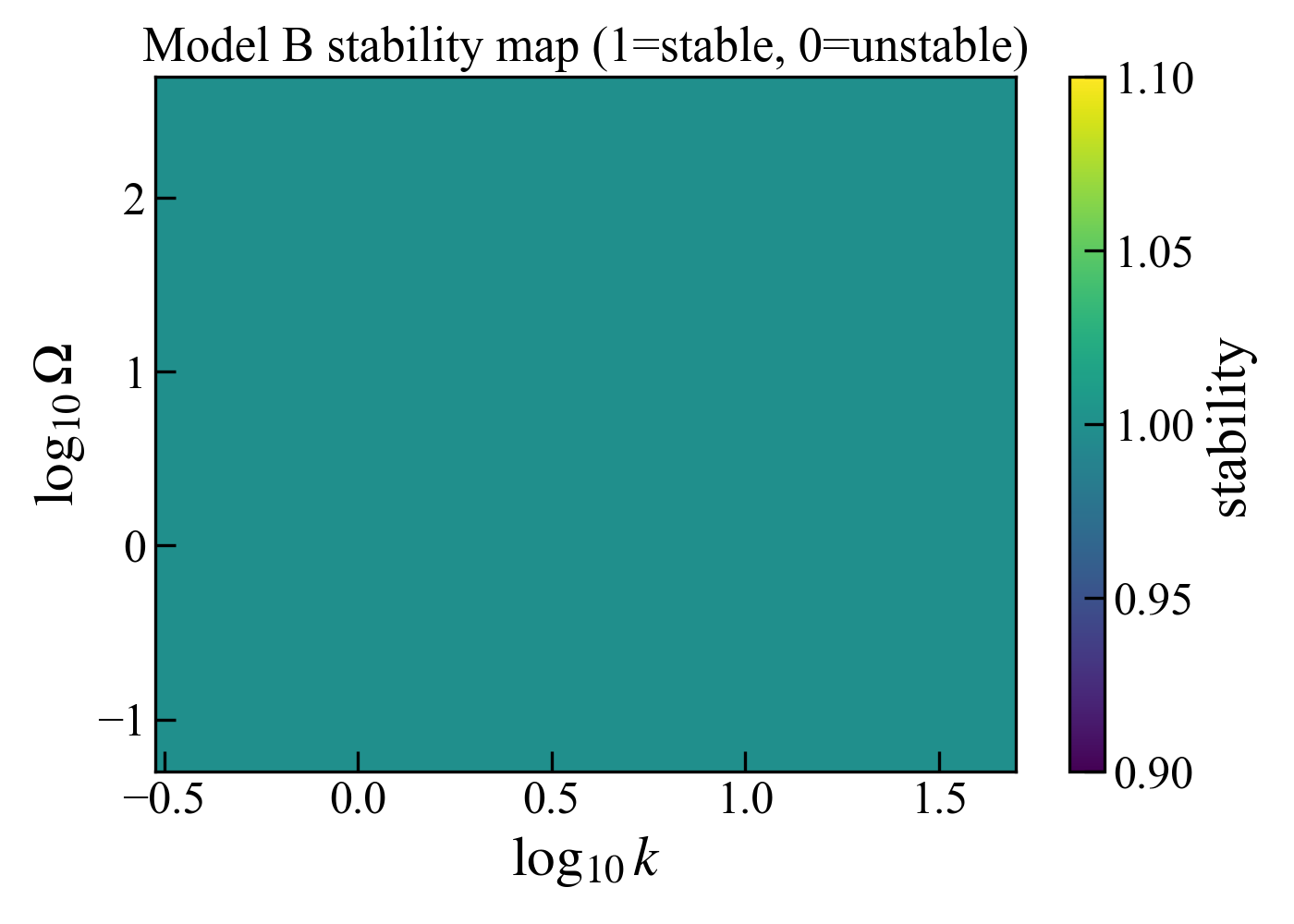}}
\end{minipage}\hfill
\begin{minipage}[t]{0.49\textwidth}
\centering
\includegraphics[width=\linewidth]{\ResFigWithKsix{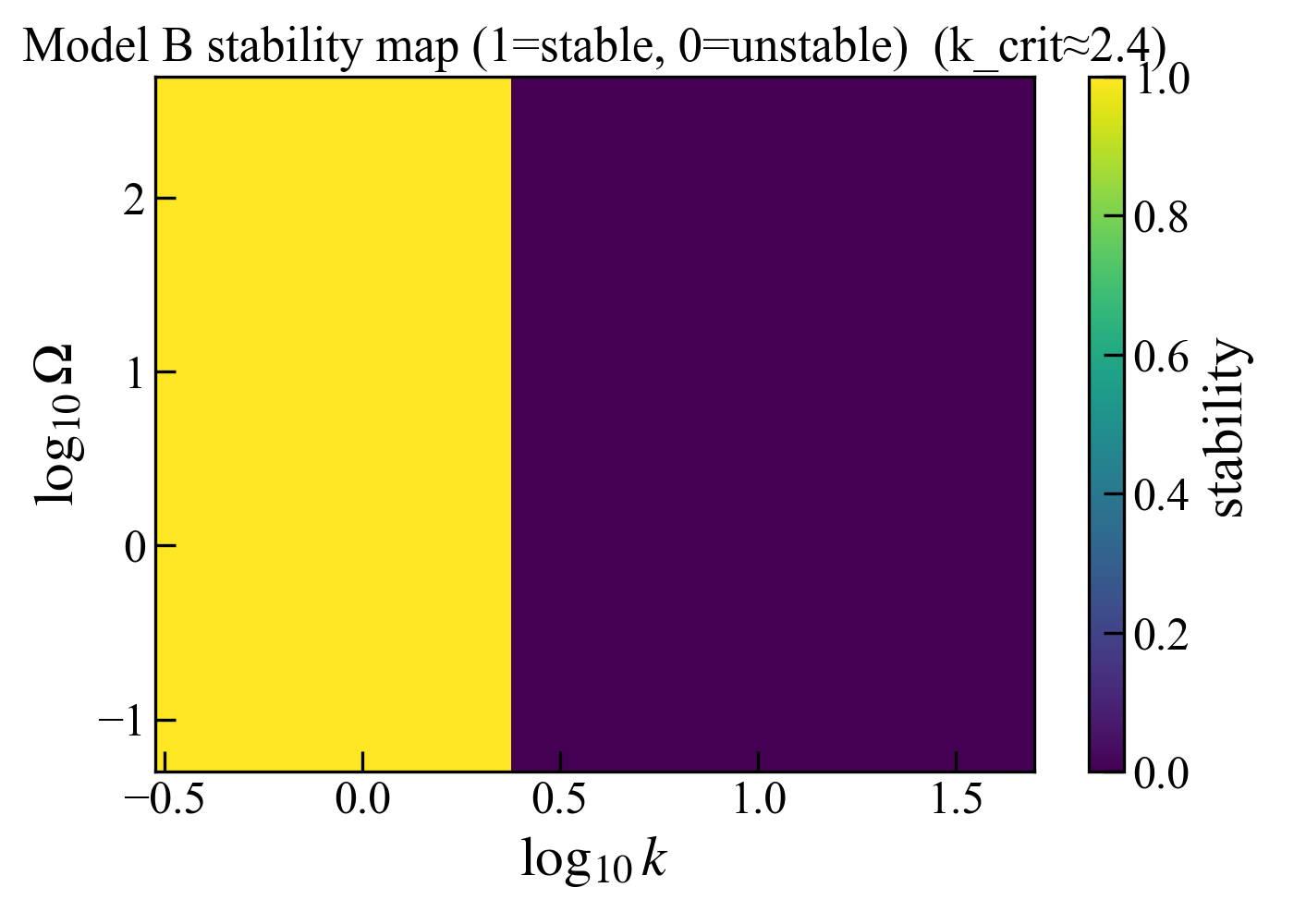}}
\end{minipage}
\caption{
Stability maps for Model~B in the $(k,\Omega)$ plane.
Left: strict $k^4$ polynomial truncation $B^{(4)}$, which is stable everywhere in the scanned domain.
Right: matched truncation $B^{(6)}$, for which stable (decaying) parameter values occur for $k<k_{\mathrm{crit}}$ and
unstable growth occurs for $k>k_{\mathrm{crit}}$.
The boundary is independent of $\Omega$ because stability is controlled by the sign of the static damping coefficient.
}
\label{fig:stability}
\end{figure*}

The corresponding stability maps are shown explicitly in Fig.~\ref{fig:stability}.
For the strict truncation $B^{(4)}$, stability holds throughout the scanned $(k,\Omega)$ domain.
For the matched truncation $B^{(6)}$, stability depends only on the sign of the real damping coefficient,
so the stable/unstable partition is independent of forcing frequency $\Omega$ (the vertical axis).
For Eq.~(\ref{eq:analysis_params}) and the matched coefficients Eq.~(\ref{eq:analysis_B6}) we find
\begin{equation}
k_{\mathrm{crit}}\simeq 2.396,
\label{eq:kcrit_value}
\end{equation}
consistent with the root of $\eta+B_1k^2+B_2k^4=0$ (in terms of $k^2$).

\HLTXT_BLUE{The finite-$k$ instability of the matched $k^6$ truncation should
be viewed in the broader context of the known Burnett-instability problem and
its regularization literature. In the present work, however, $B^{(6)}$ is not
intended as a coefficient-complete Burnett equation for the microscopic
rough-sphere model. It is a low-$k$-matched diagnostic polynomial surrogate.
Its instability is used to expose a structural failure mode that can arise when
the rational eliminated-spin kernel is replaced by a finite polynomial in
$k$.}

\HLTXT_ORANGE{We do not claim that this finite-$k$ instability of the reduced
$B^{(6)}$ benchmark is directly observed in the EDMD datasets of
Sec.~\ref{sec:edmd_benchmarks}. In the present paper, $B^{(6)}$ serves as a
reduced-model diagnostic example showing how a low-$k$-matched finite
polynomial can fail qualitatively, whereas the EDMD section addresses
observability and model discrimination in the stable sampled response window.}

\paragraph*{Why the rational kernel and a finite polynomial truncation are not asymptotically equivalent.}
Introduce the Model~D damping kernel
\begin{equation}
K_D(k)
:=
\eta k^2+\frac{\eta_r(\beta+\gamma)k^4}{4\eta_r+(\beta+\gamma)k^2}.
\label{eq:KD_def}
\end{equation}
Then
\begin{equation}
K_D(k)
=
(\eta+\eta_r)k^2
-\frac{4\eta_r^2 k^2}{4\eta_r+(\beta+\gamma)k^2}
=
(\eta+\eta_r)k^2
-\frac{4\eta_r^2}{\beta+\gamma}
+O(k^{-2}),
\qquad k\to\infty.
\label{eq:KD_largek}
\end{equation}
Thus the eliminated-spin closure returns to an effective $k^2$ scaling at
large wavenumber, with only a subleading constant shift.
By contrast, any finite polynomial truncation
\begin{equation}
P_N(k^2)=\eta k^2+\sum_{m=2}^{N} a_m k^{2m}
\label{eq:PN_def}
\end{equation}
satisfies $P_N(k^2)\sim a_N k^{2N}$ as $k\to\infty$.
If $a_N>0$, the damping grows faster than $K_D(k)$ and the model becomes
asymptotically over-damped; if $a_N<0$, then $P_N(k^2)$ is eventually
negative and the corresponding one-pole closure becomes unstable at
sufficiently large $k$.
Hence no finite polynomial truncation can match both the low-$k$ expansion
and the large-$k$ structure of the eliminated-spin kernel.
The stable but over-damped behavior of $B^{(4)}$ and the finite-$k$
instability of $B^{(6)}$ are therefore representative consequences of
approximating a rational kernel by a finite polynomial.

\subsection{Setting 1: free decay and instantaneous decay rate}

\begin{figure*}[t]
\centering
\includegraphics[width=0.315\linewidth]{\ResFigNoKsix{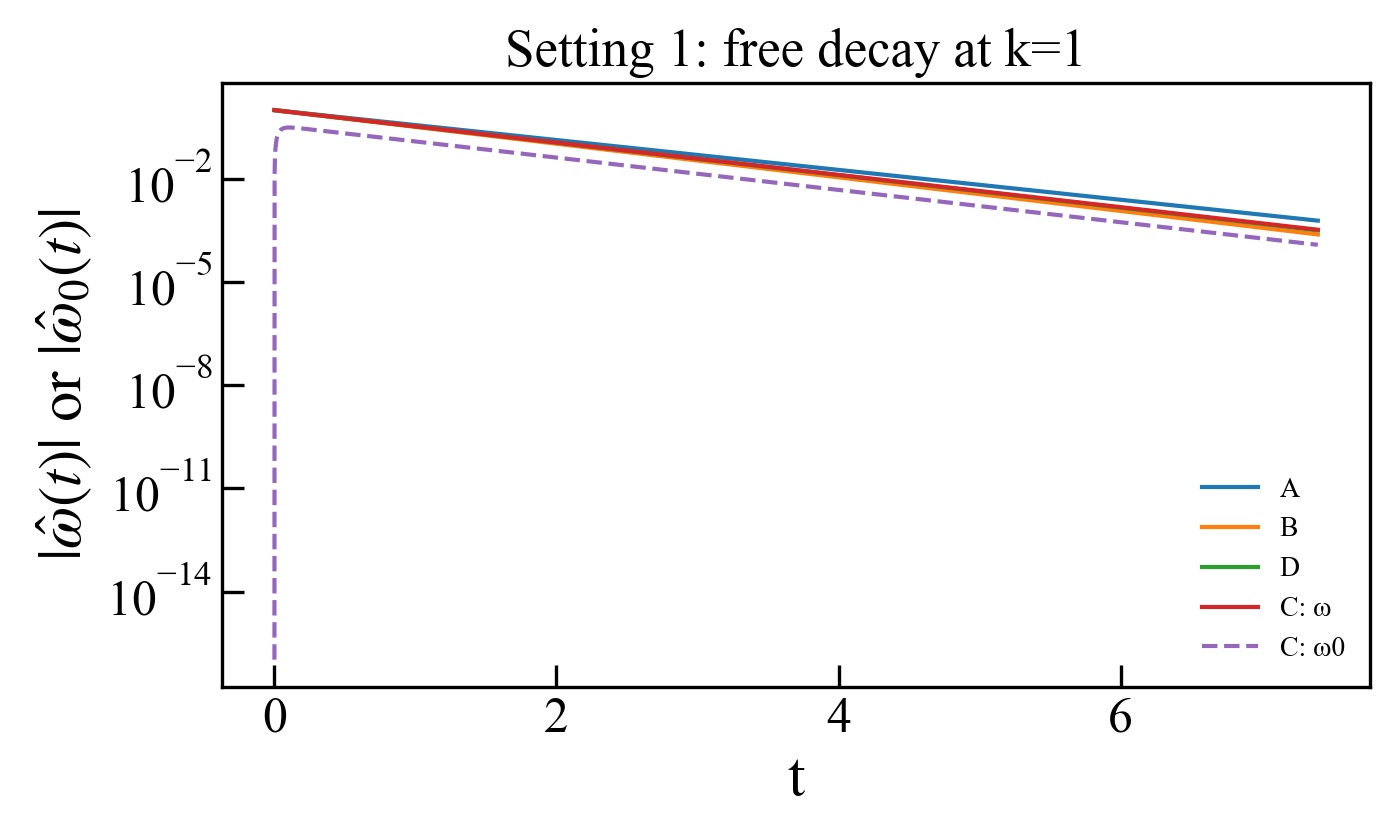}}\hfill
\includegraphics[width=0.315\linewidth]{\ResFigNoKsix{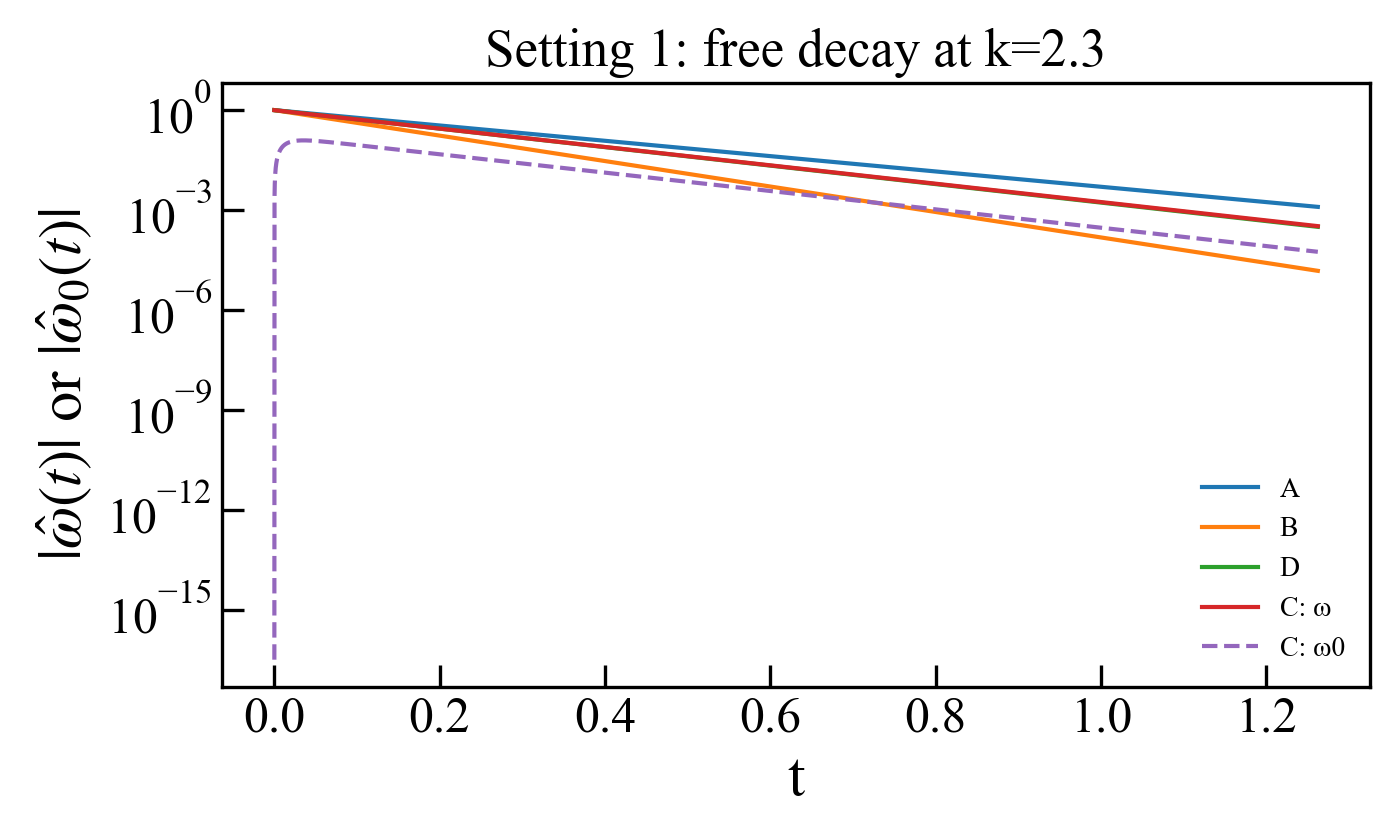}}\hfill
\includegraphics[width=0.315\linewidth]{\ResFigNoKsix{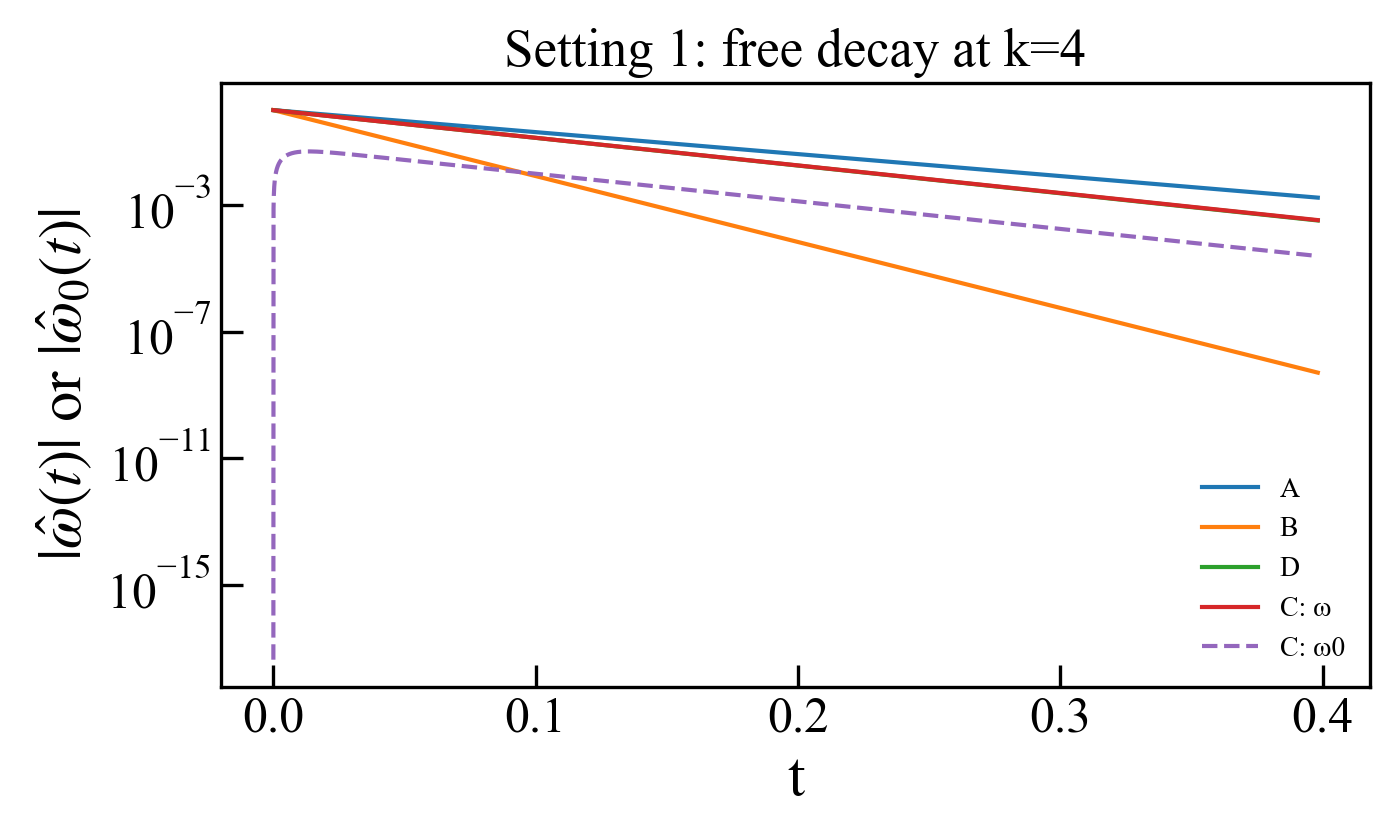}}\\[1.2ex]
\includegraphics[width=0.315\linewidth]{\ResFigWithKsix{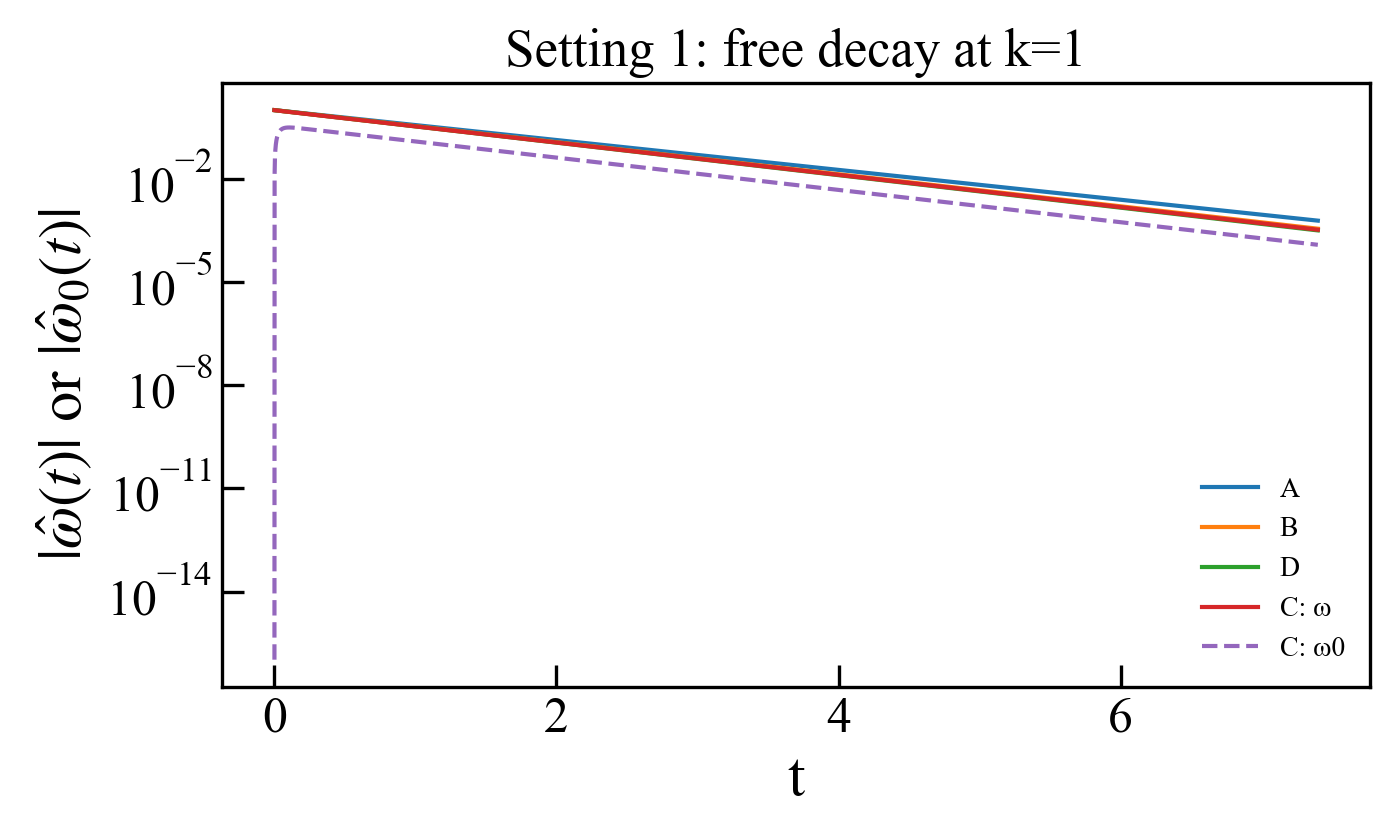}}\hfill
\includegraphics[width=0.315\linewidth]{\ResFigWithKsix{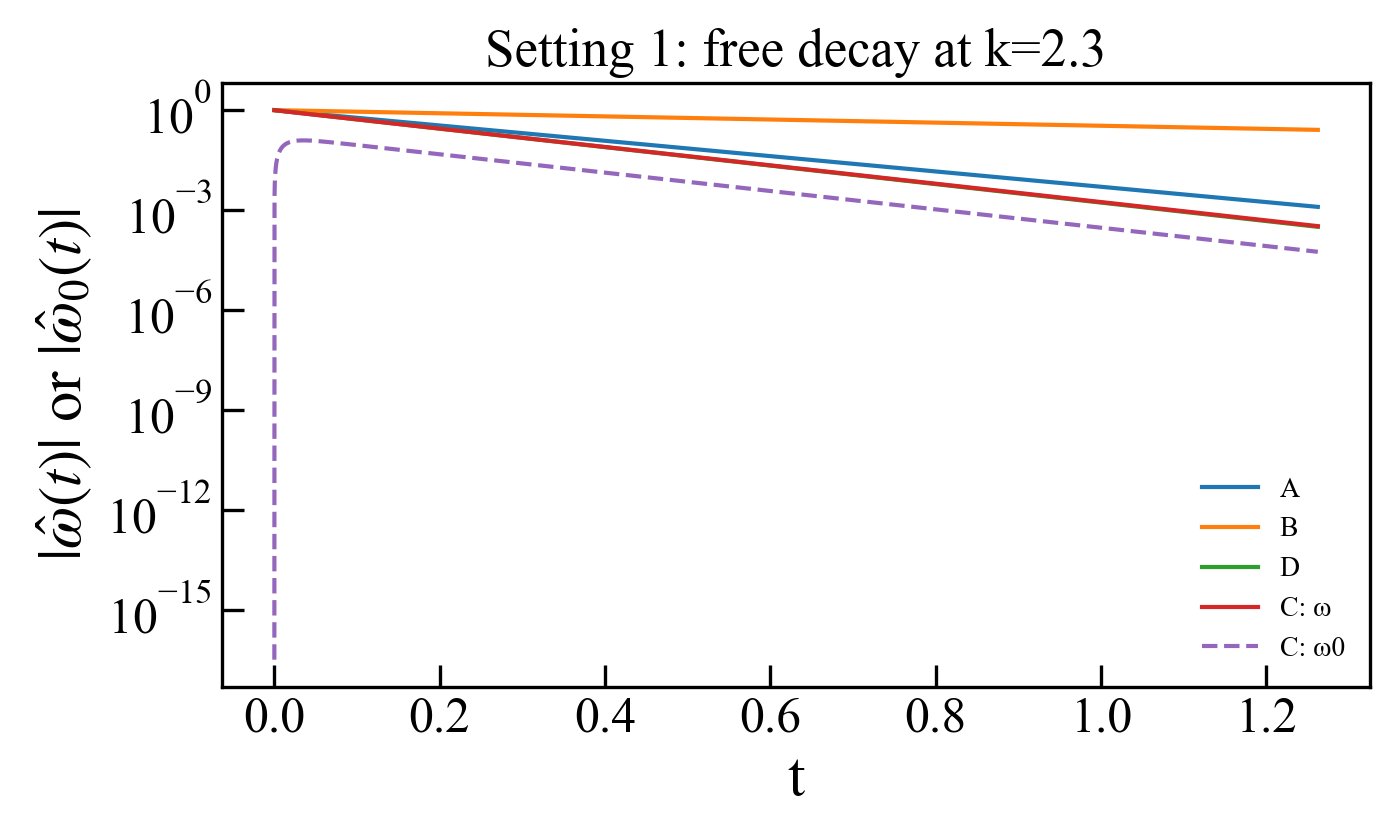}}\hfill
\includegraphics[width=0.315\linewidth]{\ResFigWithKsix{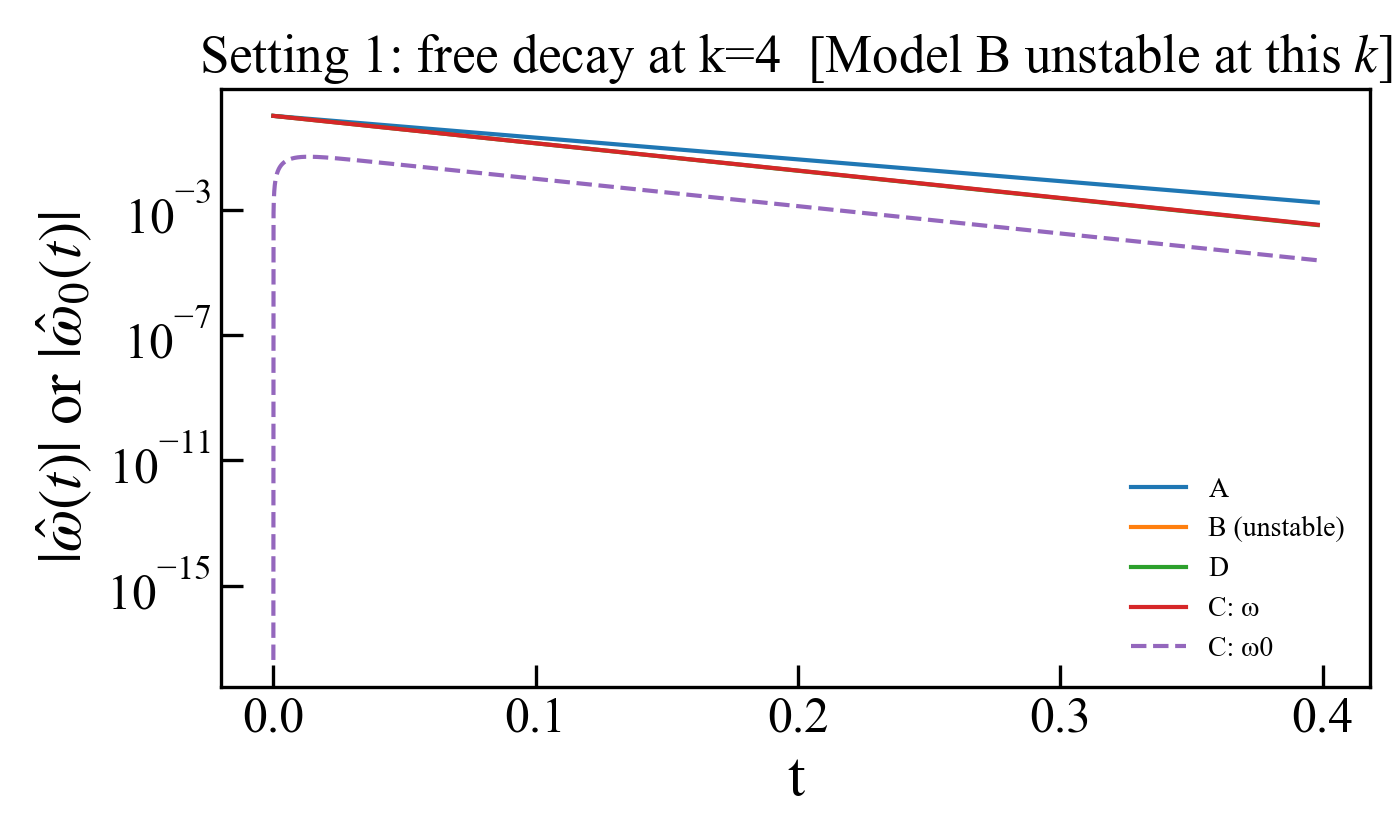}}
\caption{
Setting~1 (free decay): time evolution of the transverse vorticity mode amplitude $|\zeta(t;k)|$.
Columns correspond to $k=1$, $k=2.3\simeq 0.96\,k_{\mathrm{crit}}$, and $k=4>k_{\mathrm{crit}}$ (for $B^{(6)}$).
Top row: strict $k^4$ polynomial truncation $B^{(4)}$ (without the $k^6$ term).
Bottom row: matched truncation $B^{(6)}$ (with the $k^6$ term).
The top row stays stable but becomes increasingly over-damped at large $k$, whereas the bottom row becomes weakly damped
near $k_{\mathrm{crit}}$ and unstable beyond it.
Model~C exhibits a short fast-spin transient and then follows Model~D closely in both cases.
}
\label{fig:setting1_decay}
\end{figure*}

We first consider free decay (Setting~1), with initial conditions
\begin{equation}
\vortmode(0;k)=1,\qquad \omega_0(0;k)=0,\qquad f=g=0,
\label{eq:setting1_ic}
\end{equation}
and track the mode amplitude $|\zeta(t;k)|$.
Figure~\ref{fig:setting1_decay} shows three representative wavenumbers:
$k=1$ (deep in the stable low-$k$ regime), $k=2.3$ (just below $k_{\mathrm{crit}}$ for $B^{(6)}$), and $k=4$
(beyond $k_{\mathrm{crit}}$ for $B^{(6)}$).
The top row uses the strict truncation $B^{(4)}$, and the bottom row uses the matched truncation $B^{(6)}$.

Two signatures are immediately visible.
First, in both rows Model~C (explicit spin) exhibits a short initial transient associated with the fast pole,
after which the decay closely follows Model~D.
Second, the two polynomial surrogate variants fail in qualitatively different ways:
$B^{(4)}$ remains stable but becomes increasingly over-damped as $k$ grows,
whereas $B^{(6)}$ slows dramatically as $k$ approaches $k_{\mathrm{crit}}$ and becomes unstable for $k>k_{\mathrm{crit}}$.

\begin{figure*}[t]
\centering
\includegraphics[width=0.315\linewidth]{\ResFigNoKsix{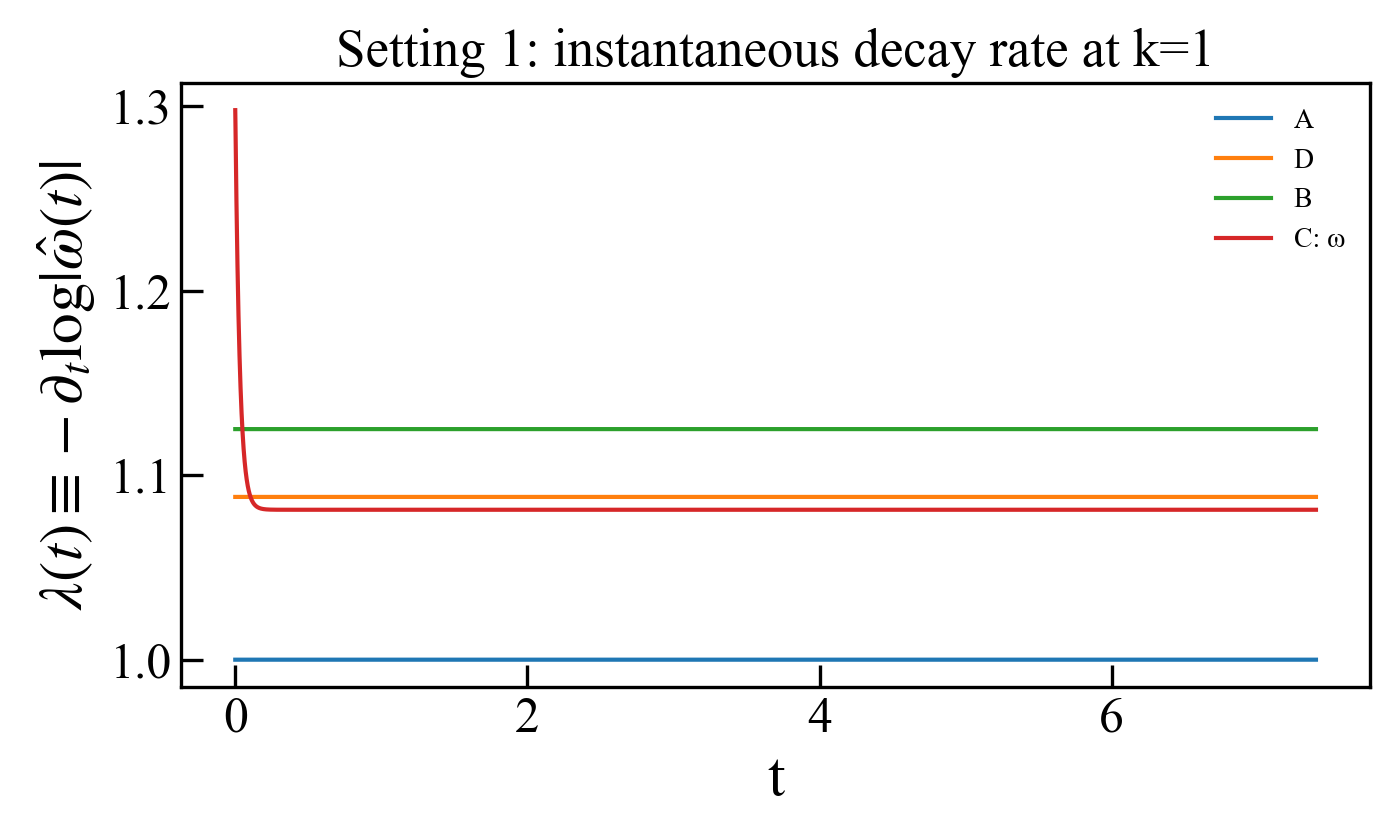}}\hfill
\includegraphics[width=0.315\linewidth]{\ResFigNoKsix{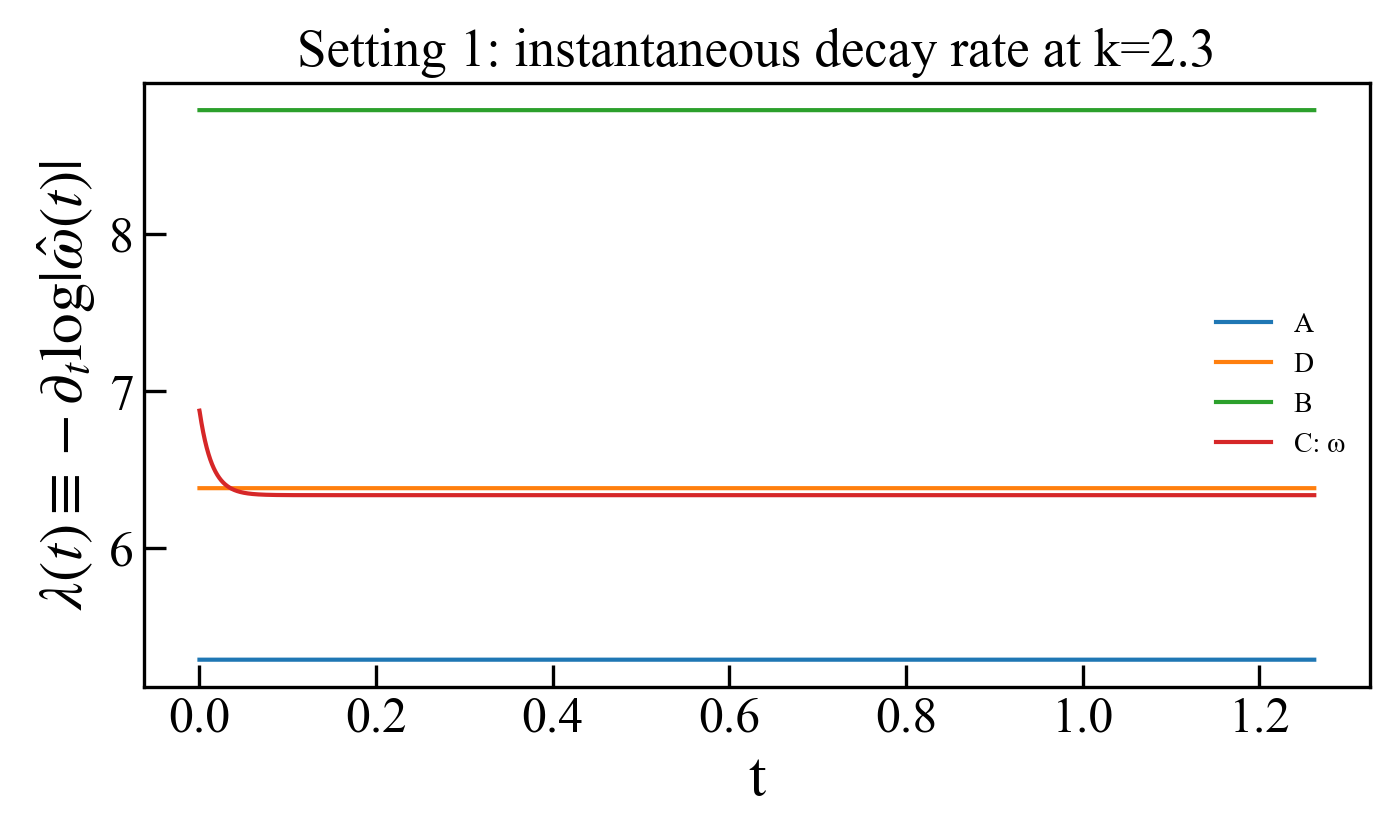}}\hfill
\includegraphics[width=0.315\linewidth]{\ResFigNoKsix{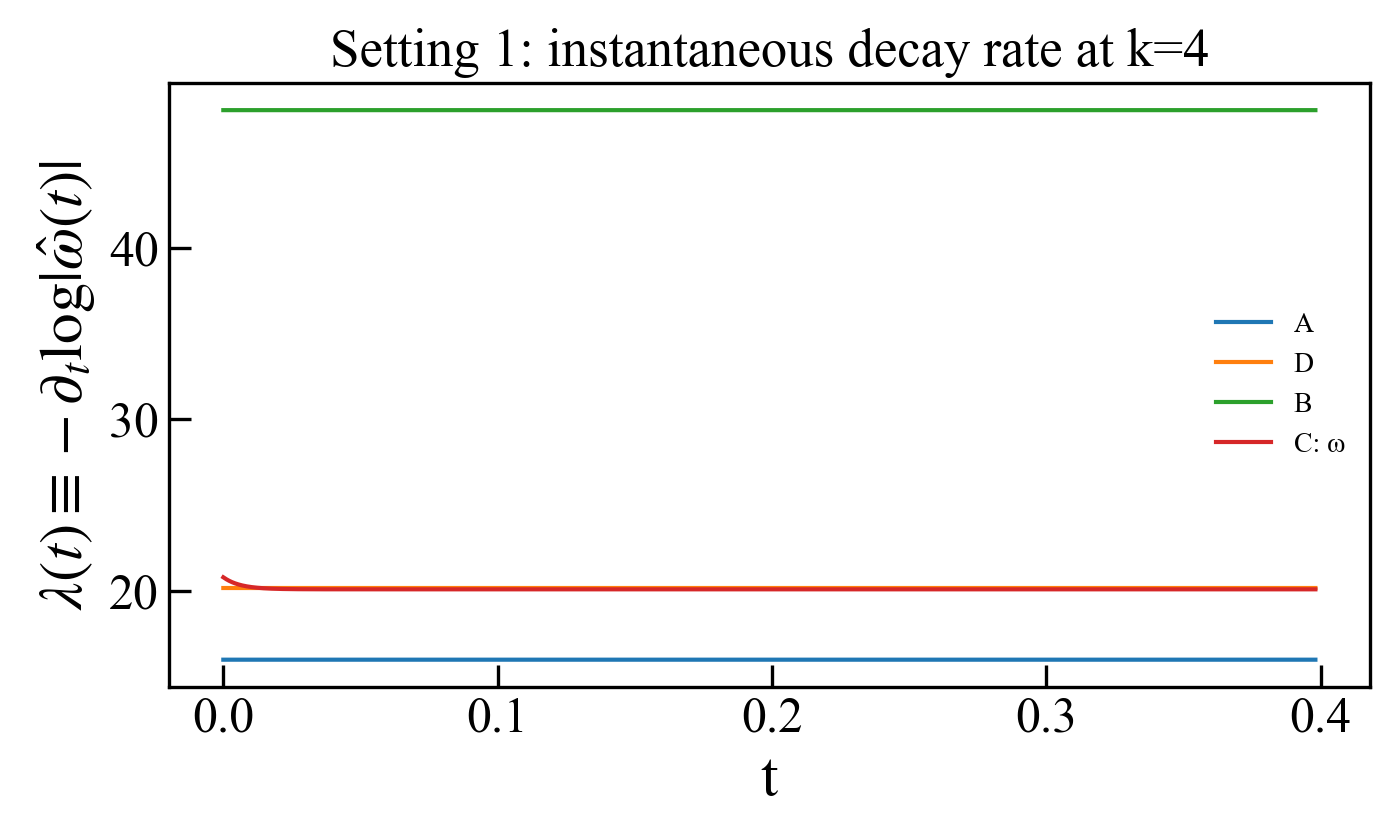}}\\[1.2ex]
\includegraphics[width=0.315\linewidth]{\ResFigWithKsix{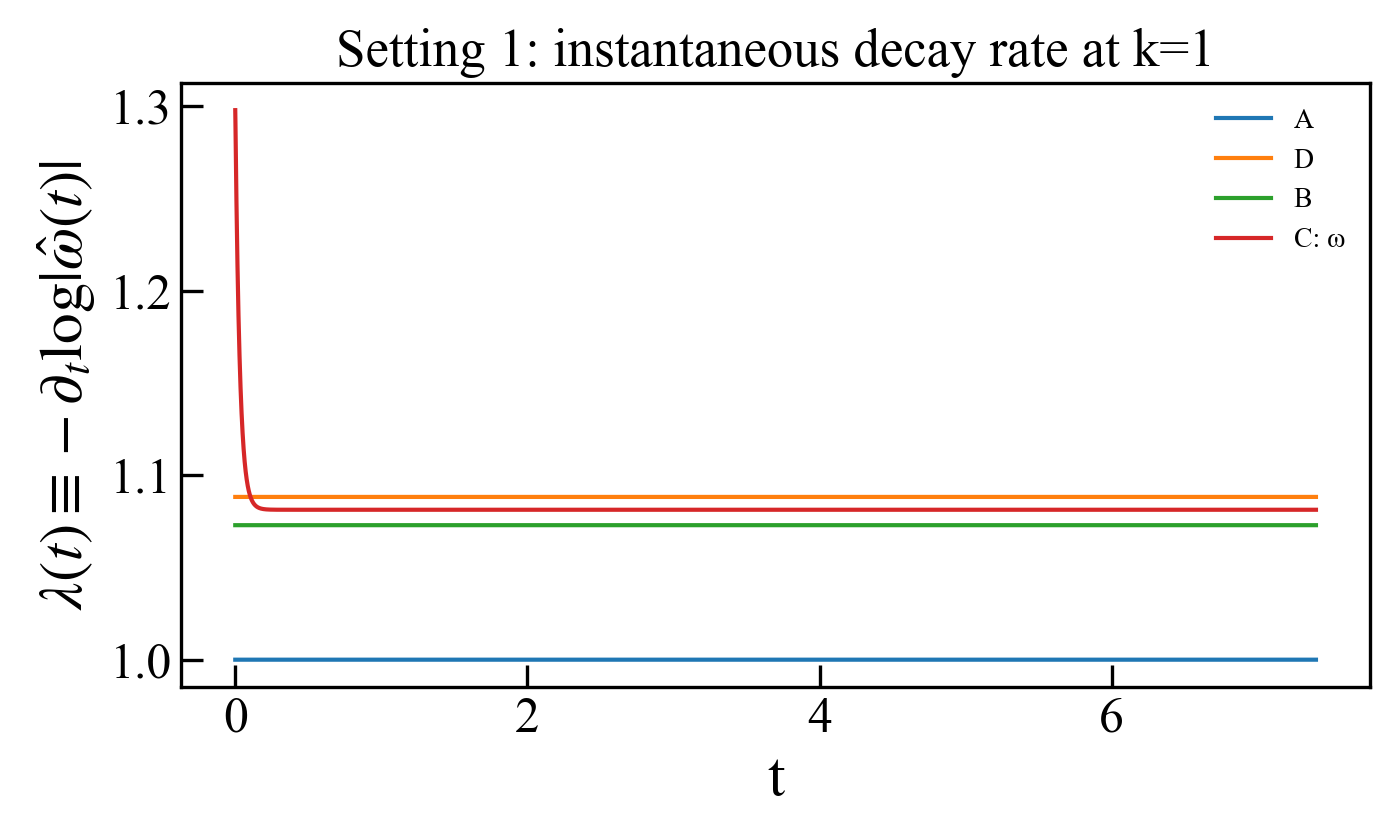}}\hfill
\includegraphics[width=0.315\linewidth]{\ResFigWithKsix{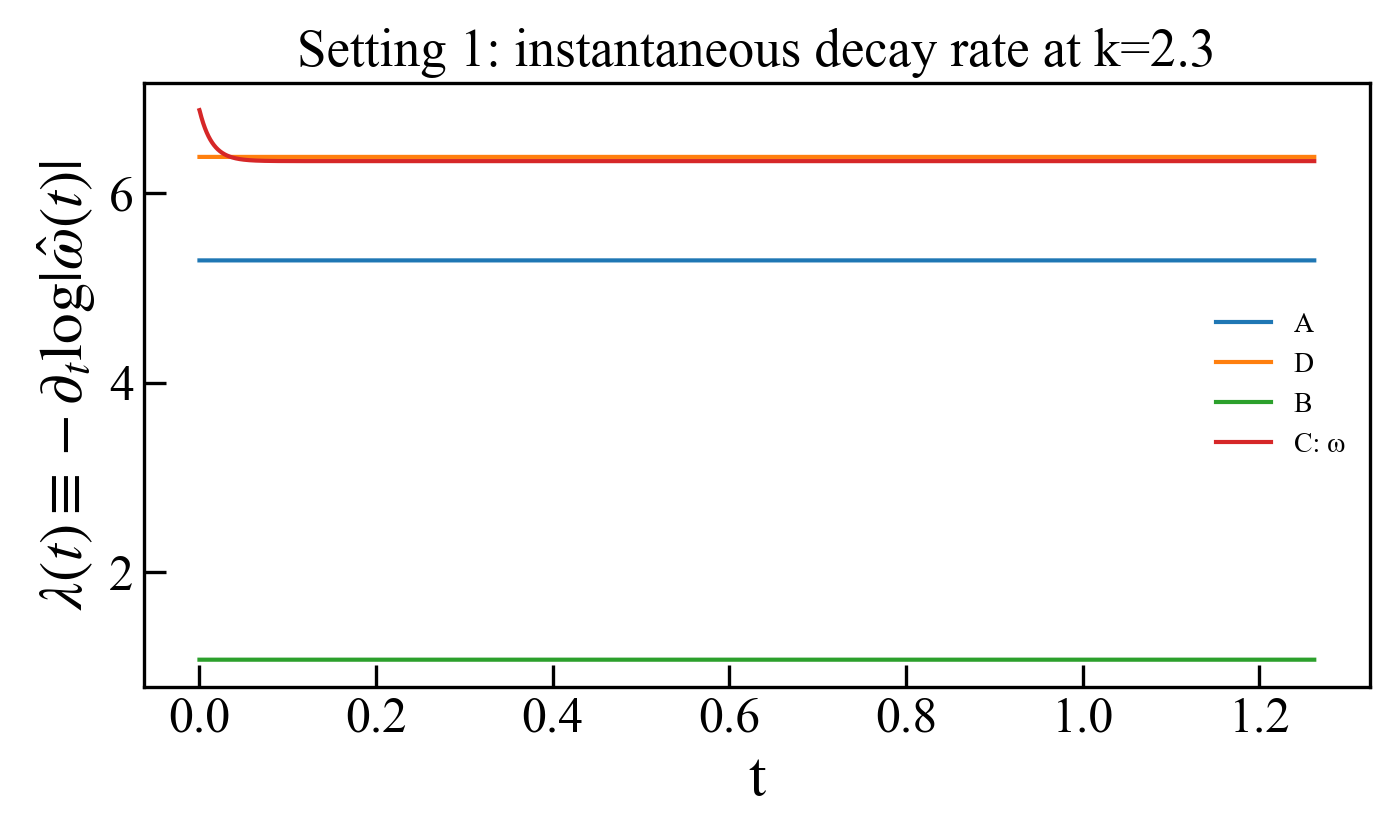}}\hfill
\includegraphics[width=0.315\linewidth]{\ResFigWithKsix{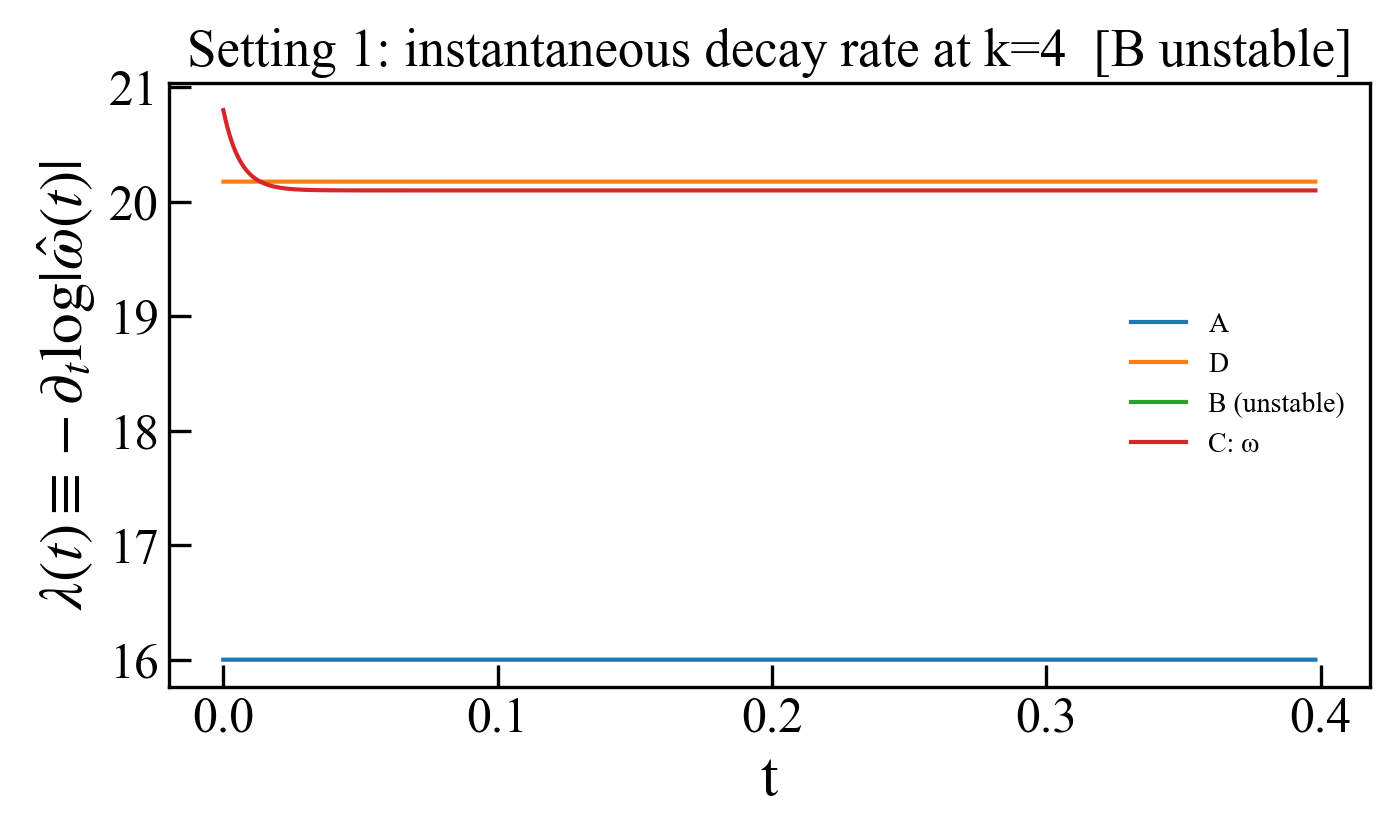}}
\caption{
Setting~1: instantaneous decay rate $\lambda(t;k)$ defined in Eq.~(\ref{eq:inst_decay_rate}).
Columns correspond to the same $k$ values as in Fig.~\ref{fig:setting1_decay}.
Top row: $B^{(4)}$.
Bottom row: $B^{(6)}$.
A and D show the expected one-pole behavior ($\lambda$ essentially constant), while Model~C shows a short-time transient
associated with the fast pole and then converges to the slow rate.
The contrast between the two polynomial surrogate variants is again clear: stable but increasingly rapid damping for $B^{(4)}$ versus
near-critical slowing and loss of stability for $B^{(6)}$.
}
\label{fig:setting1_rate}
\end{figure*}

To quantify the ``curvature'' of decay and make multi-pole behavior visible in the time domain,
we also plot the instantaneous decay rate
\begin{equation}
\lambda(t;k)\equiv -\partial_t \ln |\zeta(t;k)|.
\label{eq:inst_decay_rate}
\end{equation}
For a true one-pole model, $\lambda(t;k)$ is constant (pure exponential decay).
Figure~\ref{fig:setting1_rate} shows that Models A and D have essentially constant $\lambda$,
whereas Model~C exhibits a short-time variation in $\lambda$ before converging to the slow-branch rate.
For $B^{(4)}$, the decay rate remains positive but grows strongly with $k$, reflecting increasingly rapid single-exponential damping.
For $B^{(6)}$, the rate becomes anomalously small near $k_{\mathrm{crit}}$ before stability is lost. 

\subsection{Setting 2: harmonic forcing and frequency response}

\begin{figure*}[t]
\centering
\includegraphics[width=0.315\linewidth]{\ResFigNoKsix{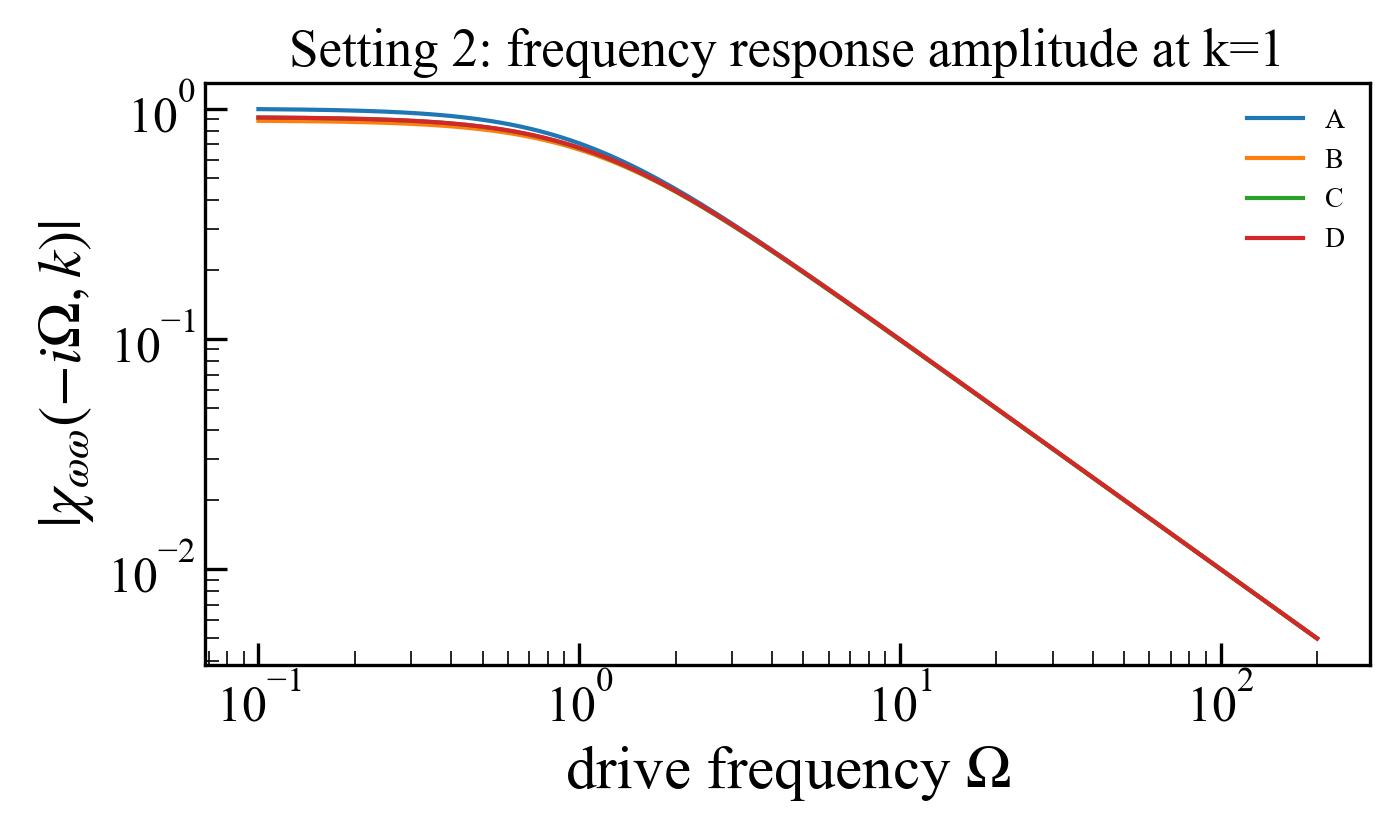}}\hfill
\includegraphics[width=0.315\linewidth]{\ResFigNoKsix{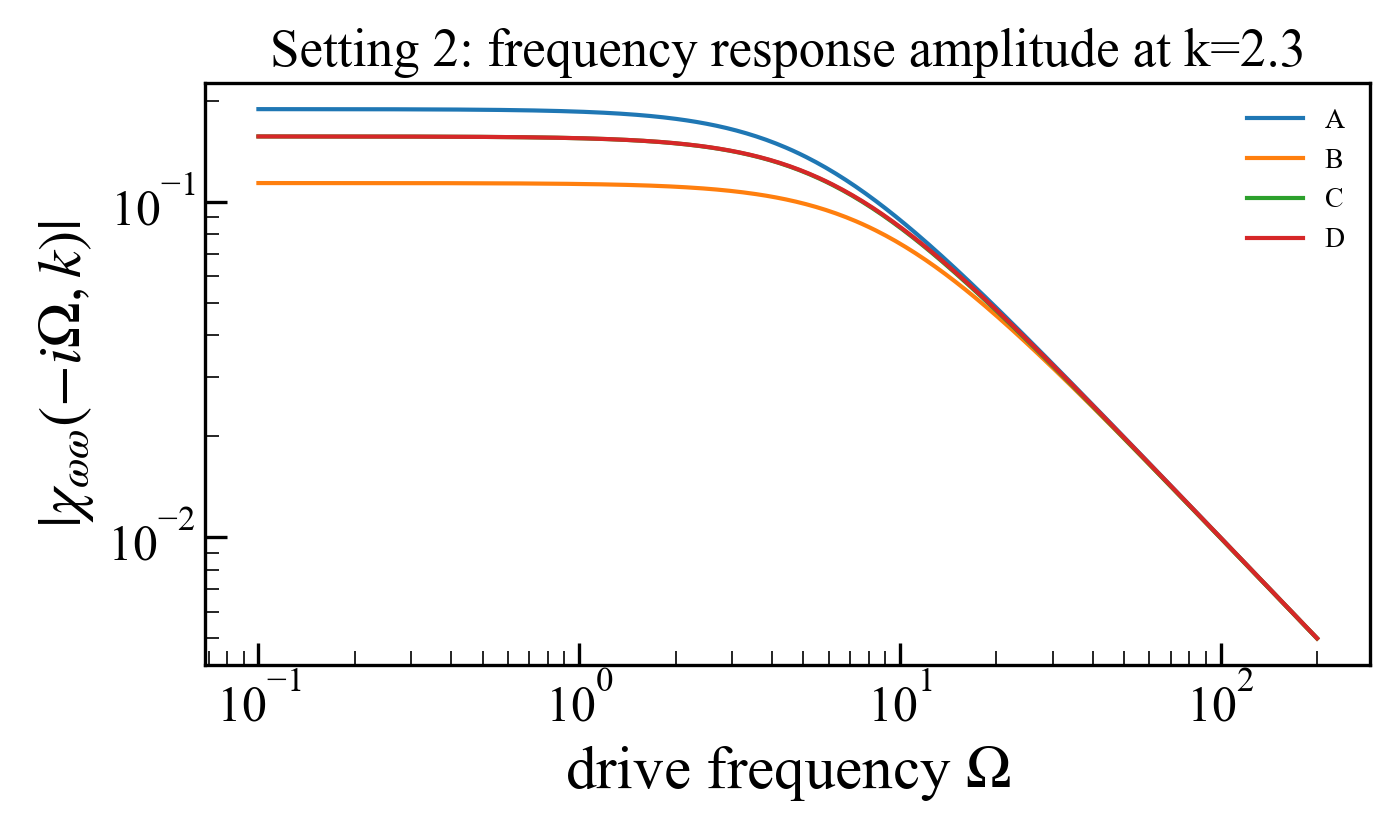}}\hfill
\includegraphics[width=0.315\linewidth]{\ResFigNoKsix{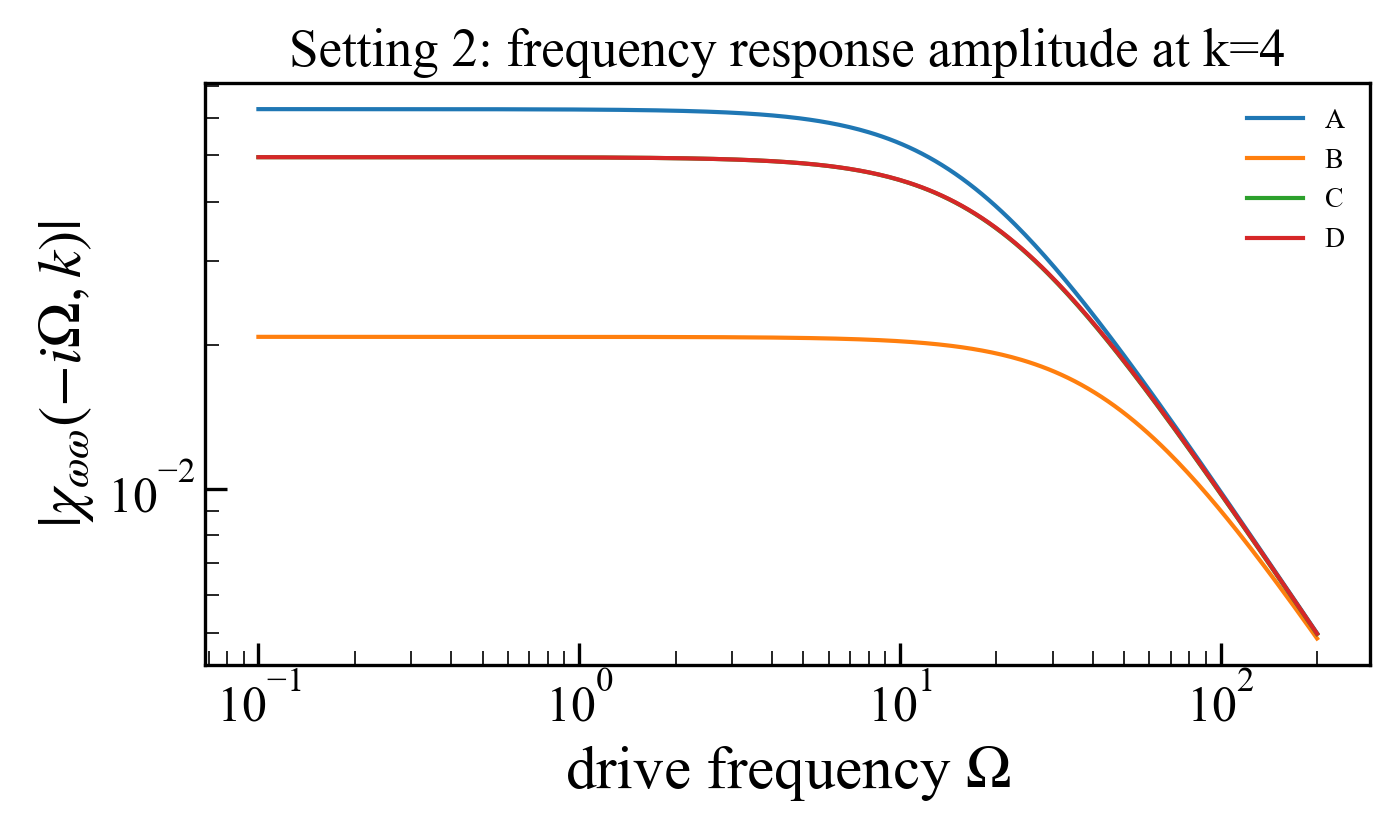}}\\[1.2ex]
\includegraphics[width=0.315\linewidth]{\ResFigWithKsix{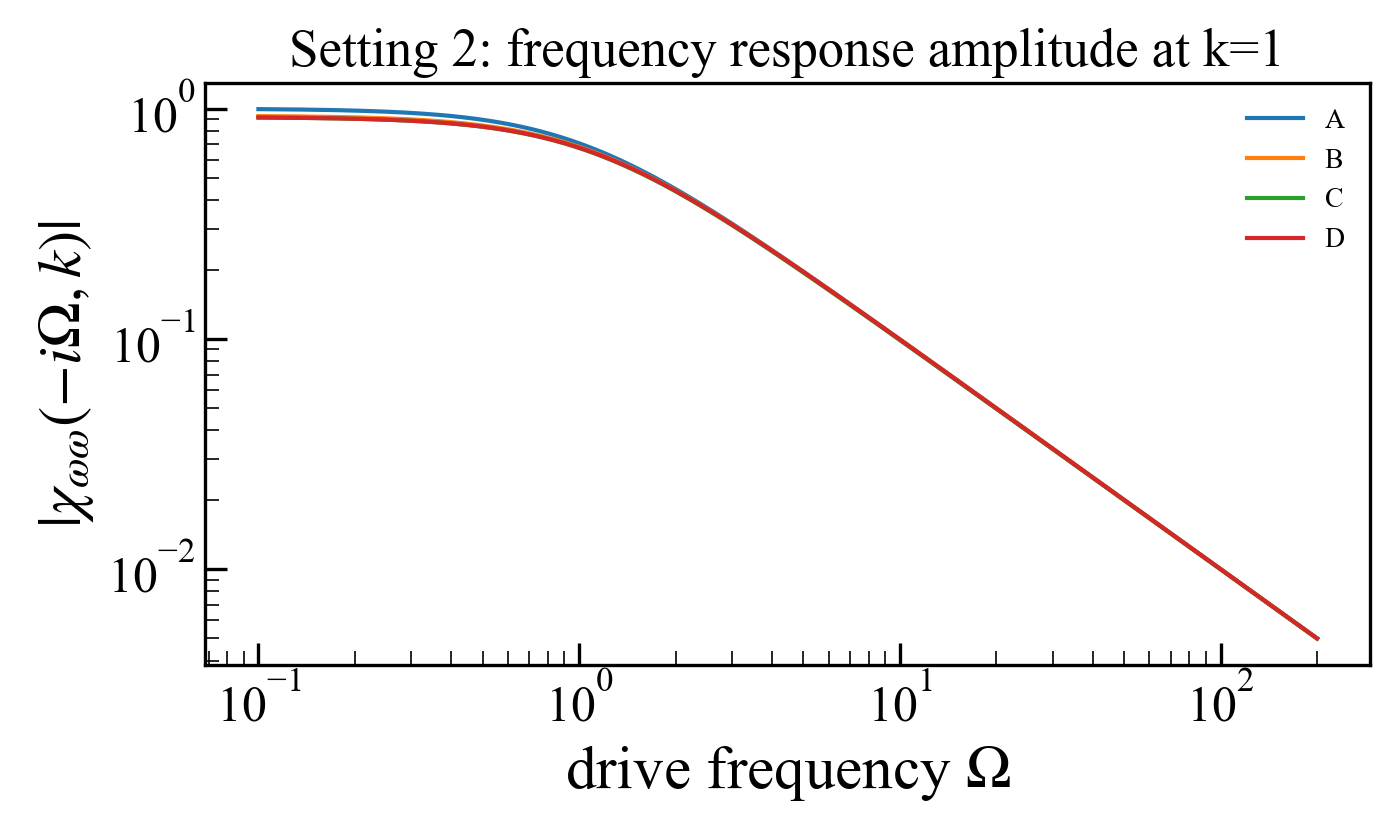}}\hfill
\includegraphics[width=0.315\linewidth]{\ResFigWithKsix{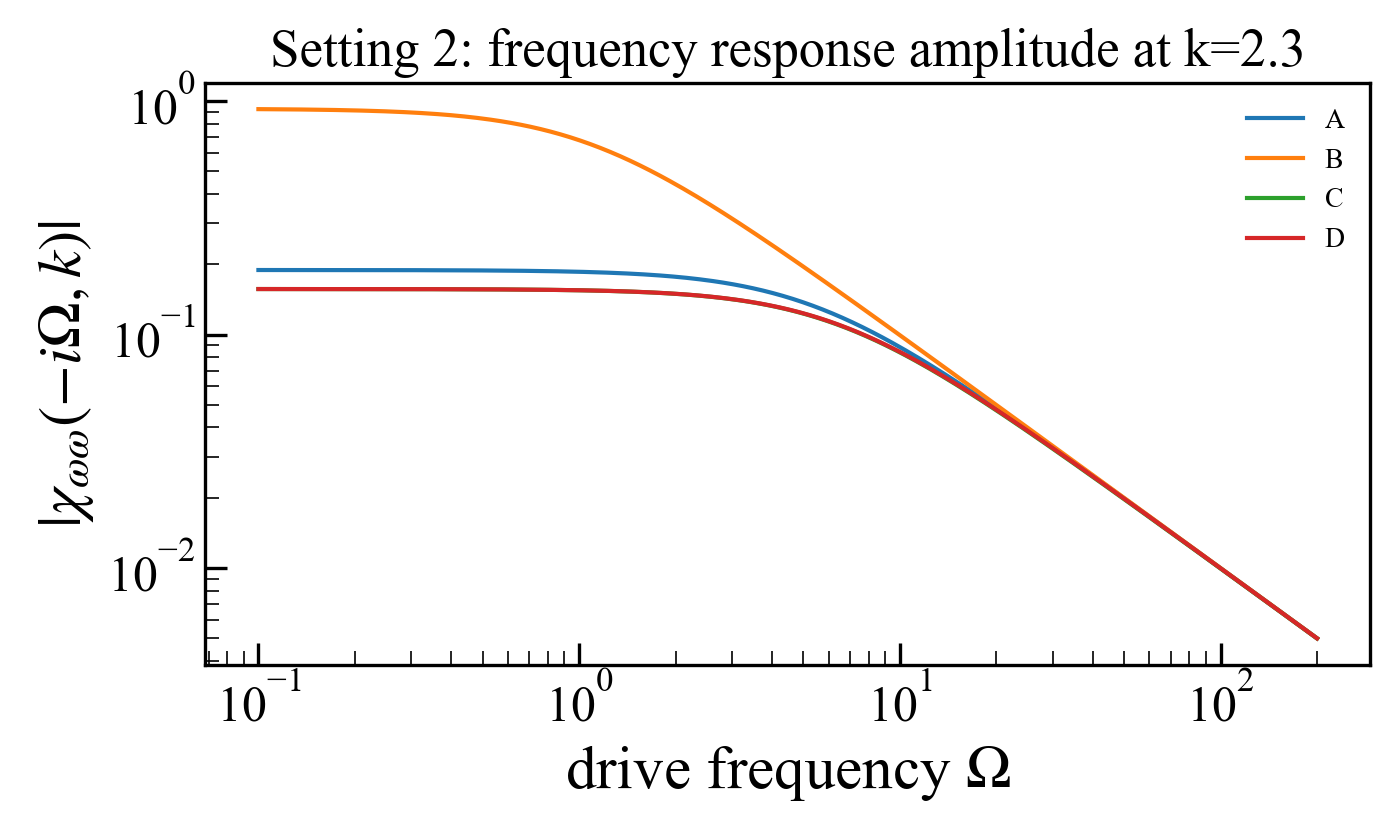}}\hfill
\includegraphics[width=0.315\linewidth]{\ResFigWithKsix{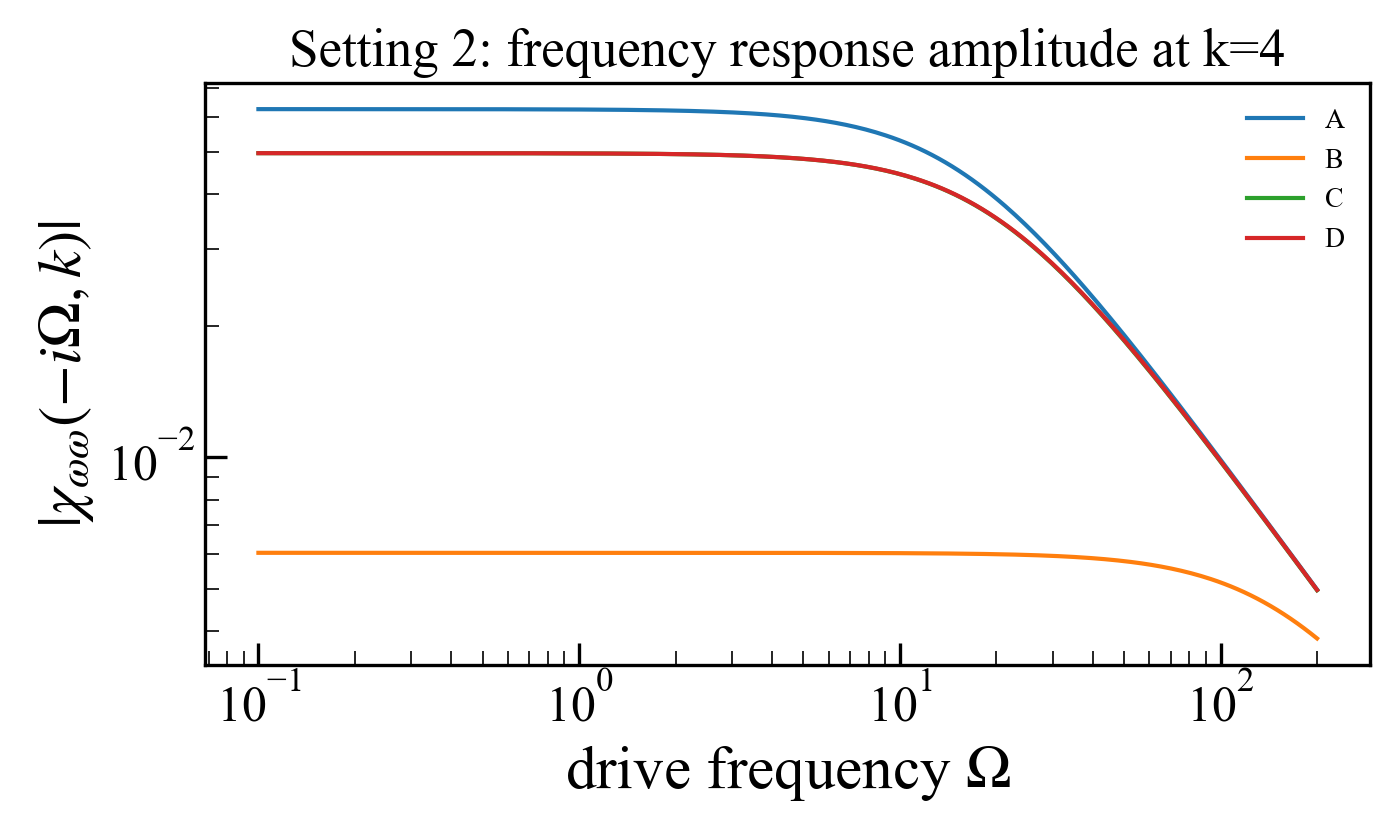}}
\caption{
Setting~2: Bode amplitude $|\chi_{\zeta\zeta}(-i\Omega,k)|$.
Columns correspond to $k=1$, $k=2.3\simeq 0.96\,k_{\mathrm{crit}}$, and $k=4>k_{\mathrm{crit}}$ (for $B^{(6)}$).
Top row: $B^{(4)}$.
Bottom row: $B^{(6)}$.
The strict truncation remains stable but becomes increasingly suppressed at large $k$, while the matched truncation shows the
largest diagnostic separation near $k_{\mathrm{crit}}$ at low $\Omega$ and ceases to admit a physical steady response for $k>k_{\mathrm{crit}}$.
}
\label{fig:bode_amp}
\end{figure*}

\begin{figure*}[t]
\centering
\includegraphics[width=0.315\linewidth]{\ResFigNoKsix{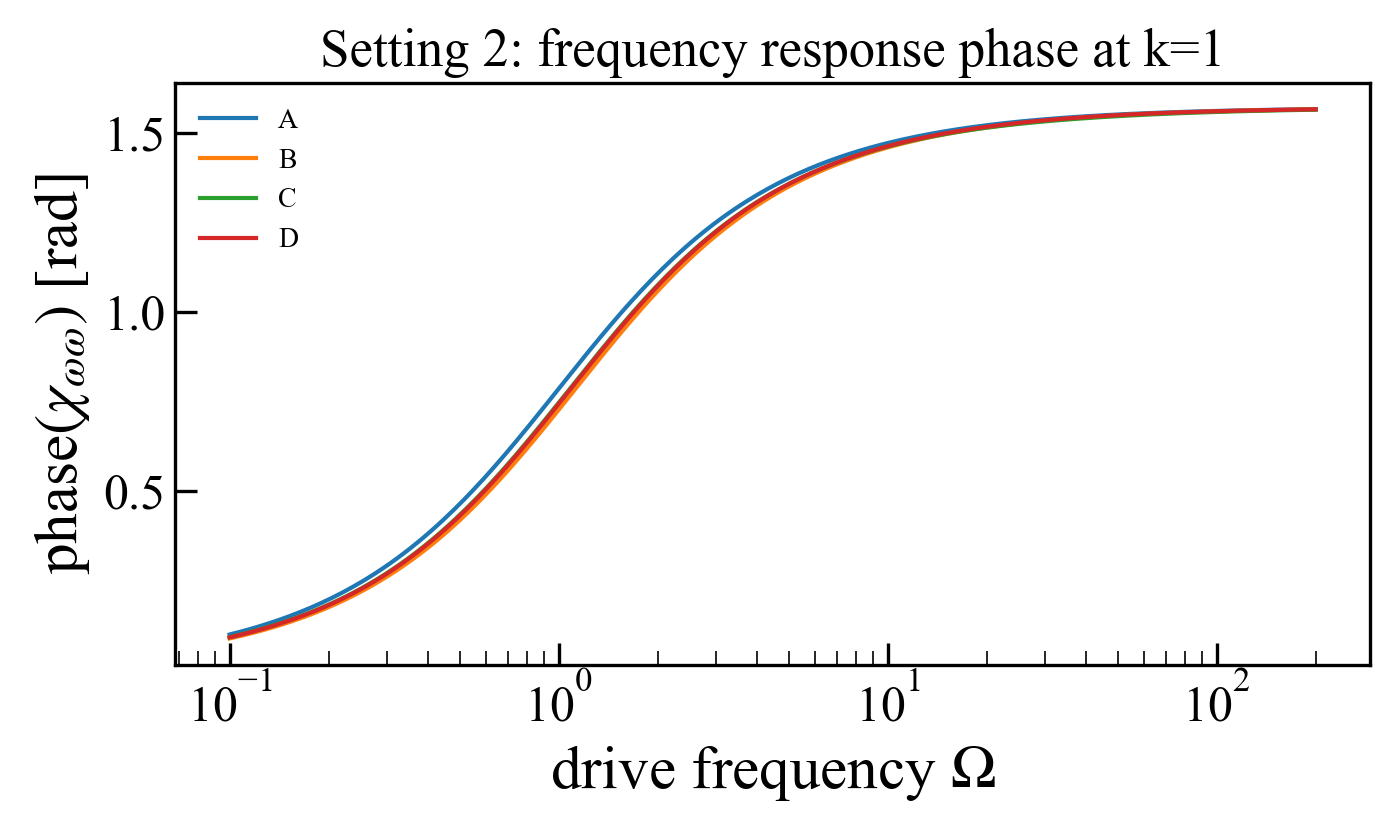}}\hfill
\includegraphics[width=0.315\linewidth]{\ResFigNoKsix{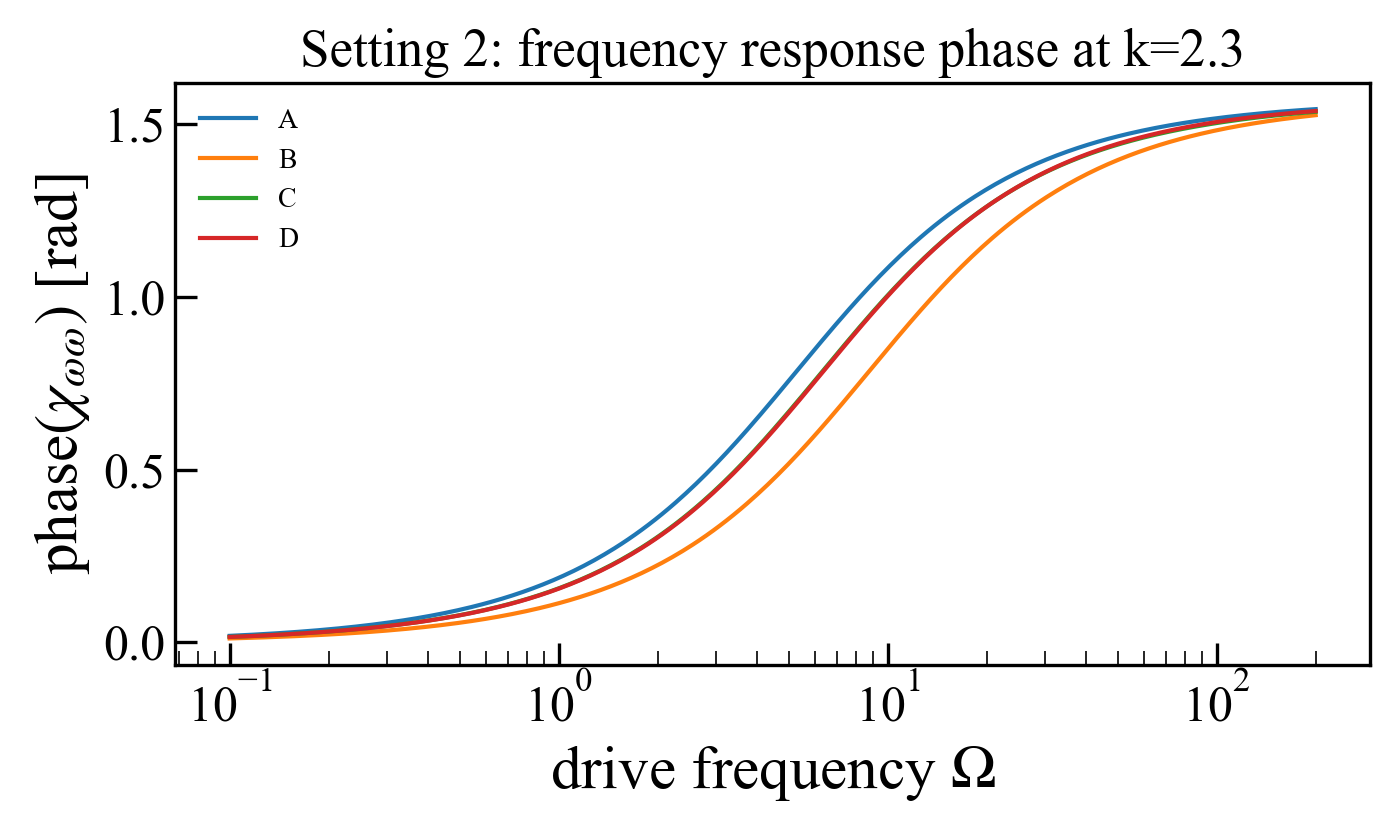}}\hfill
\includegraphics[width=0.315\linewidth]{\ResFigNoKsix{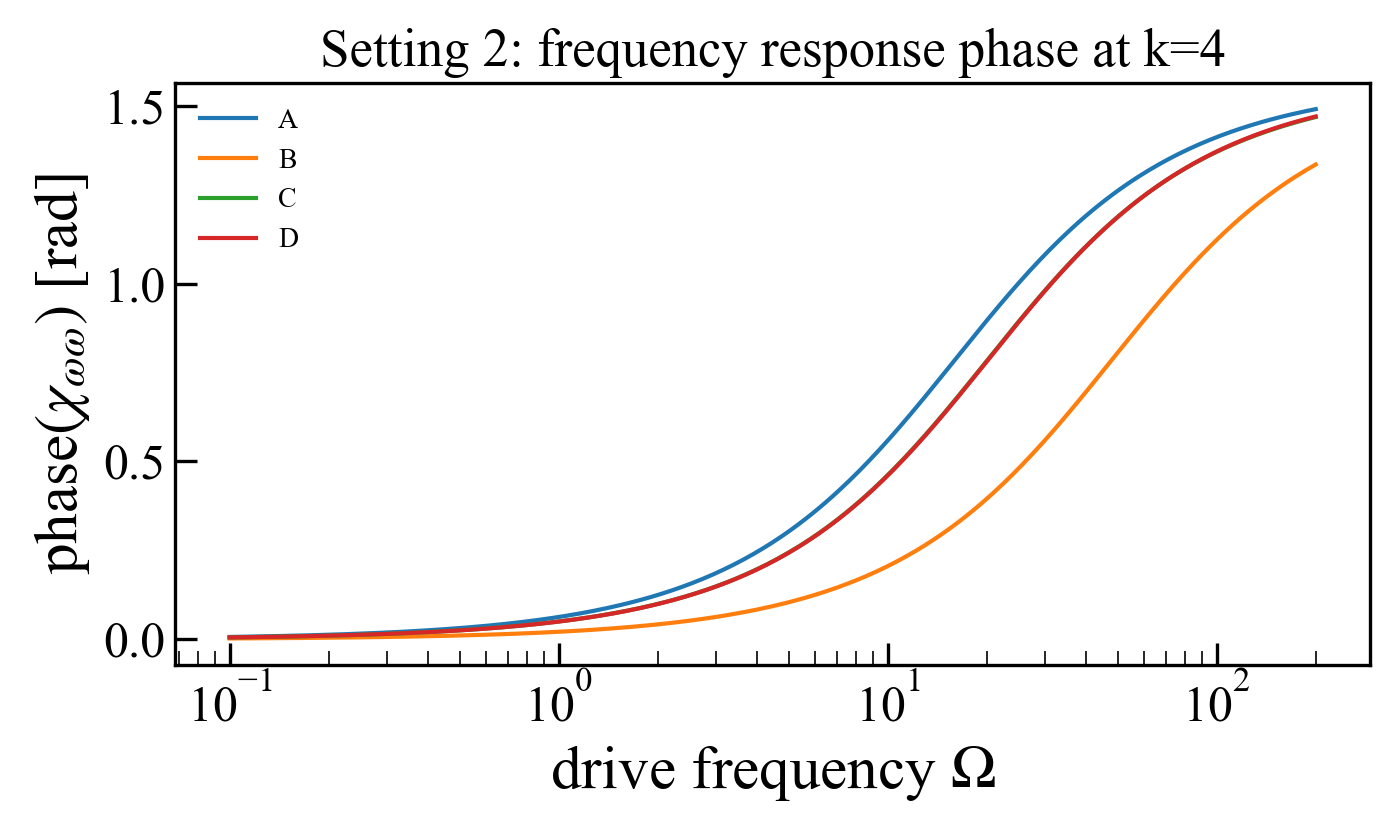}}\\[1.2ex]
\includegraphics[width=0.315\linewidth]{\ResFigWithKsix{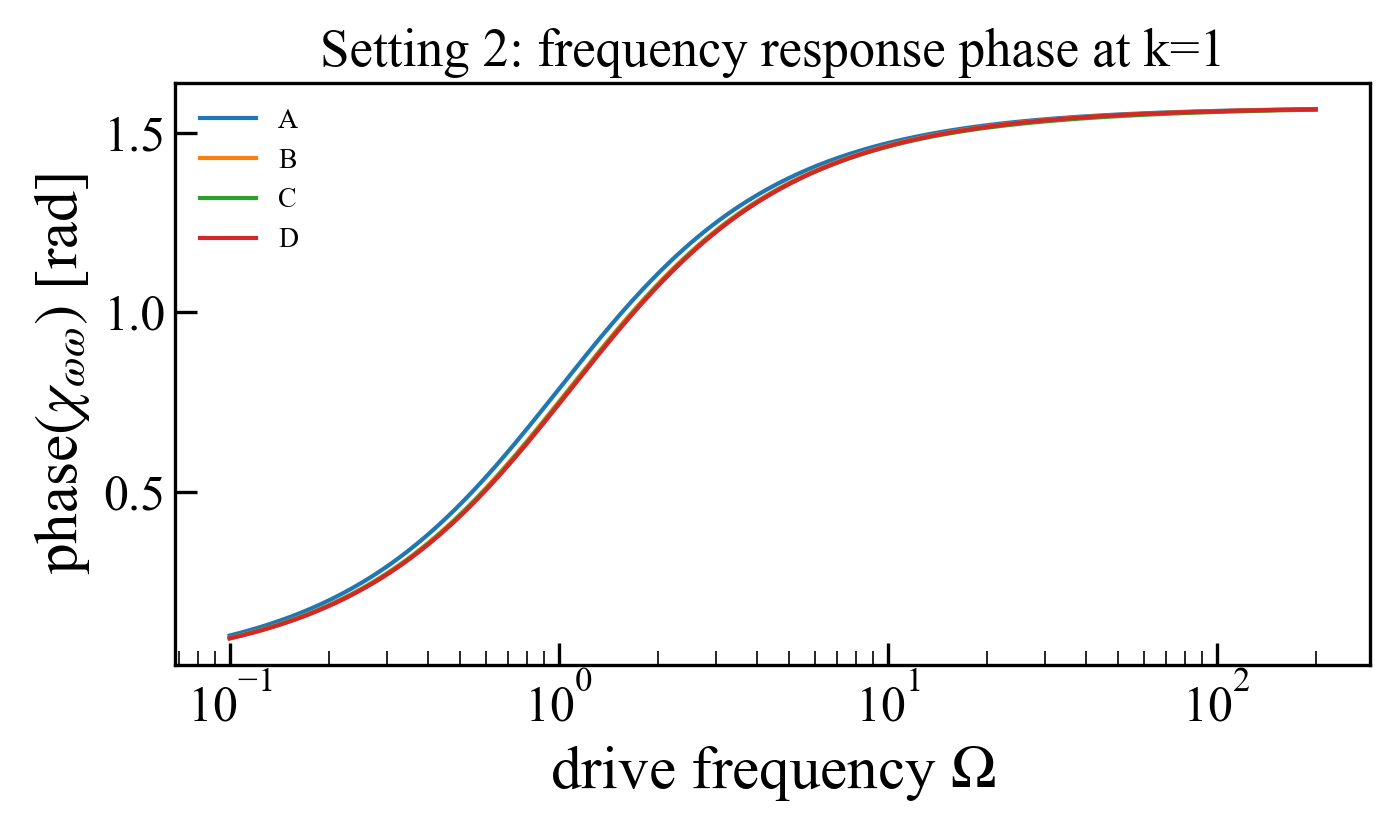}}\hfill
\includegraphics[width=0.315\linewidth]{\ResFigWithKsix{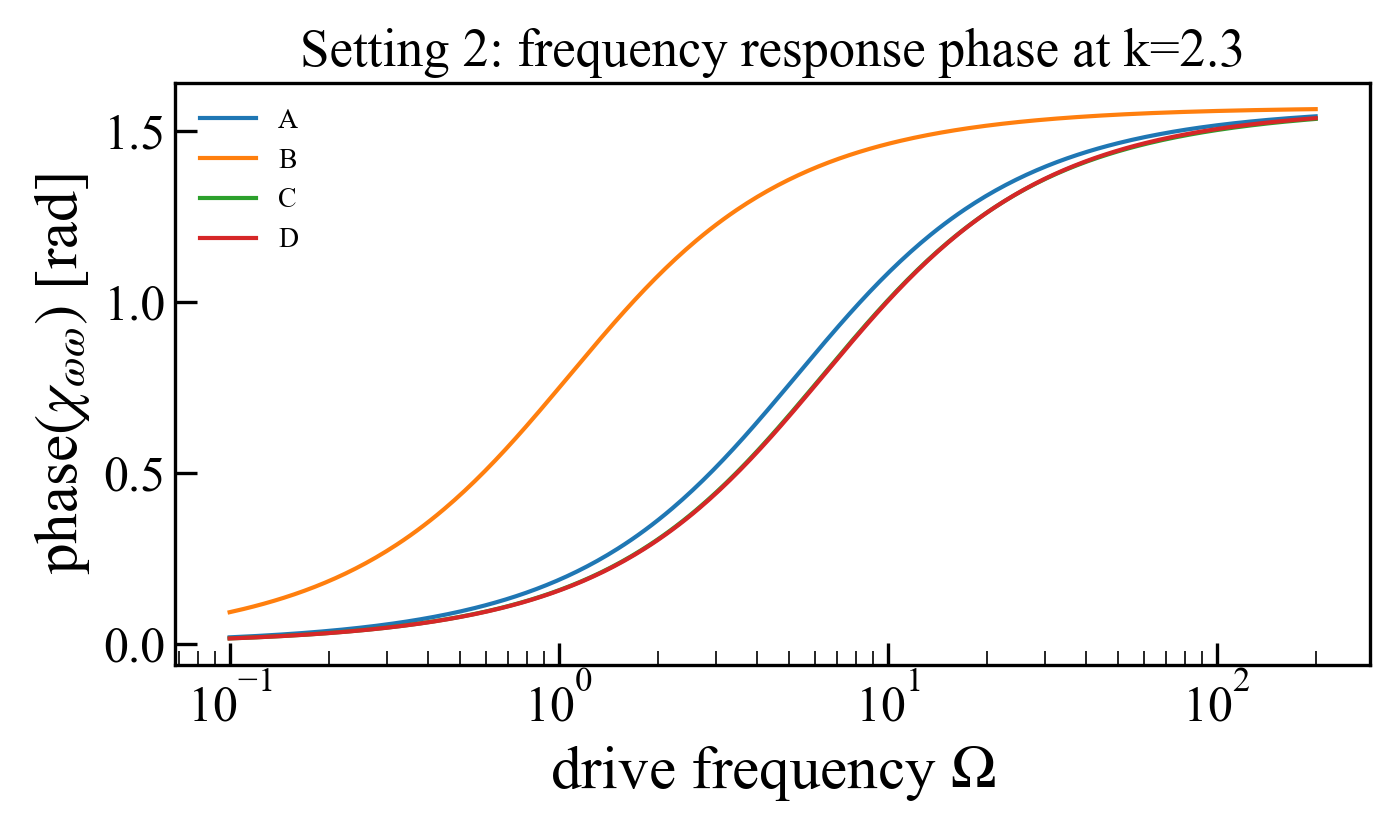}}\hfill
\includegraphics[width=0.315\linewidth]{\ResFigWithKsix{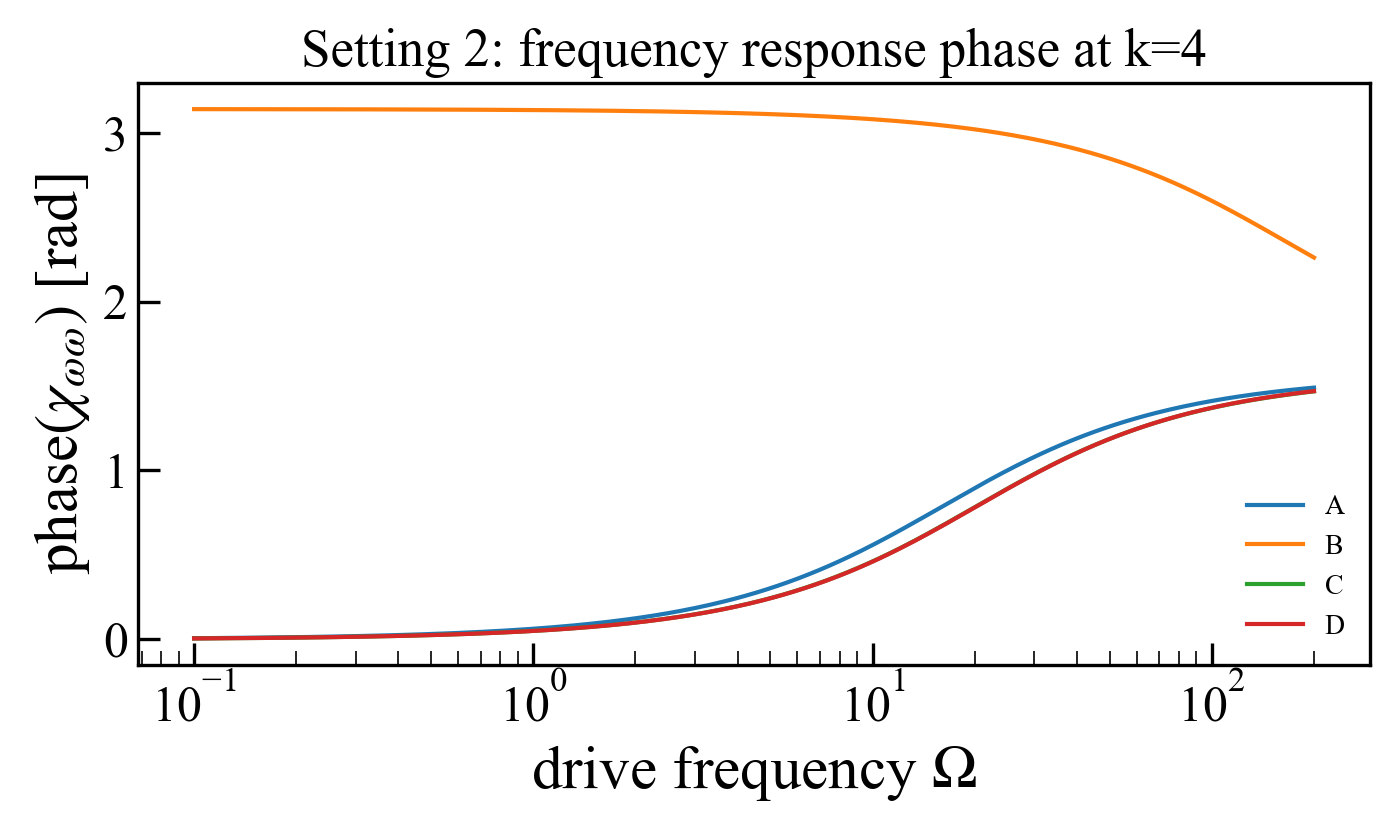}}
\caption{
Setting~2: Bode phase $\arg\chi_{\zeta\zeta}(-i\Omega,k)$ for the same three $k$ values as in Fig.~\ref{fig:bode_amp}.
Top row: $B^{(4)}$.
Bottom row: $B^{(6)}$.
The matched truncation develops a large phase shift relative to the eliminated-spin theory near $k_{\mathrm{crit}}$ at intermediate
$\Omega$, whereas the strict truncation stays stable and shows a different, high-$k$ phase-deformation pattern.
}
\label{fig:bode_phase}
\end{figure*}

We next apply harmonic forcing in the vorticity channel (Setting~2),
\begin{equation}
f(t;k)=F_0 e^{-i\Omega t},\qquad g(t;k)=0,
\label{eq:setting2_forcing}
\end{equation}
and measure the steady-state response
$\zeta(t;k)\sim \chi_{\zeta\zeta}(-i\Omega,k)\,F_0\,e^{-i\Omega t}$.
Figures~\ref{fig:bode_amp} and \ref{fig:bode_phase} show the Bode amplitude $|\chi_{\zeta\zeta}(-i\Omega,k)|$
and phase $\arg\chi_{\zeta\zeta}(-i\Omega,k)$ for the same three $k$ values as above,
again with $B^{(4)}$ in the top row and $B^{(6)}$ in the bottom row.

At $k=1$ all stable models behave similarly (as expected from the shared low-$k$ structure).
At $k=2.3\simeq 0.96\,k_{\mathrm{crit}}$, the two polynomial surrogate variants already separate:
$B^{(4)}$ remains stable but differs appreciably from Models C/D, whereas $B^{(6)}$ exhibits the near-cancellation of its
polynomial damping coefficient and deviates much more strongly at low and moderate frequencies.
At high frequency, all stable models tend toward the inertial scaling $|\chi|\sim 1/(\rho\Omega)$.
At $k=4>k_{\mathrm{crit}}$, $B^{(4)}$ still admits a steady response but with a strongly suppressed low-frequency amplitude
relative to D, whereas $B^{(6)}$ admits no physical steady response because it is linearly unstable.

\subsection{Diagnostic maps in the $(k,\Omega)$ plane}

Finally, we summarize the diagnostic power of the response functions by scanning the $(k,\Omega)$ plane on a log--log grid
($k\in[0.3,50]$, $\Omega\in[0.05,500]$) and comparing complex responses.
For a model pair $X$--$Y$ we define three comparison metrics:
\begin{align}
\mathrm{RelErr}_{X|Y}(k,\Omega) &\equiv \frac{|\chi_X-\chi_Y|}{|\chi_Y|+\varepsilon},\label{eq:relerr_def}\\
\Delta\phi_{X|Y}(k,\Omega) &\equiv \left|\mathrm{wrap}\bigl(\arg\chi_X-\arg\chi_Y\bigr)\right|,\label{eq:phasediff_def}\\
\mathrm{AmpRatio}_{X|Y}(k,\Omega) &\equiv \log_{10}\!\left(\frac{|\chi_X|}{|\chi_Y|+\varepsilon}\right),\label{eq:ampratio_def}
\end{align}
with a small $\varepsilon$ to regularize division by zero.
When the matched variant $B^{(6)}$ is used, we mask the region $k>k_{\mathrm{crit}}$ where Model~B is unstable and does not admit a
steady response; no masking is needed for $B^{(4)}$.

\begin{figure*}[t]
\centering
\includegraphics[width=0.32\linewidth]{\ResFig{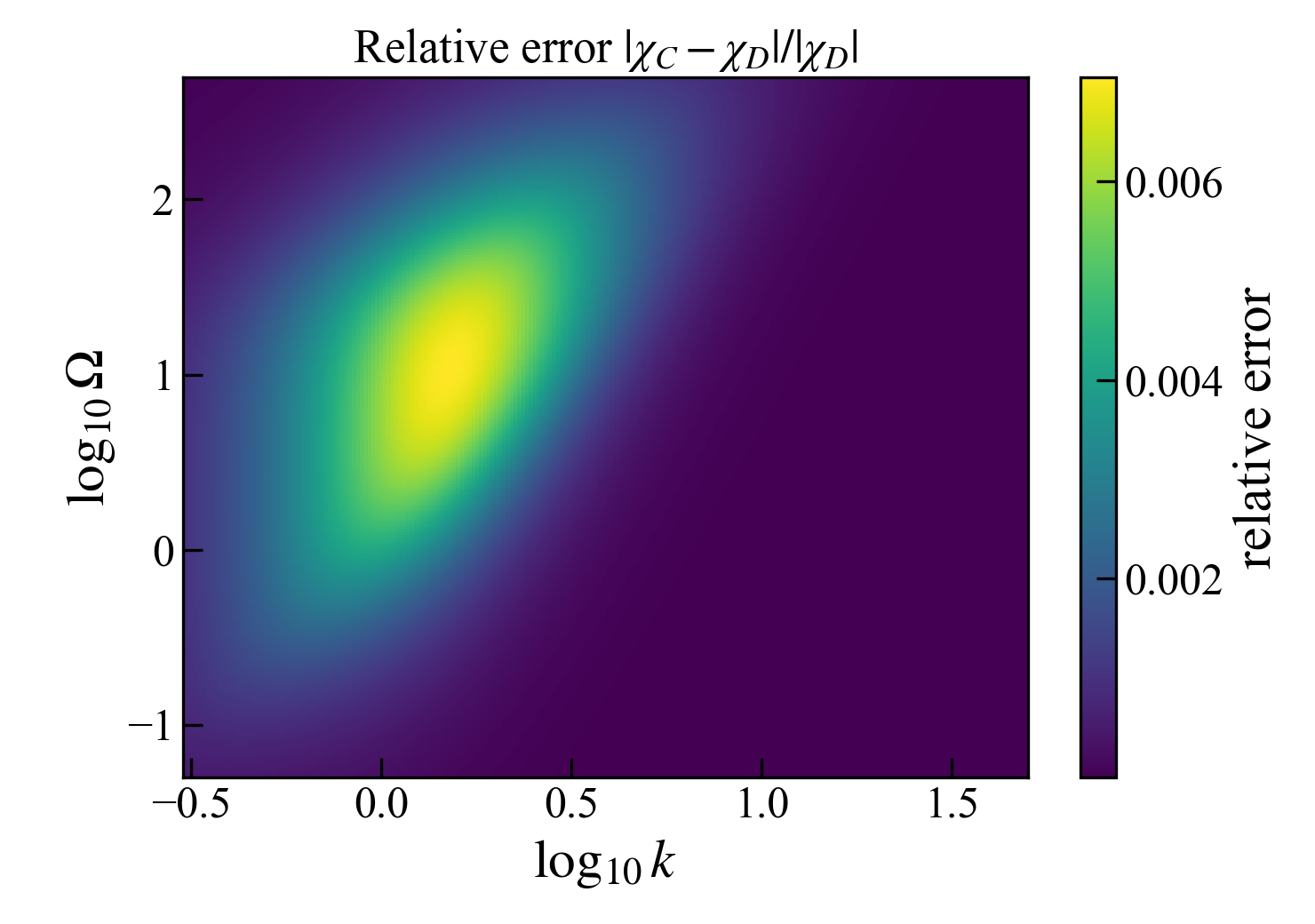}}\hfill
\includegraphics[width=0.32\linewidth]{\ResFig{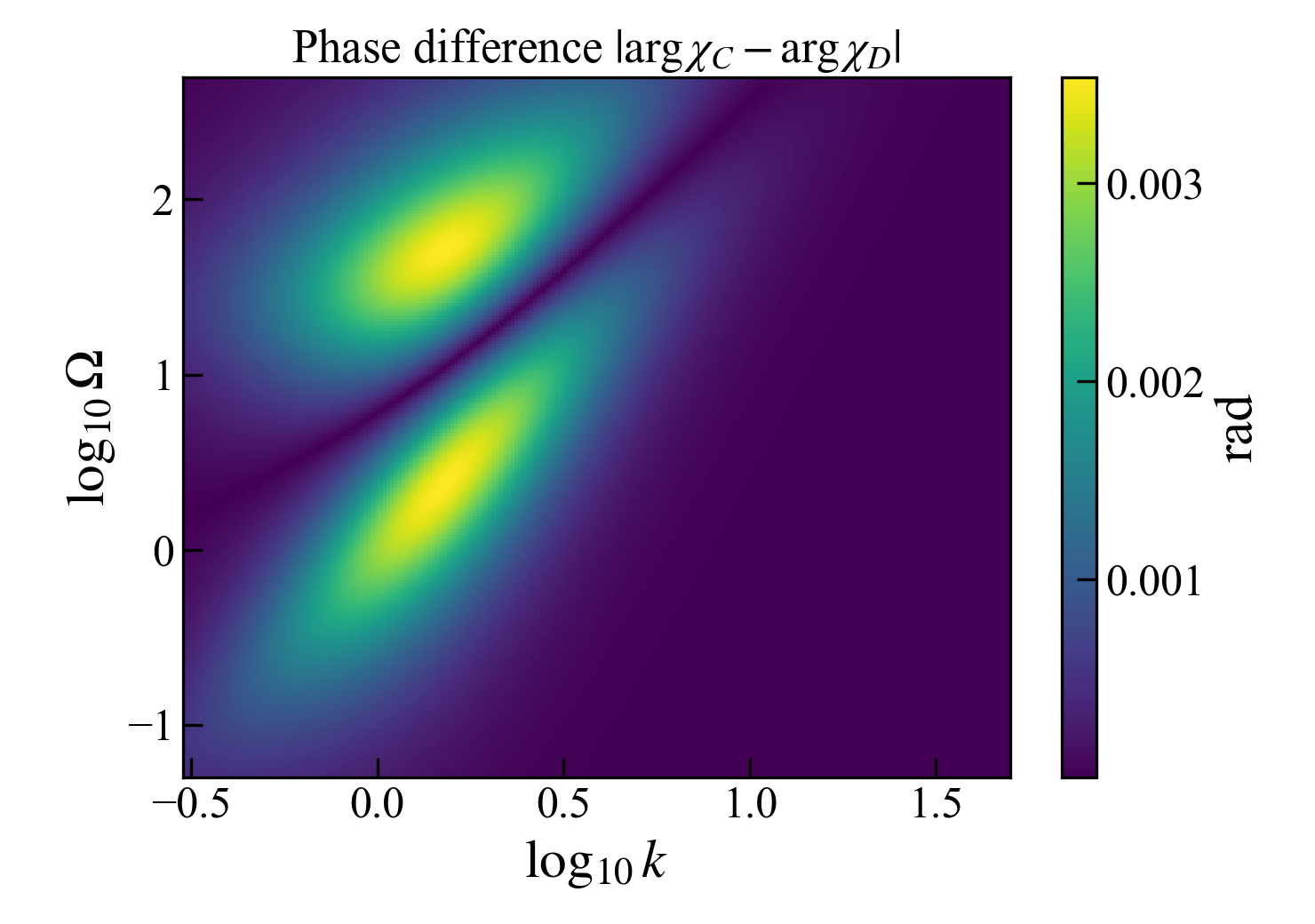}}\hfill
\includegraphics[width=0.32\linewidth]{\ResFig{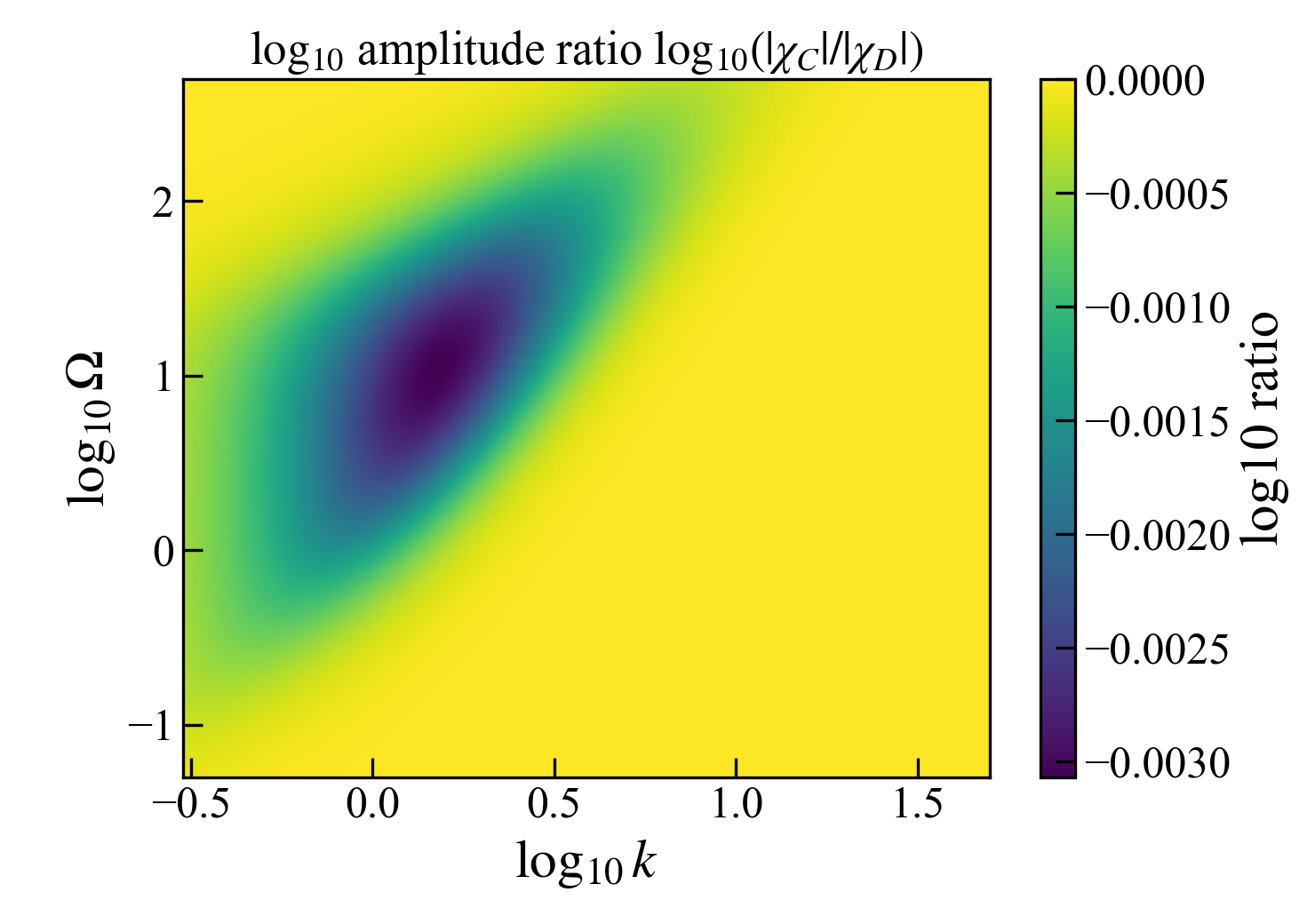}}
\caption{
Diagnostic maps comparing Model~C (explicit internal spin) to Model~D (adiabatic elimination):
relative error [Eq.~(\ref{eq:relerr_def})], phase difference Eq.~(\ref{eq:phasediff_def}), and amplitude ratio Eq.~(\ref{eq:ampratio_def})
(left to right).
Across the scanned $(k,\Omega)$ domain the discrepancy remains below the percent level for the parameter set Eq.~(\ref{eq:analysis_params}), demonstrating the accuracy of fast-spin elimination in this regime.
}
\label{fig:map_CD}
\end{figure*}

\paragraph*{C vs D (explicit spin vs adiabatic elimination).}
Figure~\ref{fig:map_CD} shows that, for Eq.~(\ref{eq:analysis_params}), the eliminated-spin theory (D)
approximates the explicit-spin response (C) to high accuracy over the entire scanned domain:
the maximum relative error is $\max\mathrm{RelErr}_{\mathrm{C|D}}\simeq 7.0\times 10^{-3}$
(at $k\simeq 1.50$, $\Omega\simeq 10.9$), and the maximum phase difference is below $4\times 10^{-3}$~rad.
This quantitatively corroborates the fast-spin elimination regime for the chosen parameters.

\begin{figure*}[t]
\centering
\includegraphics[width=0.315\linewidth]{\ResFigNoKsix{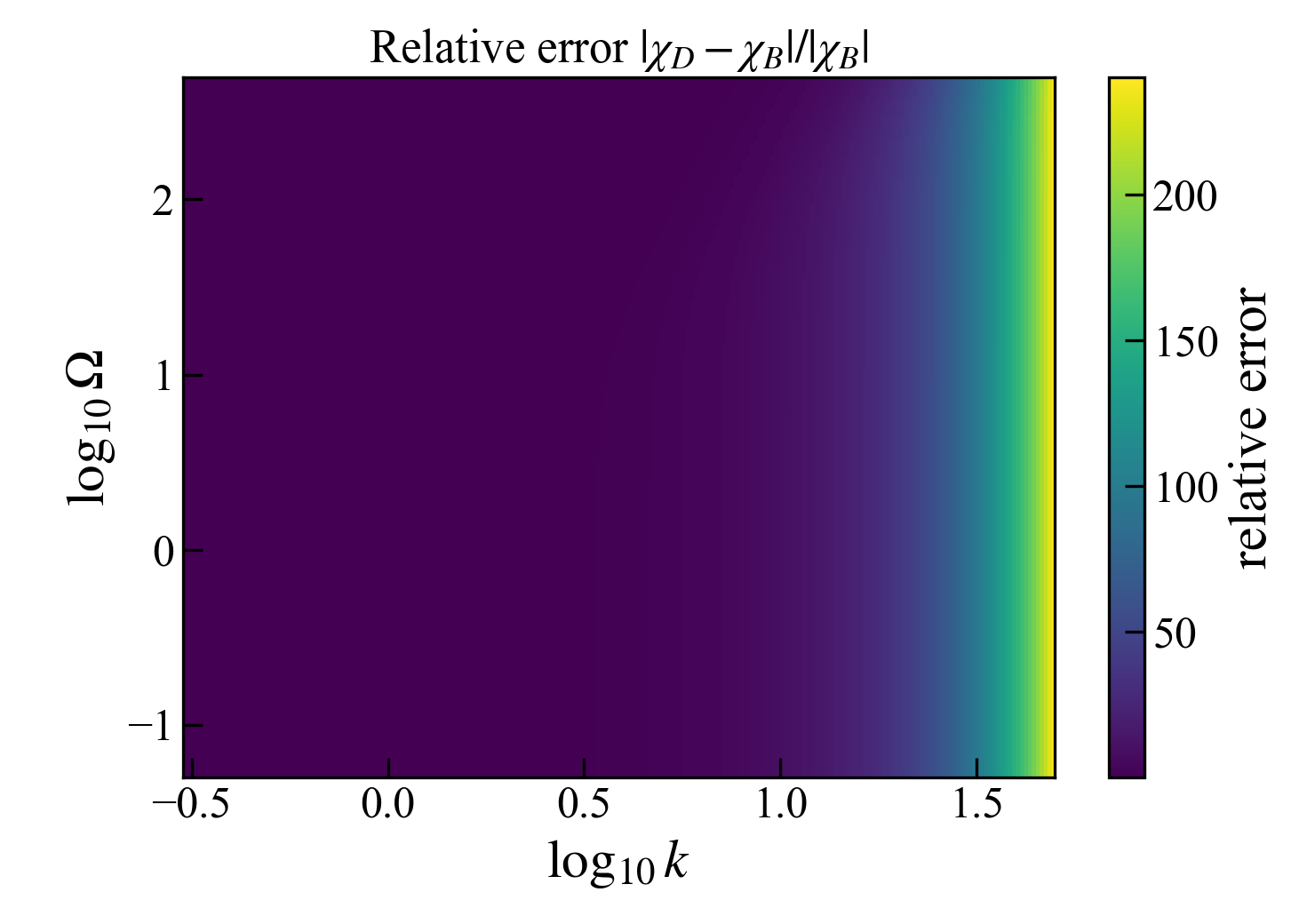}}\hfill
\includegraphics[width=0.315\linewidth]{\ResFigNoKsix{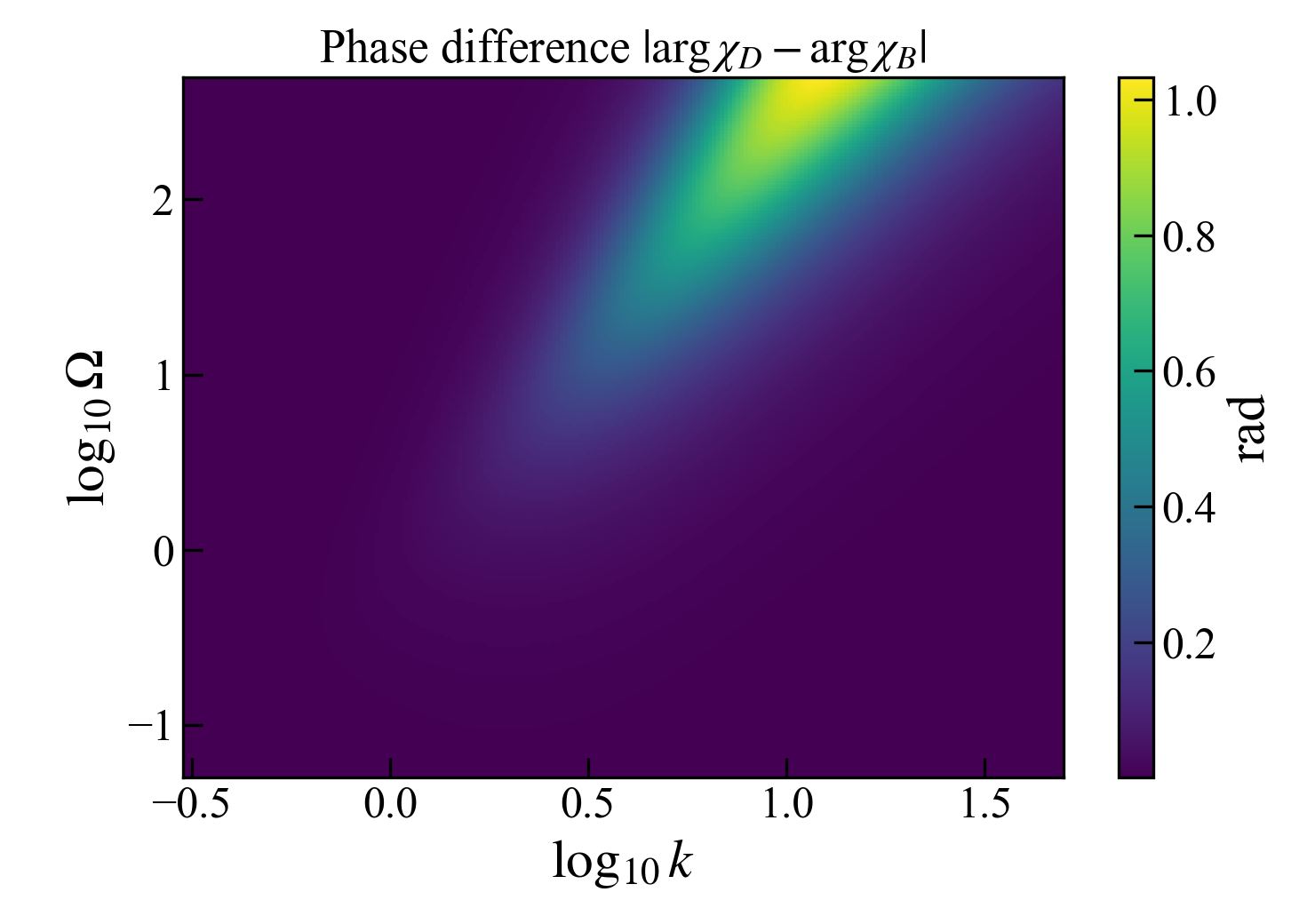}}\hfill
\includegraphics[width=0.315\linewidth]{\ResFigNoKsix{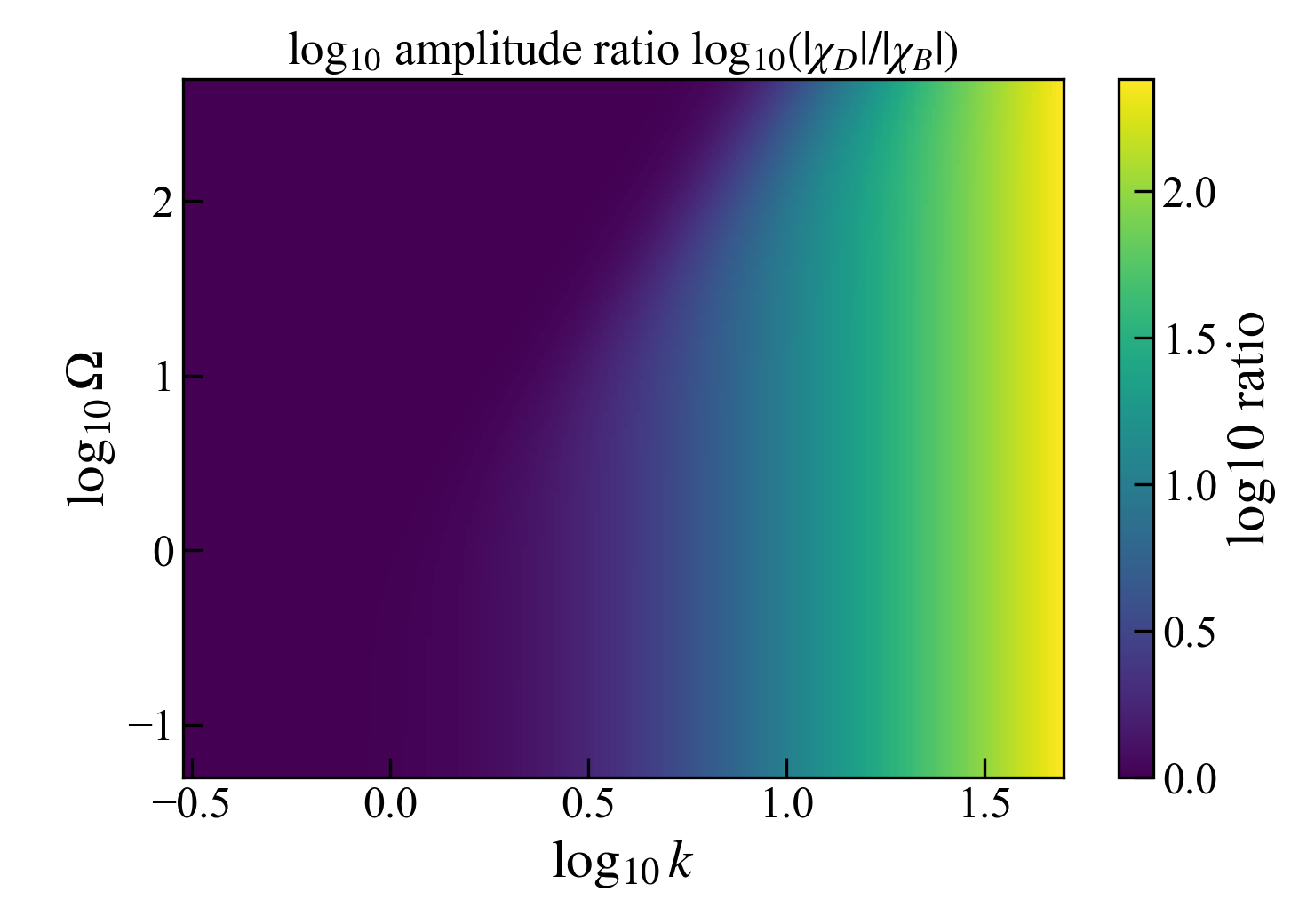}}\\[1.2ex]
\includegraphics[width=0.315\linewidth]{\ResFigWithKsix{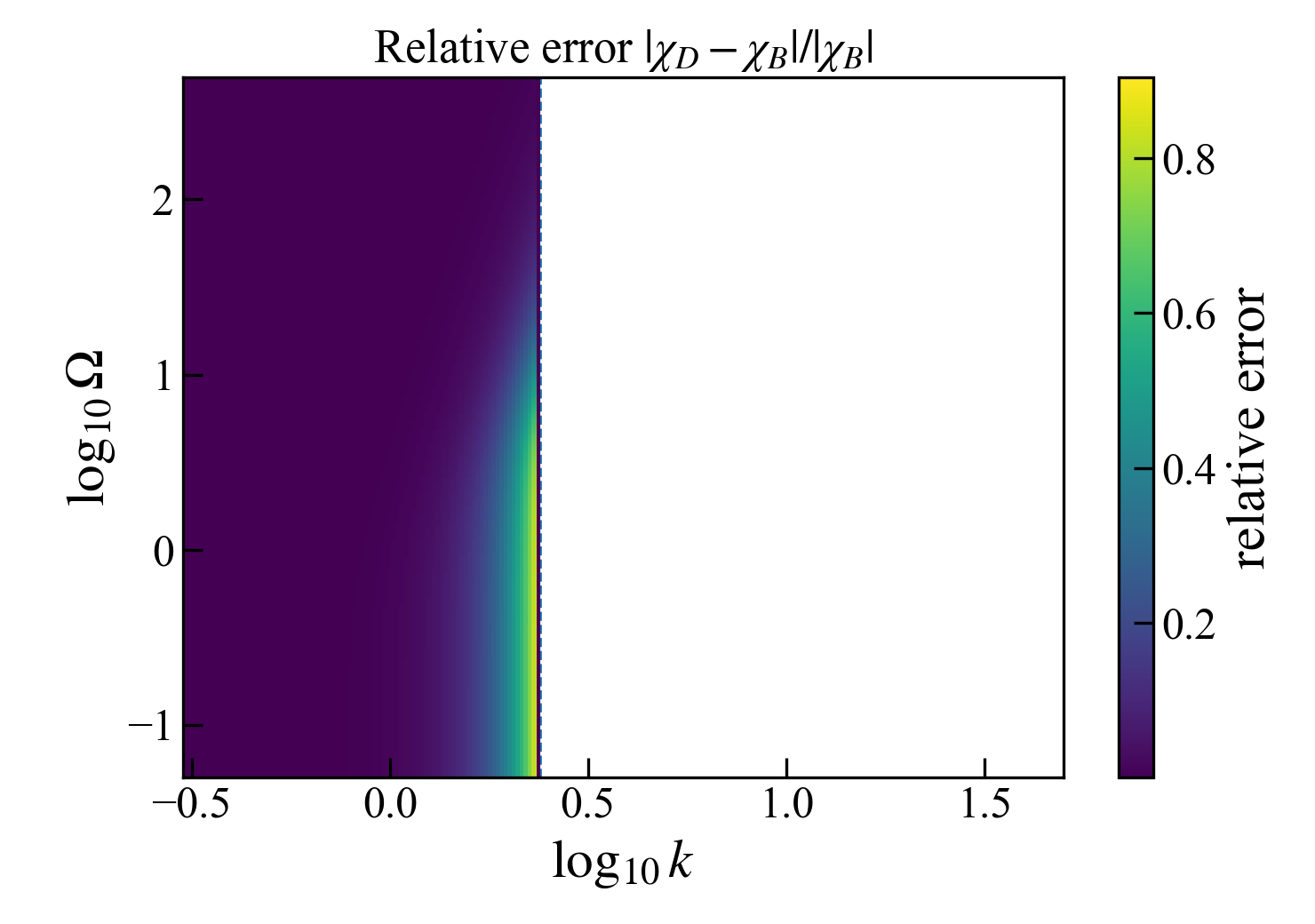}}\hfill
\includegraphics[width=0.315\linewidth]{\ResFigWithKsix{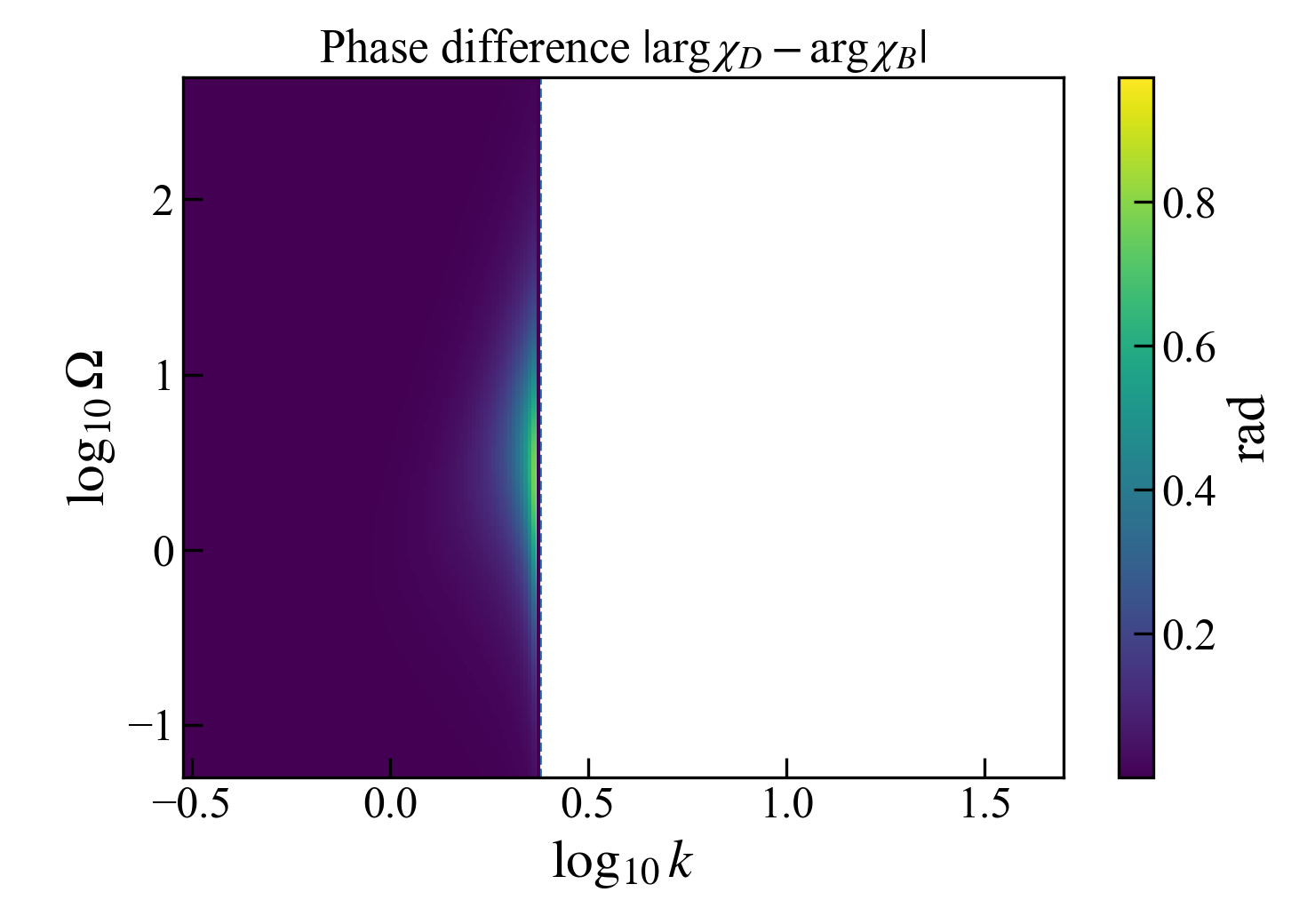}}\hfill
\includegraphics[width=0.315\linewidth]{\ResFigWithKsix{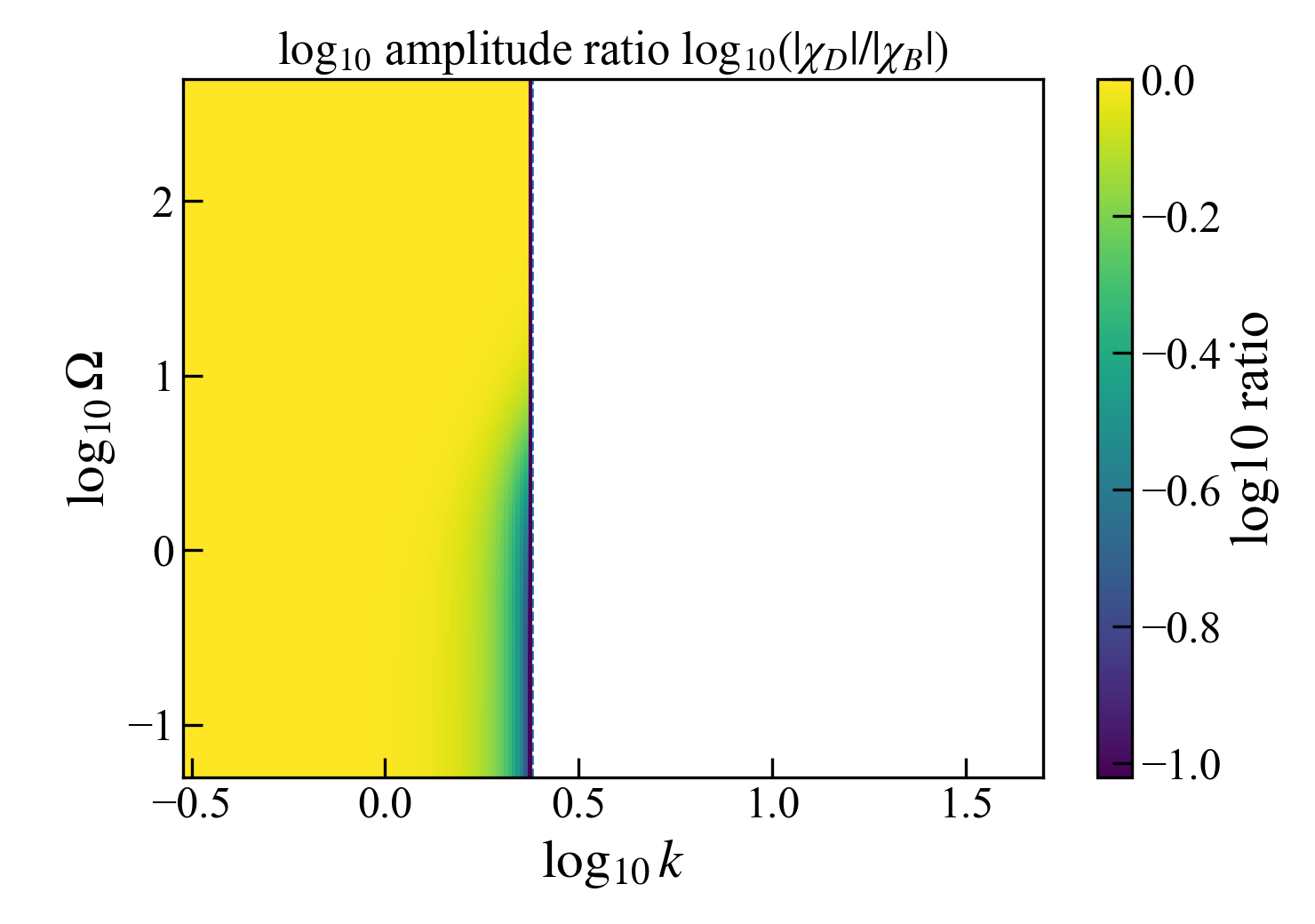}}
\caption{
Diagnostic maps comparing Model~D (adiabatically eliminated internal spin; rational kernel) to Model~B.
Columns show relative error, phase difference, and amplitude ratio.
Top row: strict $k^4$ polynomial truncation $B^{(4)}$ (without the $k^6$ term), for which the discrepancy grows toward large $k$ because
Model~B remains stable but becomes increasingly over-damped.
Bottom row: matched truncation $B^{(6)}$ (with the $k^6$ term), for which the unstable region $k>k_{\mathrm{crit}}$ is masked and the
strongest diagnostic separation occurs near $k_{\mathrm{crit}}$ in both amplitude (low $\Omega$) and phase (intermediate $\Omega$).
}
\label{fig:map_DB}
\end{figure*}

\begin{figure*}[t]
\centering
\includegraphics[width=0.315\linewidth]{\ResFigNoKsix{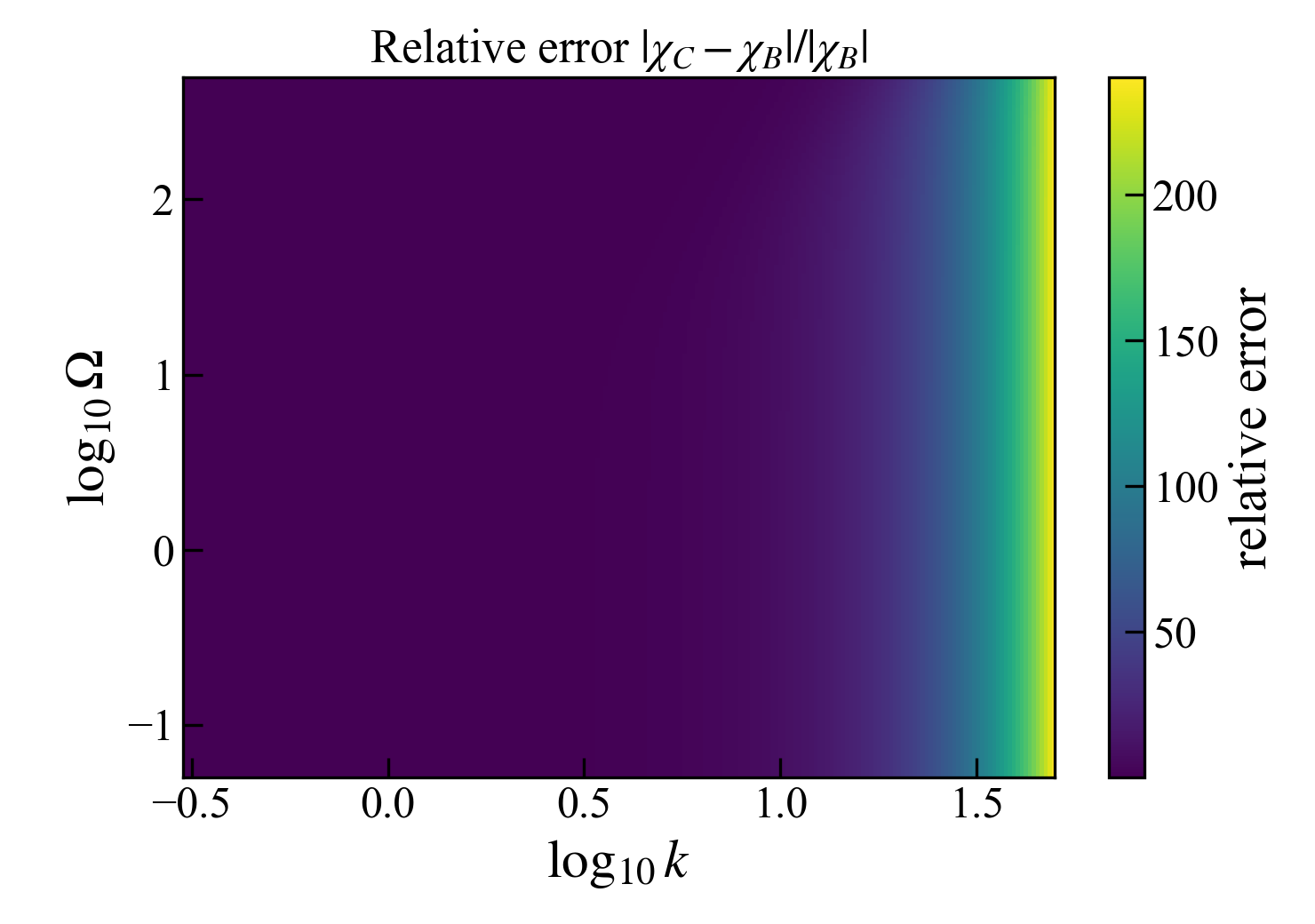}}\hfill
\includegraphics[width=0.315\linewidth]{\ResFigNoKsix{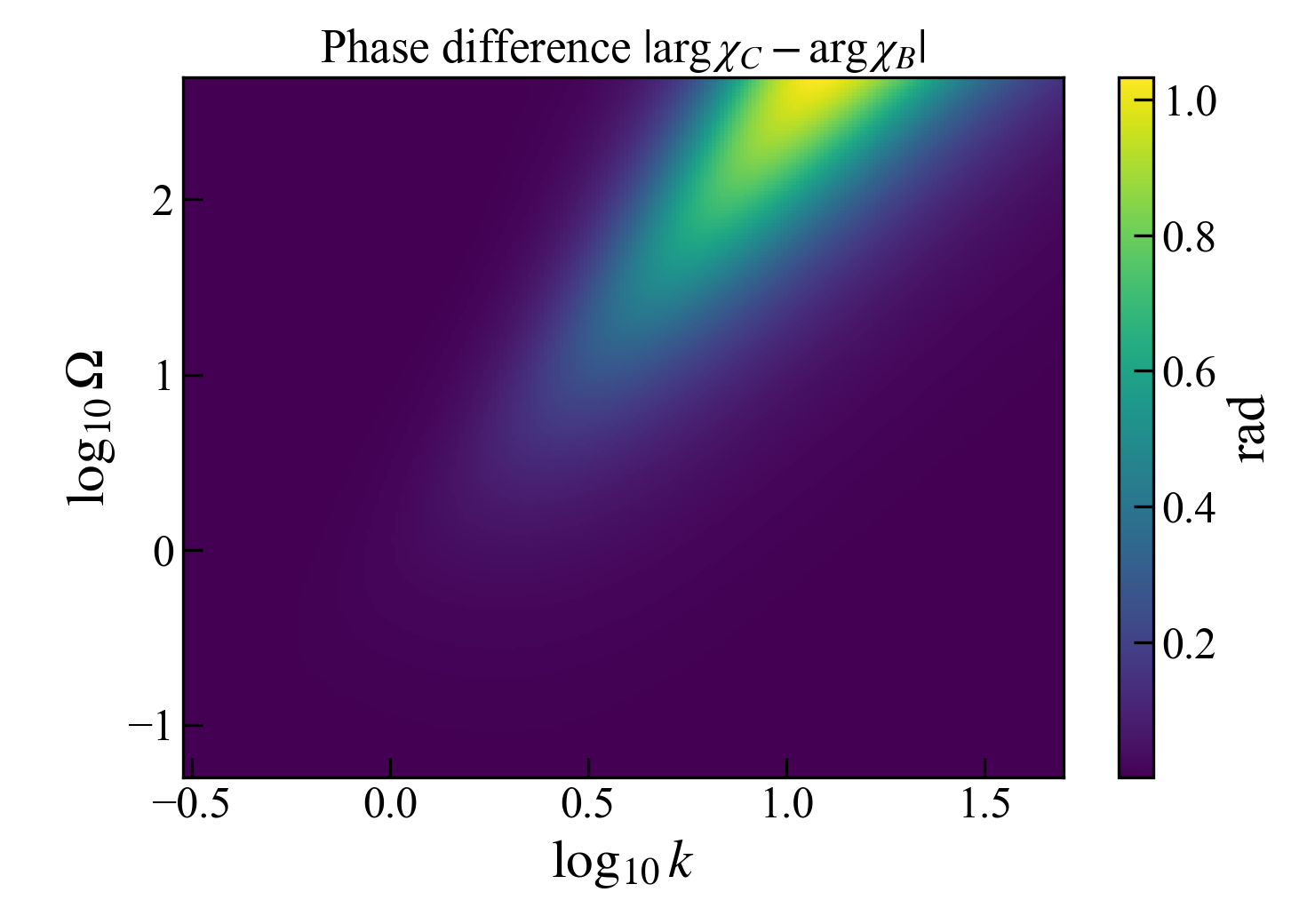}}\hfill
\includegraphics[width=0.315\linewidth]{\ResFigNoKsix{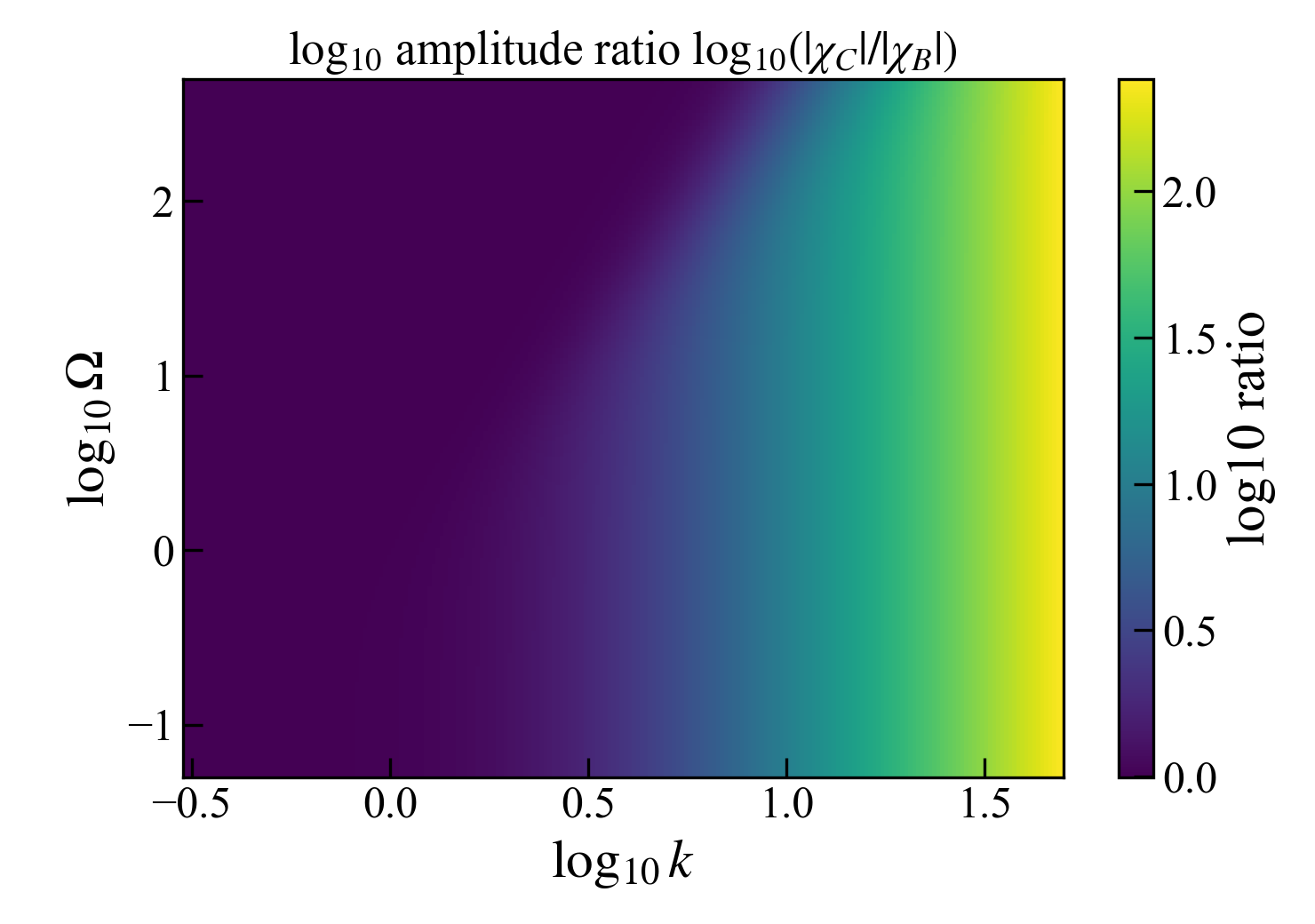}}\\[1.2ex]
\includegraphics[width=0.315\linewidth]{\ResFigWithKsix{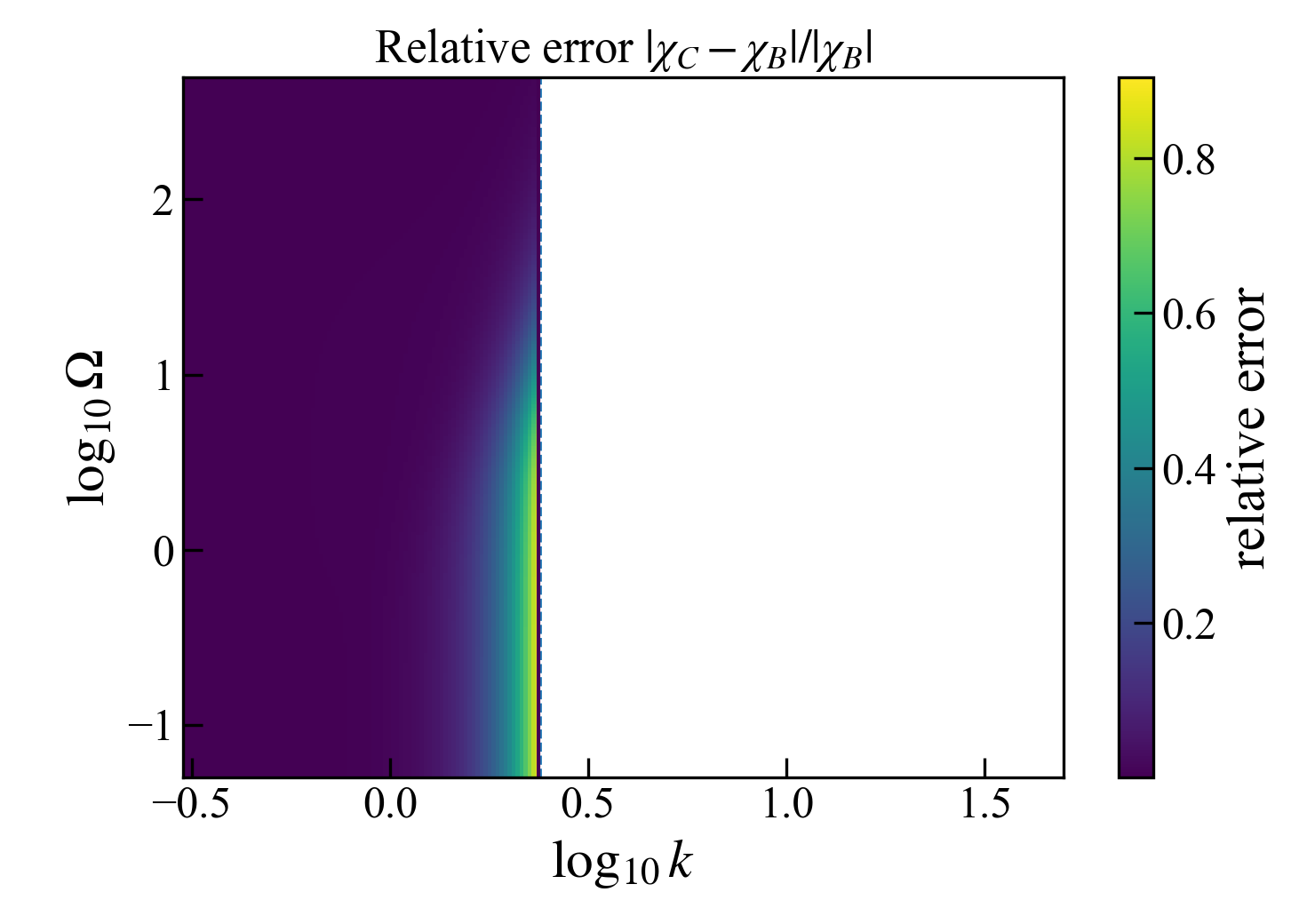}}\hfill
\includegraphics[width=0.315\linewidth]{\ResFigWithKsix{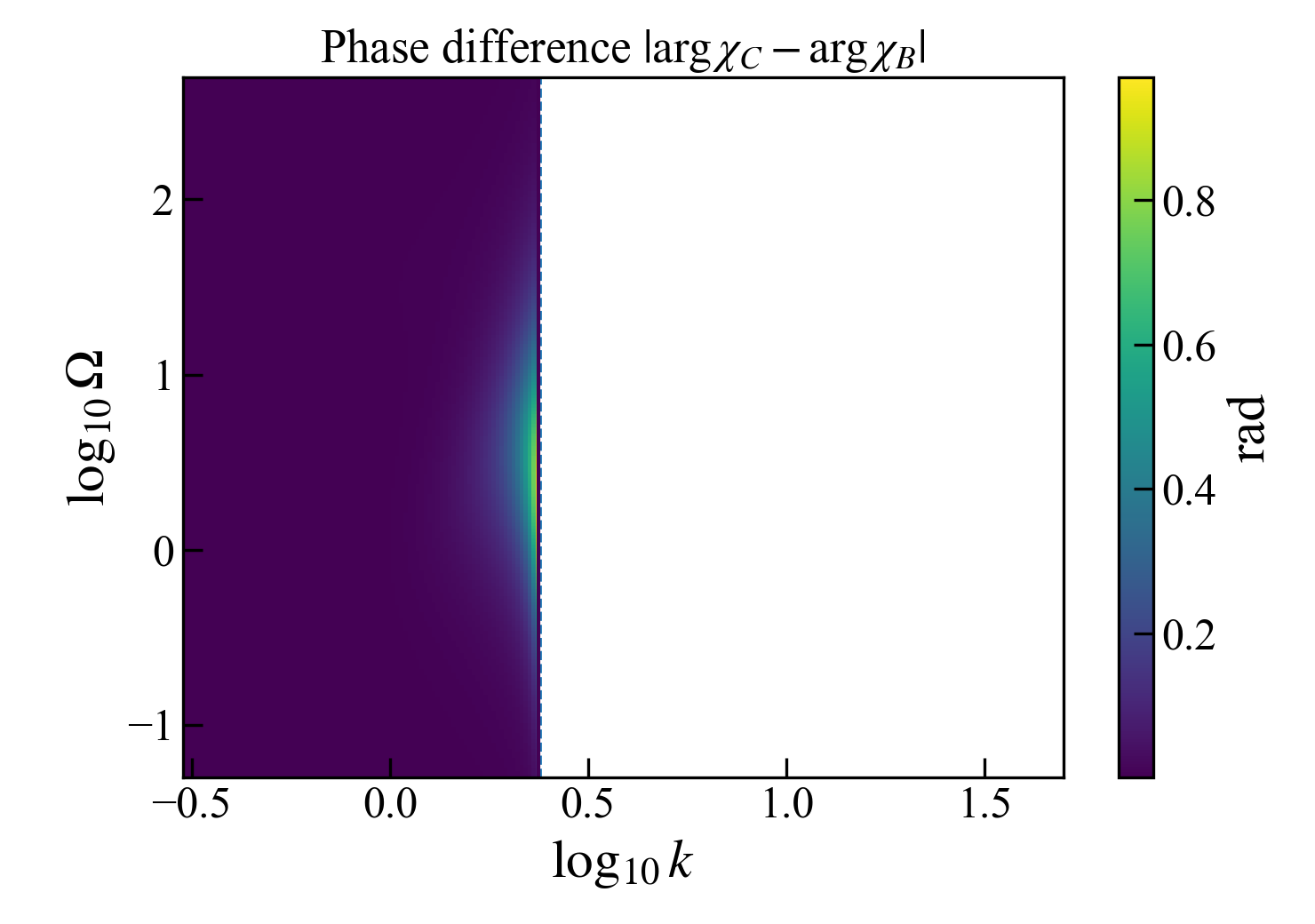}}\hfill
\includegraphics[width=0.315\linewidth]{\ResFigWithKsix{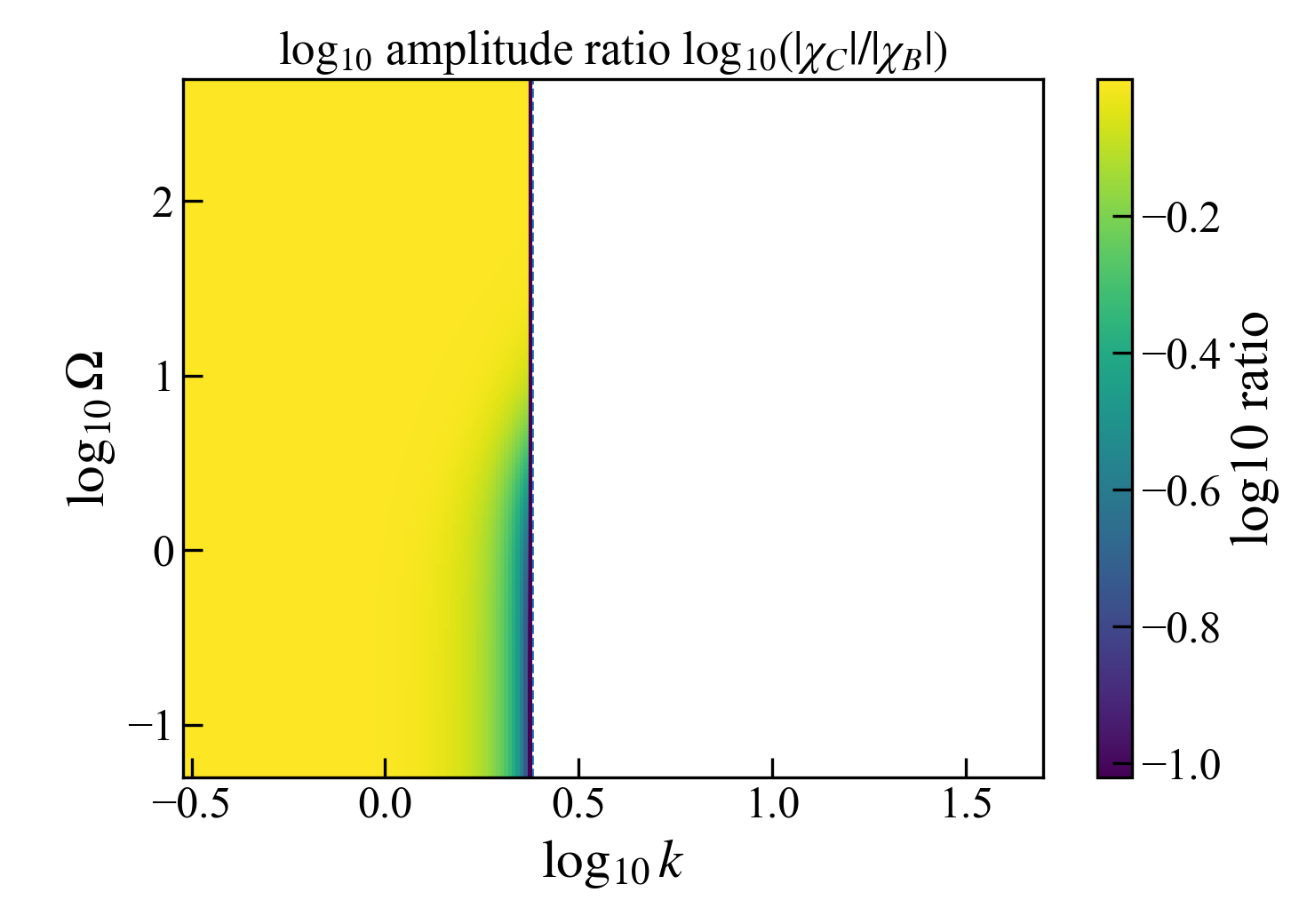}}
\caption{
Diagnostic maps comparing Model~C (explicit spin retained) to Model~B.
Columns show relative error, phase difference, and amplitude ratio.
Top row: $B^{(4)}$.
Bottom row: $B^{(6)}$.
Because Model~C and Model~D are close in the present fast-spin regime (Fig.~\ref{fig:map_CD}),
the C--B maps closely resemble the D--B maps for both polynomial surrogate variants, emphasizing that the dominant diagnostic separation here is
between a rational (spin-eliminated) kernel and a finite polynomial surrogate.
}
\label{fig:map_CB}
\end{figure*}

\paragraph*{D vs B and C vs B (rational kernel vs polynomial truncation).}
Figures~\ref{fig:map_DB} and \ref{fig:map_CB} compare the eliminated-spin kernel (D) and the explicit-spin response (C)
to the polynomial surrogate (B) for both polynomial surrogate variants:
the top row uses the strict truncation $B^{(4)}$, and the bottom row uses the matched truncation $B^{(6)}$.
The maps reveal two distinct ways in which a polynomial approximation can fail.

(i) For the strict truncation $B^{(4)}$, the discrepancy grows monotonically toward large $k$ and low $\Omega$.
For D--$B^{(4)}$ we find $\max\mathrm{RelErr}\simeq 2.40\times 10^{2}$ at $(k,\Omega)=(50,0.05)$,
together with $\log_{10}(|\chi_D|/|\chi_{B^{(4)}}|)\simeq 2.38$.
This reflects the fact that the strict $k^4$ polynomial remains stable but becomes increasingly over-damped relative to the rational kernel.

(ii) For the matched truncation $B^{(6)}$, the added negative $k^6$ term shifts the strongest separation to the near-critical band
$k\lesssim k_{\mathrm{crit}}$.
For D--$B^{(6)}$ we find $\max\mathrm{RelErr}\simeq 0.905$ at $(k,\Omega)\simeq(2.34,0.05)$,
together with $\log_{10}(|\chi_D|/|\chi_{B^{(6)}}|)\simeq -1.02$, i.e.\ $|\chi_D|/|\chi_{B^{(6)}}|\simeq 0.095$.
The phase difference is maximized at intermediate frequencies near the same $k$:
$\max\Delta\phi\simeq 0.97$~rad at $(k,\Omega)\simeq(2.34,2.0)$.
Thus, depending on the truncation and measurement protocol, large-$k$ amplitude suppression and near-critical amplitude/phase anomalies
provide complementary diagnostics.

\subsection{Additional integrated response diagnostics}
\label{sec:additional_integrated_diagnostics}
To complement the pointwise $(k,\Omega)$ maps above, we also introduce two aggregated diagnostics.
The first compresses the full frequency-dependent response at fixed $k$ into a small set of scalar distances;
this makes it easier to visualize how the separation between response classes evolves with wavenumber.
The second varies the spin-relaxation timescale proxy
\begin{equation}
\tau_s \equiv \frac{\rho J}{4\eta_r},
\label{eq:tau_s_def_additional}
\end{equation}
and quantifies when the explicit-spin two-pole structure of Model~C becomes macroscopically distinguishable from the eliminated-spin one-pole theory~D.

\begin{figure*}[t]
\centering
\includegraphics[width=0.675\textwidth]{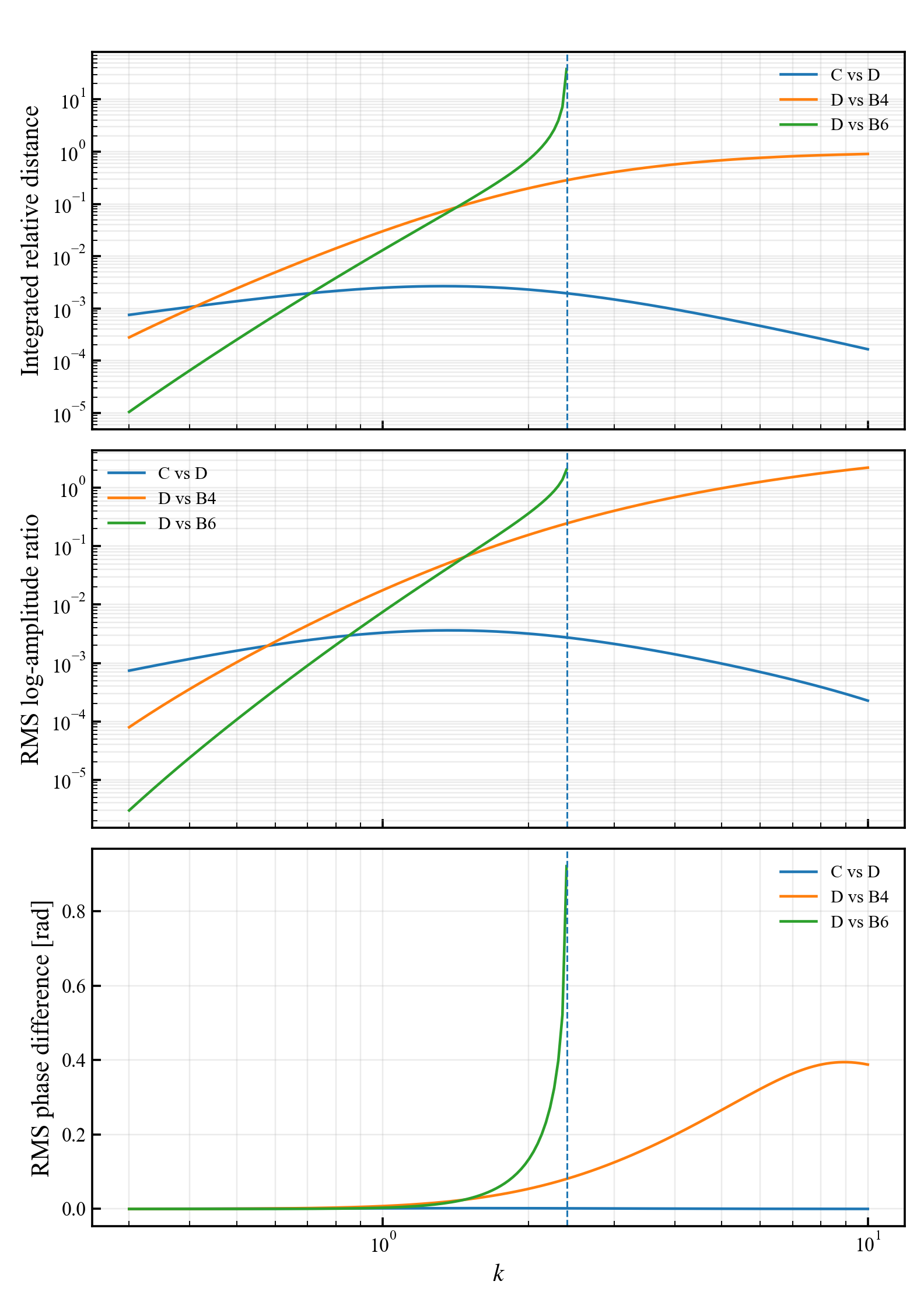}
\caption{
Additional aggregated response diagnostics versus wavenumber.
Top: integrated relative distance $\mathcal{D}_{X\to Y}(k)$ defined in Eq.~(\ref{eq:integrated_response_distance}).
Middle: RMS log-amplitude ratio $\mathcal{A}_{X\to Y}(k)$ defined in Eq.~(\ref{eq:rms_log_amp_ratio}).
Bottom: RMS phase difference $\Phi_{X\to Y}(k)$ defined in Eq.~(\ref{eq:rms_phase_distance}).
The dashed vertical line marks $k_{\mathrm{crit}}$ for the matched truncation $\mathrm{B}^{(6)}$.
Model~C remains close to Model~D across the full scan, whereas D--$\mathrm{B}^{(4)}$ separates gradually with increasing $k$ and D--$\mathrm{B}^{(6)}$ exhibits a sharp near-critical blow-up as $k\to k_{\mathrm{crit}}^{-}$.
}
\label{fig:response_distance_vs_k}
\end{figure*}

Let
\begin{equation}
\chi_X(k,\Omega) \equiv \chi^{(X)}_{\zeta\zeta}(-i\Omega,k)
\end{equation}
denote the transverse response of model $X\in\{\mathrm{A},\mathrm{B}^{(4)},\mathrm{B}^{(6)},\mathrm{C},\mathrm{D}\}$,
evaluated on a logarithmically spaced frequency set $\{\Omega_j\}_{j=1}^{N_\Omega}$.
For an ordered pair $(X,Y)$ we define the frequency-aggregated distances
\begin{align}
\mathcal{D}_{X\to Y}(k)
&\equiv
\frac{\left[\sum_{j=1}^{N_\Omega}\left|\chi_Y(k,\Omega_j)-\chi_X(k,\Omega_j)\right|^2\right]^{1/2}}
{\left[\sum_{j=1}^{N_\Omega}\left|\chi_X(k,\Omega_j)\right|^2\right]^{1/2}},
\label{eq:integrated_response_distance}\\
\mathcal{A}_{X\to Y}(k)
&\equiv
\left[\frac{1}{N_\Omega}\sum_{j=1}^{N_\Omega}
\left(\ln\frac{|\chi_Y(k,\Omega_j)|+\varepsilon}{|\chi_X(k,\Omega_j)|+\varepsilon}\right)^2\right]^{1/2},
\label{eq:rms_log_amp_ratio}\\
\Phi_{X\to Y}(k)
&\equiv
\left[\frac{1}{N_\Omega}\sum_{j=1}^{N_\Omega}
\mathrm{wrap}\!\left(\arg\chi_Y(k,\Omega_j)-\arg\chi_X(k,\Omega_j)\right)^2\right]^{1/2},
\label{eq:rms_phase_distance}
\end{align}
where $\varepsilon$ is a tiny regularizer and $\mathrm{wrap}(\cdot)$ maps phase differences to $(-\pi,\pi]$.
Thus, the top panel of Fig.~\ref{fig:response_distance_vs_k} shows the integrated relative distance $\mathcal{D}_{X\to Y}$,
the middle panel shows the RMS log-amplitude ratio $\mathcal{A}_{X\to Y}$,
and the bottom panel shows the RMS phase difference $\Phi_{X\to Y}$.

Figure~\ref{fig:response_distance_vs_k} sharpens the distinction between the response classes.
The C--D curve remains at the $10^{-3}$ level throughout the scanned range, confirming that the present parameter set lies in a fast-spin regime in which adiabatic elimination is highly accurate.
By contrast, D--$\mathrm{B}^{(4)}$ separates gradually as $k$ increases: the strict $k^4$ truncation stays stable, but it drifts monotonically away from the rational kernel because its damping grows too rapidly at large $k$.
The behavior of D--$\mathrm{B}^{(6)}$ is qualitatively different.
At low $k$ the matched truncation closely tracks D, but as $k$ approaches the finite-$k$ threshold $k_{\mathrm{crit}}$ from below, all three aggregated distances rise sharply.
This shows that the dominant pathology of the matched polynomial truncation is not merely the appearance of a $k^4$ correction, but the near-critical cancellation induced once the $k^6$ term is retained.
For $\mathrm{B}^{(6)}$ the curves are shown only on the stable side $k<k_{\mathrm{crit}}$.

\begin{figure*}[t]
\centering
\includegraphics[width=0.62\textwidth]{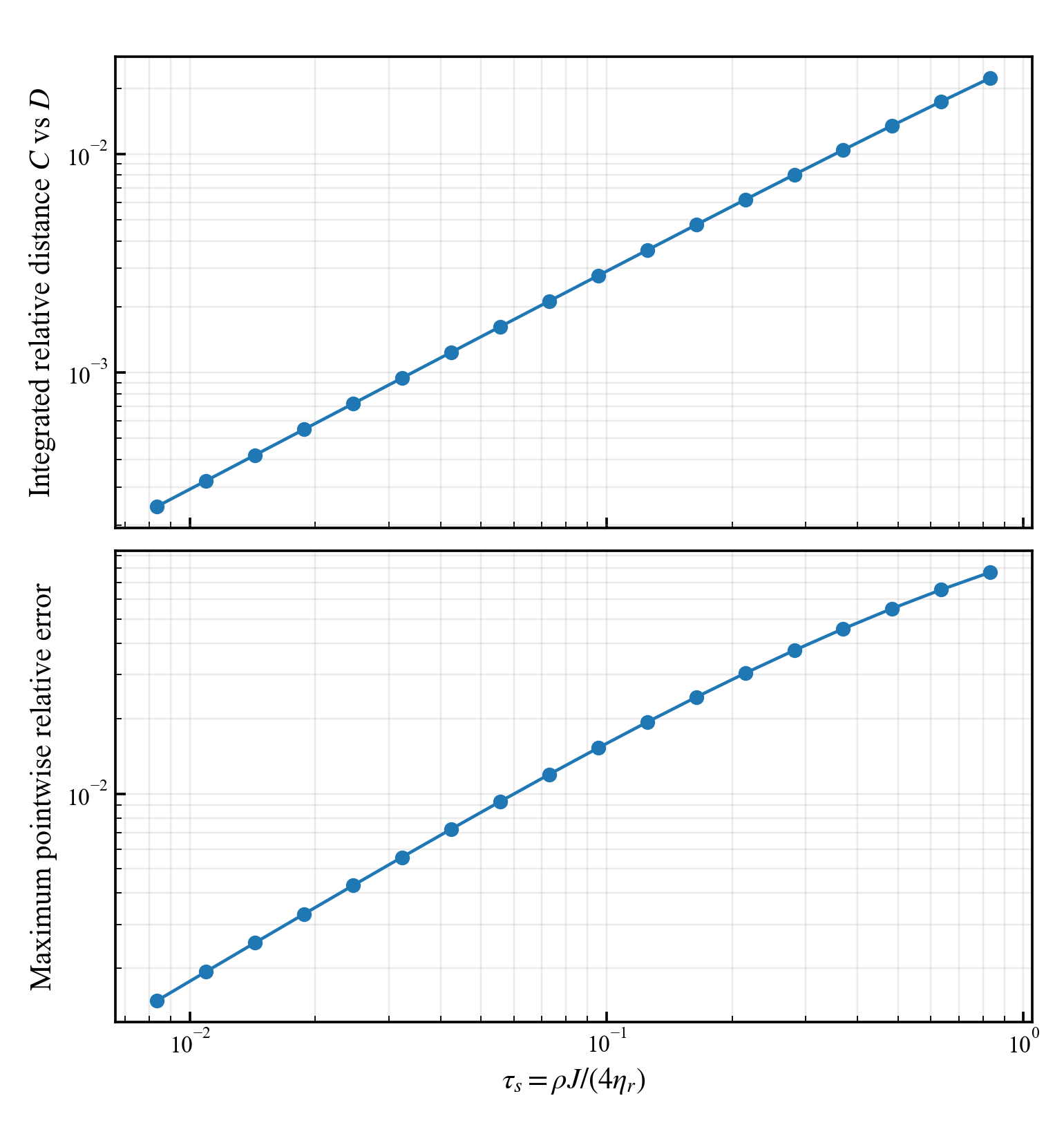}
\caption{
Sensitivity of the C--D separation to the spin-relaxation timescale proxy $\tau_s=\rho J/(4\eta_r)$.
Top: integrated relative distance $\mathcal{D}^{(J)}_{\mathrm{C}\to\mathrm{D}}$ defined in Eq.~(\ref{eq:J_sweep_integrated_distance}).
Bottom: maximum pointwise relative error $E^{(J)}_{\infty,\mathrm{C}\to\mathrm{D}}$ defined in Eq.~(\ref{eq:J_sweep_maxrel}).
As $J$ (equivalently $\tau_s$) increases, the explicit-spin model C separates monotonically from the eliminated-spin model D, quantitatively illustrating when the second relaxation channel becomes more visible in transverse linear response.
}
\label{fig:J_sweep_CD_distance}
\end{figure*}

A complementary question is why Models~C and D are so close in the present data.
To answer this, we sweep the microinertia $J$ while keeping $(\rho,\eta,\eta_r,\beta+\gamma)$ fixed, and we measure the resulting C--D separation over a logarithmic $(k,\Omega)$ grid $\{k_m\}_{m=1}^{N_k}\times\{\Omega_j\}_{j=1}^{N_\Omega}$.
The two quantities shown in Fig.~\ref{fig:J_sweep_CD_distance} are
\begin{align}
\mathcal{D}^{(J)}_{\mathrm{C}\to\mathrm{D}}
&\equiv
\frac{\left[\sum_{m=1}^{N_k}\sum_{j=1}^{N_\Omega}
\left|\chi_{\mathrm{D}}(k_m,\Omega_j;J)-\chi_{\mathrm{C}}(k_m,\Omega_j;J)\right|^2\right]^{1/2}}
{\left[\sum_{m=1}^{N_k}\sum_{j=1}^{N_\Omega}
\left|\chi_{\mathrm{C}}(k_m,\Omega_j;J)\right|^2\right]^{1/2}},
\label{eq:J_sweep_integrated_distance}\\
E^{(J)}_{\infty,\mathrm{C}\to\mathrm{D}}
&\equiv
\max_{m,j}
\frac{\left|\chi_{\mathrm{D}}(k_m,\Omega_j;J)-\chi_{\mathrm{C}}(k_m,\Omega_j;J)\right|}
{\left|\chi_{\mathrm{C}}(k_m,\Omega_j;J)\right|+\varepsilon}.
\label{eq:J_sweep_maxrel}
\end{align}
For the present scan, increasing $\tau_s$ from $8.3\times 10^{-3}$ to $8.3\times 10^{-1}$ increases
$\mathcal{D}^{(J)}_{\mathrm{C}\to\mathrm{D}}$ monotonically from $2.4\times 10^{-4}$ to $2.2\times 10^{-2}$,
while the maximum pointwise relative error rises from $1.5\times 10^{-3}$ to $7.7\times 10^{-2}$.
The trend is nearly monotone on the log--log plot, confirming that the explicit-spin two-pole structure becomes progressively more visible as the spin relaxation time moves toward macroscopic response times.
In this sense, the present parameter choice is not ``too weak'' to distinguish the models;
it is deliberately located in the regime where the fast mode is still present in principle but remains only weakly observable in coarse transverse response data.

\section{Microscopic EDMD observability benchmarks}\label{sec:edmd_benchmarks}
The benchmarks of Sec.~\ref{sec:analysis} were controlled tests of the reduced
linear models. We now complement them by direct many-particle
event-driven molecular dynamics (EDMD) simulations of the elastic perfectly
rough hard-sphere model used throughout the companion derivation paper
\cite{Tsuzuki2026CompanionRetainedSpin}. The purpose of this section is
deliberately narrower than full model validation: unlike the companion paper,
which is coefficient- and closure-oriented, we use EDMD here as an
observability benchmark for the response diagnostics. Instead, we ask which of
the transverse-sector observables identified in
Secs.~\ref{sec:translineardiagno} and \ref{sec:analysis} can be extracted
reliably from noisy microscopic data.

\HLTXT_BLUE{The EDMD calculations below should not be read as a substitute for
continuum DNS or laboratory measurements. They are microscopic observability
benchmarks designed to test whether the transverse-response observables
proposed in Secs.~\ref{sec:translineardiagno} and \ref{sec:analysis} can be
extracted from many-particle data. Direct comparison with DNS and experiments
remains a necessary next step for assessing the broader predictive range of the
proposed diagnostics.}

\HLTXT_ORANGE{The EDMD benchmarks below should therefore not be interpreted as
a direct search for the $B^{(6)}$ instability. Instead, they ask which response
observables are measurable in a microscopic system and whether those
observables discriminate retained-spin dynamics, instantaneous elimination,
and stable one-field kernel shapes.}

Throughout this section we use
three-dimensional periodic-box simulations with $N=8192$ particles at packing
fraction $\phi=0.03$, particle radius $R=0.35$, mass $m=1$, and reduced moment
of inertia $K=4I/(ma^2)=0.4$. We monitor the single-mode observables
\begin{align}
\hat{u}_x(k,t) &= \avg{v_x e^{-iky}}, \\
\hat{\omega}_z(k,t) &= \avg{\omega_z e^{-iky}}, \\
\hat{\zeta}_z(k,t) &= -ik\,\hat{u}_x(k,t),
\end{align}
with discrete transverse wavenumber $k=2\pi n/L_y$ set by the mode index $n$.
For harmonic forcing we impose
\begin{equation}
 a_x(y,t)=a_0\cos(ky)\cos(\Omega t),
 \label{eq:edmd_harmonic_forcing}
\end{equation}
and extract the late-time first-harmonic response by lock-in averaging. The
quantity plotted below is the coherent ensemble mean
$\avg{\hat{q}/\hat{f}}_{\mathrm{seed}}$ of the complex response normalized by the
measured forcing mode $\hat{f}$.

\paragraph*{Convention note.}
In Secs.~\ref{sec:translineardiagno} and \ref{sec:analysis} we wrote
harmonic modes as $e^{st+i\mathbf{k}\cdot\mathbf{x}}$ and set $s=-i\Omega$.
In the EDMD/lock-in analysis below we instead use the signal-processing
convention $q(t)=\Re[\tilde q(\Omega)e^{i\Omega t}]$, so that a delayed
response appears with a negative phase in the measured complex amplitudes.
The two conventions differ only by the sign convention for the complex
frequency (equivalently, by complex conjugation of the response
representation), and no physical conclusion depends on this choice.

\subsection{Free decay: late-time one-pole hydrodynamics}
\label{sec:edmd_free_decay}

\begin{figure*}[t]
\centering
\includegraphics[width=0.95\textwidth]{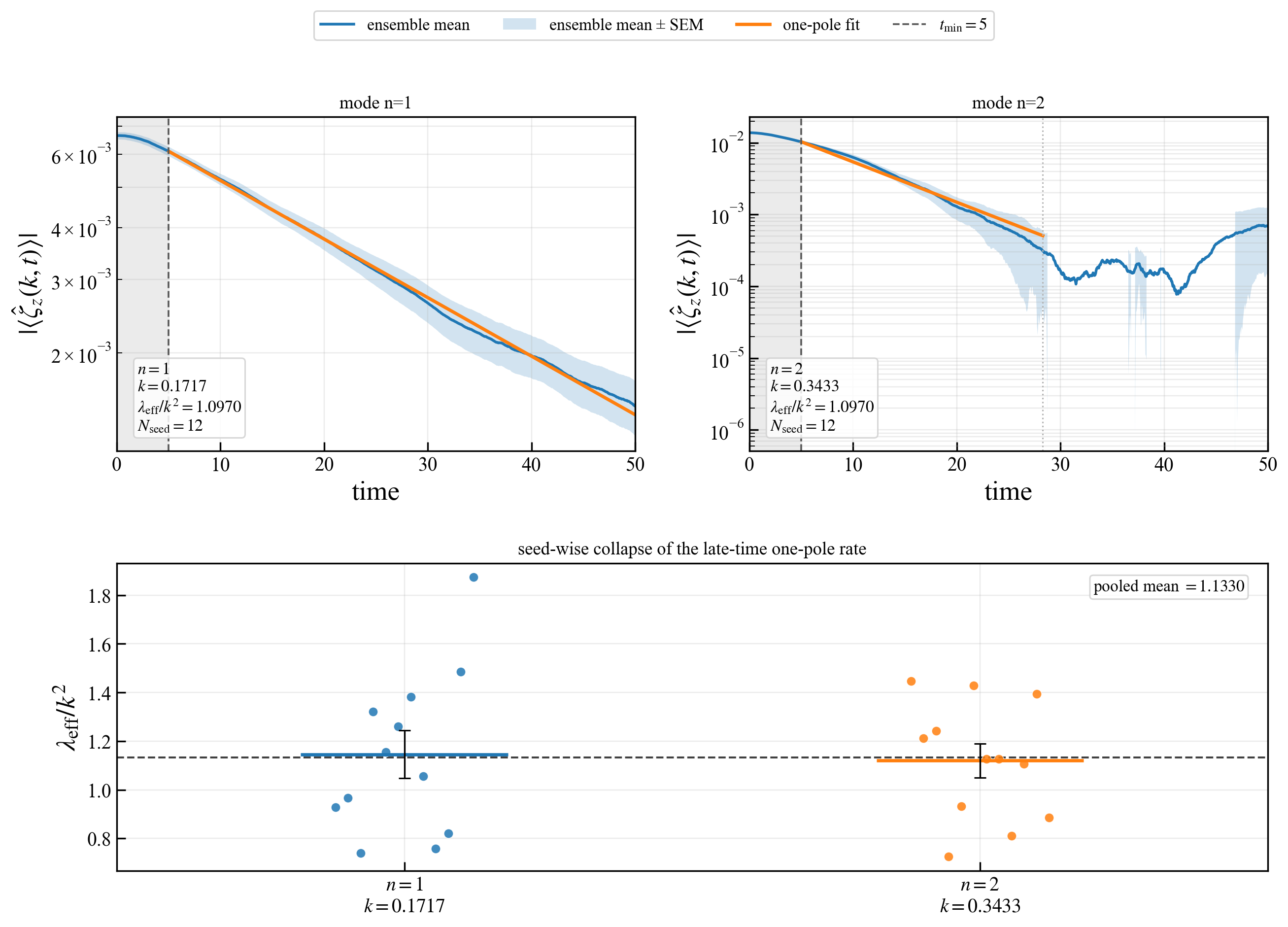}
\caption{
Free-decay EDMD benchmark in the transverse sector. Top row: ensemble-averaged
vorticity-mode amplitudes $|\avg{\hat{\zeta}_z(k,t)}|$ for the two lowest
transverse modes, shown together with the late-time one-pole fit used in the
second-stage analysis. The dashed vertical line marks the fit onset
$t_{\min}=5$, and the dotted line in the $n=2$ panel marks the automatic
amplitude-threshold cutoff of the fit window. Bottom: seed-wise collapse of the
late-time rate $\lambda_{\mathrm{eff}}/k^2$. The ensemble mean is therefore
consistent with a single diffusive pole, while the seed-wise scatter confirms
that the same $k^2$ scaling is recovered independently in the two modes.
}
\label{fig:edmd_free_decay}
\end{figure*}

We first consider unforced decay of a single transverse mode. The initial state
contains a weak sinusoidal velocity perturbation but no imposed mean spin mode,
and we ensemble-average over 12 statistically independent runs for each of the
two lowest transverse modes after phase-aligning $\hat{\zeta}_z(k,0)$. The
result is shown in Fig.~\ref{fig:edmd_free_decay}. Two robust features emerge.
First, once the early kinetic transient has passed, the ensemble-mean vorticity
mode $|\avg{\hat{\zeta}_z(k,t)}|$ is well described by a single exponential over
the fitting window. Specifically, over the post-transient fitting window 
we define the late-time effective decay rate $\lambda_{\mathrm{eff}}(k)$ by
\[
\bigl|\langle \hat{\zeta}_z(k,t)\rangle\bigr|
\approx A_k e^{-\lambda_{\mathrm{eff}}(k)t}.
\]
Equivalently, $\lambda_{\mathrm{eff}}$ is the approximately constant late-time 
value approached by the instantaneous decay rate $\lambda(t;k)$ defined in
Eq.~(\ref{eq:inst_decay_rate}). Second, the fitted decay rates collapse when normalized by
$k^2$: the one-pole fits to the ensemble-averaged runs give
$\lambda_{\mathrm{eff}}/k^2\simeq 1.097$ for both $n=1$ and $n=2$, while the
seed-wise estimates cluster around a common pooled mean of order unity.
Microscopically, the free-decay EDMD therefore supports the late-time
hydrodynamic one-pole sector very cleanly. At the same time, under these
conditions the spin-channel signal remains much weaker than the vorticity
channel, which motivates the forced-response tests below.

\subsection{Harmonic forcing: coherent linear responses of $u_x$ and $\zeta_z$}
\label{sec:edmd_harmonic_uvzeta}

\begin{figure*}[t]
\centering
\includegraphics[width=0.95\textwidth]{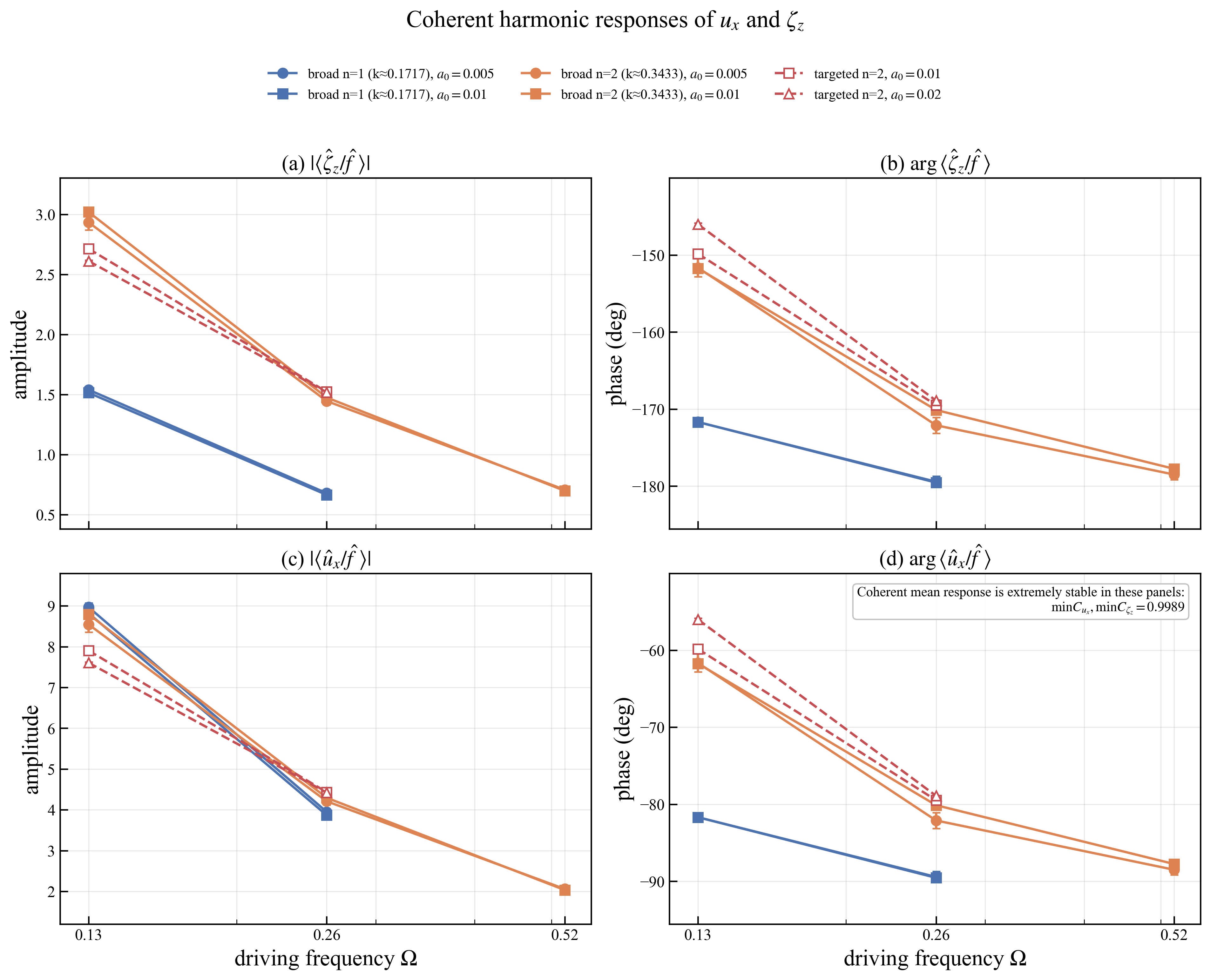}
\caption{
Coherent harmonic responses of $u_x$ and $\zeta_z$ extracted from the EDMD
campaigns. Blue and orange solid curves: broad survey with modes $n=1$ and
$n=2$ and forcing amplitudes $a_0=0.005$ and $0.01$ (8 seeds per point).
Red dashed curves: targeted campaign at the $n=2$ mode with $a_0=0.01$ and $0.02$
(33 seeds per point). Error bars are propagated from the real and imaginary
SEMs of the complex ensemble mean. The $u_x$ and $\zeta_z$ channels are both
highly coherent across the full sweep, and their amplitudes/phases exhibit the
same one-pole trend toward $|\chi|\sim \Omega^{-1}$ with phase approaching
$-90^\circ$ for $u_x$ and $-180^\circ$ for $\zeta_z$.
}
\label{fig:edmd_harmonic_main}
\end{figure*}

We next drive the system with Eq.~(\ref{eq:edmd_harmonic_forcing}) and examine the
coherent first-harmonic responses of $u_x$ and $\zeta_z$. The broad survey uses
modes $n=1$ and $n=2$ with forcing amplitudes $a_0=0.005$ and $0.01$; for
mode $n=2$ we additionally carried out a targeted campaign with a longer
lock-in window and 33 seeds at $a_0=0.01$ and $0.02$. Figure~\ref{fig:edmd_harmonic_main}
collects the resulting ensemble-mean amplitudes and phases. Across all displayed
points, the coherent responses of $u_x$ and $\zeta_z$ are extremely stable:
the minimum coherence is $0.9989$ in both channels. The broad survey and the
targeted $n=2$ campaign show the same qualitative behavior: the amplitude
decreases with $\Omega$, while the phase approaches $-90^\circ$ for
$\avg{\hat{u}_x/\hat{f}}$ and $-180^\circ$ for
$\avg{\hat{\zeta}_z/\hat{f}}$, as expected for an overdamped one-pole
response. Moreover, the two observables are not independent. The identity
$\hat{\zeta}_z=-ik\hat{u}_x$ is satisfied numerically point-by-point, so the
$u_x$ data act mainly as a direct velocity-space cross-check while
$\zeta_z$ remains the most natural observable for comparison with the reduced
response functions.

\begin{figure*}[t]
\centering
\includegraphics[width=0.88\textwidth]{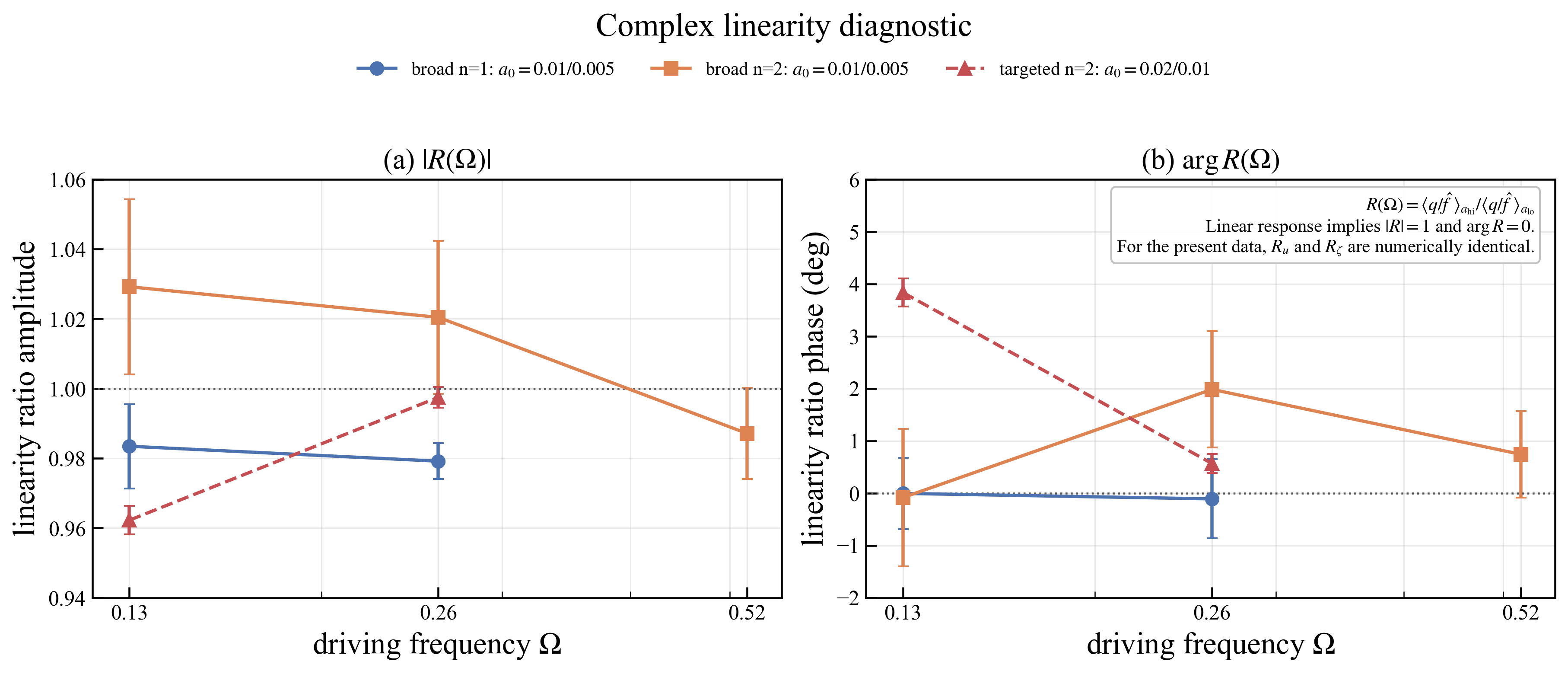}
\caption{
Complex linearity diagnostic for the $u_x$ and $\zeta_z$ responses. For each
campaign we plot the ratio $R_q(\Omega)$ defined in Eq.~(\ref{eq:edmd_linearity_ratio})
between the higher- and lower-amplitude responses. In ideal linear response,
$|R_q|=1$ and $\arg R_q=0$. For the present data, $R_{u_x}$ and $R_{\zeta_z}$
are numerically indistinguishable, so only a single curve is shown for each
campaign. The broad survey remains very close to the linear-response limit, and
the targeted campaign at the $n=2$ mode shows only a modest low-frequency deviation for
the stronger drive.
}
\label{fig:edmd_harmonic_linearity}
\end{figure*}

To make the linearity claim explicit, Fig.~\ref{fig:edmd_harmonic_linearity}
plots the complex amplitude ratio
\begin{equation}
R_q(\Omega) \equiv
\frac{\avg{\hat{q}/\hat{f}}_{a_{\mathrm{hi}}}}
     {\avg{\hat{q}/\hat{f}}_{a_{\mathrm{lo}}}},
\qquad q\in\{u_x,\zeta_z\},
\label{eq:edmd_linearity_ratio}
\end{equation}
for each two-amplitude pair. Perfect linear response implies $|R_q|=1$ and
$\arg R_q=0$. The broad survey lies within a few percent of this ideal over all
displayed frequencies, and even the stronger targeted campaign shows only a
mild low-frequency deviation at $(\Omega,a_0)=(0.13,0.02)$. We therefore
interpret the $u_x$ and $\zeta_z$ data as a coherent linear-response benchmark
for the rough-sphere EDMD system.

\subsection{Targeted spin-sensitive forcing test}
\label{sec:edmd_harmonic_spin}

\begin{figure*}[t]
\centering
\includegraphics[width=0.95\textwidth]{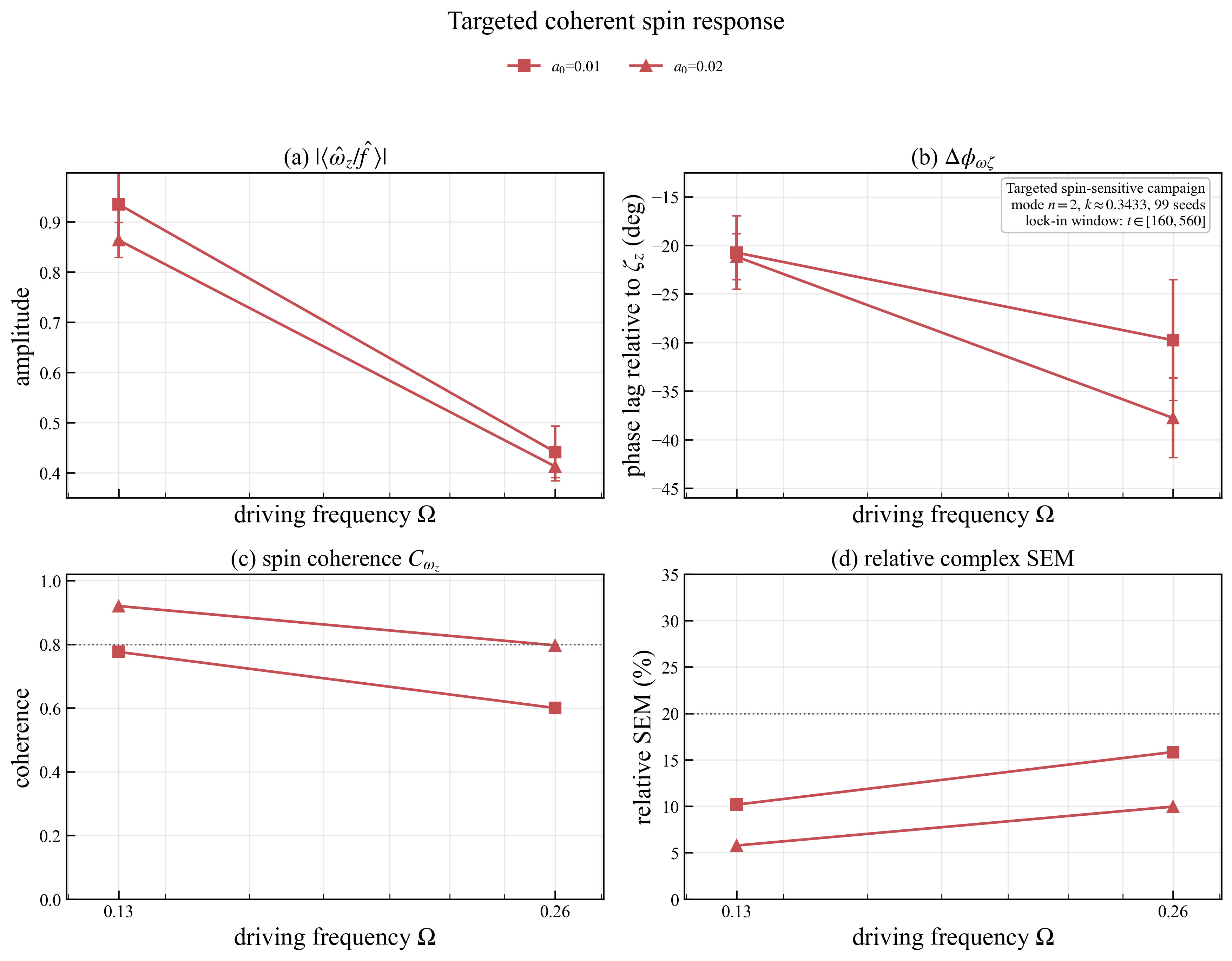}
\caption{
Targeted spin-sensitive harmonic-response campaign at the $n=2$ mode, using
99 seeds per point and a lock-in window $t\in[160,560]$.
Panel (a) shows $|\avg{\hat{\omega}_z/\hat{f}}|$; panel (b) shows the phase lag
$\Delta\phi_{\omega\zeta}$ relative to the vorticity response; panel (c)
shows the spin coherence
$C_{\omega_z}=|\avg{\hat{\omega}_z/\hat{f}}|/\avg{|\hat{\omega}_z/\hat{f}|}$;
and panel (d) shows the relative complex SEM,
$\sqrt{\mathrm{SEM}(\Re)^2+\mathrm{SEM}(\Im)^2}/|\avg{\hat{\omega}_z/\hat{f}}|$.
Error bars in panels (a) and (b) are propagated from the real and imaginary
SEMs of the complex ensemble mean. The stronger drive $a_0=0.02$ yields the
cleanest signal, with $C_{\omega_z}\simeq 0.92$ and $0.80$ at
$\Omega=0.13$ and $0.26$, respectively, and relative complex SEM values of
about $5.8\%$ and $10.0\%$. The weaker drive remains clearly nonzero but
noisier. In both cases the spin response exhibits a systematic finite phase lag
relative to vorticity.}
\label{fig:edmd_harmonic_spin}
\end{figure*}

The remaining question is whether a phase-locked spin response can be extracted
reliably at the many-particle level. For this purpose we focus first on mode
$n=2$ and use a targeted campaign with 99 seeds, a longer lock-in window
$t\in[160,560]$, and forcing amplitudes $a_0=0.01$ and $0.02$.
Figure~\ref{fig:edmd_harmonic_spin} summarizes the results. The quantity
$\avg{\hat{\omega}_z/\hat{f}}$ is now clearly nonzero and phase locked to the
driving. The stronger drive $a_0=0.02$ gives the cleanest signal, with spin
coherence $C_{\omega_z}\simeq 0.92$ at $\Omega=0.13$ and $0.80$ at
$\Omega=0.26$, and relative complex SEM values of about $5.8\%$ and $10.0\%$,
respectively. The weaker drive $a_0=0.01$ still shows a visible response, with
$C_{\omega_z}\simeq 0.78$ and $0.60$ and relative complex SEM values of about
$10.2\%$ and $15.8\%$.

Most importantly, the spin response is not merely nonzero but systematically
shifted relative to the vorticity response. The phase lag
$\Delta\phi_{\omega\zeta}\equiv
\arg\avg{\hat{\omega}_z/\hat{f}}-\arg\avg{\hat{\zeta}_z/\hat{f}}$
is about $-20.8^\circ$ and $-29.8^\circ$ for the weaker drive, and
$-21.2^\circ$ and $-37.8^\circ$ for the stronger drive, at
$\Omega=0.13$ and $0.26$, respectively. Thus the translational and
vorticity channels remain statistically cleaner, but the targeted mode-$2$
campaign now provides an unambiguous many-particle demonstration of a coherent
spin-channel response with a measurable finite lag. The model-selection use of
this lag, and its multi-$k$ extension, is taken up in the Discussion below.

\section{Discussion}\label{sec:discussion}
The results above sharpen the diagnostic message in three ways.

\paragraph*{(i) Pole counting identifies explicit retained spin, but only when the fast mode is resolvable.}
Model~C is distinguished by a second pole in
$\chi_{\zeta\zeta}^{\mathrm{C}}$ [Eq.~(\ref{eq:chi_case1})], associated with the
spin-relaxation timescale $\tau_s\sim \rho J/(4\eta_r)$. For the representative
parameter set of Sec.~\ref{sec:analysis}, this fast pole is strongly damped:
it appears as a short transient in free decay
(Figs.~\ref{fig:setting1_decay}--\ref{fig:setting1_rate}) and produces only a
sub-percent correction to the steady forced response (Fig.~\ref{fig:map_CD}).
That is precisely the regime in which adiabatic elimination succeeds.
Figure~\ref{fig:J_sweep_CD_distance} then clarifies the complementary point:
as $J$ increases, the C--D separation grows systematically, so the second
relaxation channel becomes progressively more visible in macroscopic transverse
response.

\paragraph*{(ii) Among one-pole models, the decisive distinction is rational versus polynomial kernel structure.}
The most robust separation in the present data is between the rational
eliminated-spin kernel (Model~D) and the finite polynomial surrogate
(Model~B). The two polynomial variants illustrate two distinct failure modes.
The strict $B^{(4)}$ truncation remains stable but becomes increasingly
over-damped at large $k$, because its damping grows like a polynomial rather
than returning to the $k^2$ roll-off of the rational kernel.
The matched $B^{(6)}$ truncation behaves differently: the negative $k^6$
coefficient needed to match the low-$k$ series induces near-critical
cancellation and, eventually, finite-$k$ instability
\cite{Bobylev1982,Uribe2000,JinSlemrod2001,StruchtrupTorrilhon2003,Struchtrup2005}.
\HLTXT_BLUE{This behavior is consistent with the classical observation that
finite Burnett-type truncations can develop unphysical short-wavelength
instabilities, although the present $B^{(6)}$ model is used only as a
diagnostic low-$k$-matched surrogate.}
The point is structural rather than tied to any single numerical example:
no finite polynomial can reproduce both the low-$k$ expansion and the large-$k$
asymptote of the eliminated-spin kernel.

This statement should nevertheless be read with the correct scope.
Model~B is deliberately minimal. It is not intended to represent every
regularized Burnett, Grad, or higher-moment theory. Those more elaborate
closures can modify the pole structure or restore stability by adding
relaxation channels. The role of Model~B here is narrower and more precise:
it isolates what is lost when the eliminated-spin kernel is replaced by a
finite polynomial in $k$. In that sense, the stable overdamping of $B^{(4)}$
and the near-critical pathology of $B^{(6)}$ are diagnostic examples, not a
universal theorem about all higher-order closures.

\paragraph*{(iii) The many-particle EDMD section now reaches restricted but genuine closure discrimination.}
The many-particle simulations show that several of the response-based
observables survive beyond the reduced-model setting. Free decay isolates a
clean late-time one-pole sector; broad harmonic forcing yields highly coherent
responses in $u_x$ and $\zeta_z$; and the targeted campaigns show that the spin
channel is not only observable but discriminating. What is still missing is a
full transport-coefficient inversion. The coefficient-oriented Chapman--Enskog
side of the problem is treated separately in
Ref.~\cite{Tsuzuki2026CompanionRetainedSpin}; the EDMD benchmarks here instead
show which observables already carry model-selection information in a
microscopic rough-sphere system.

\HLTXT_BLUE{Accordingly, the EDMD results should be interpreted as microscopic
evidence for the observability and discriminating power of the proposed
response functions, not as a comprehensive validation against DNS or
experimental data.}

\begin{figure*}[t]
    \centering
    \includegraphics[width=\linewidth]{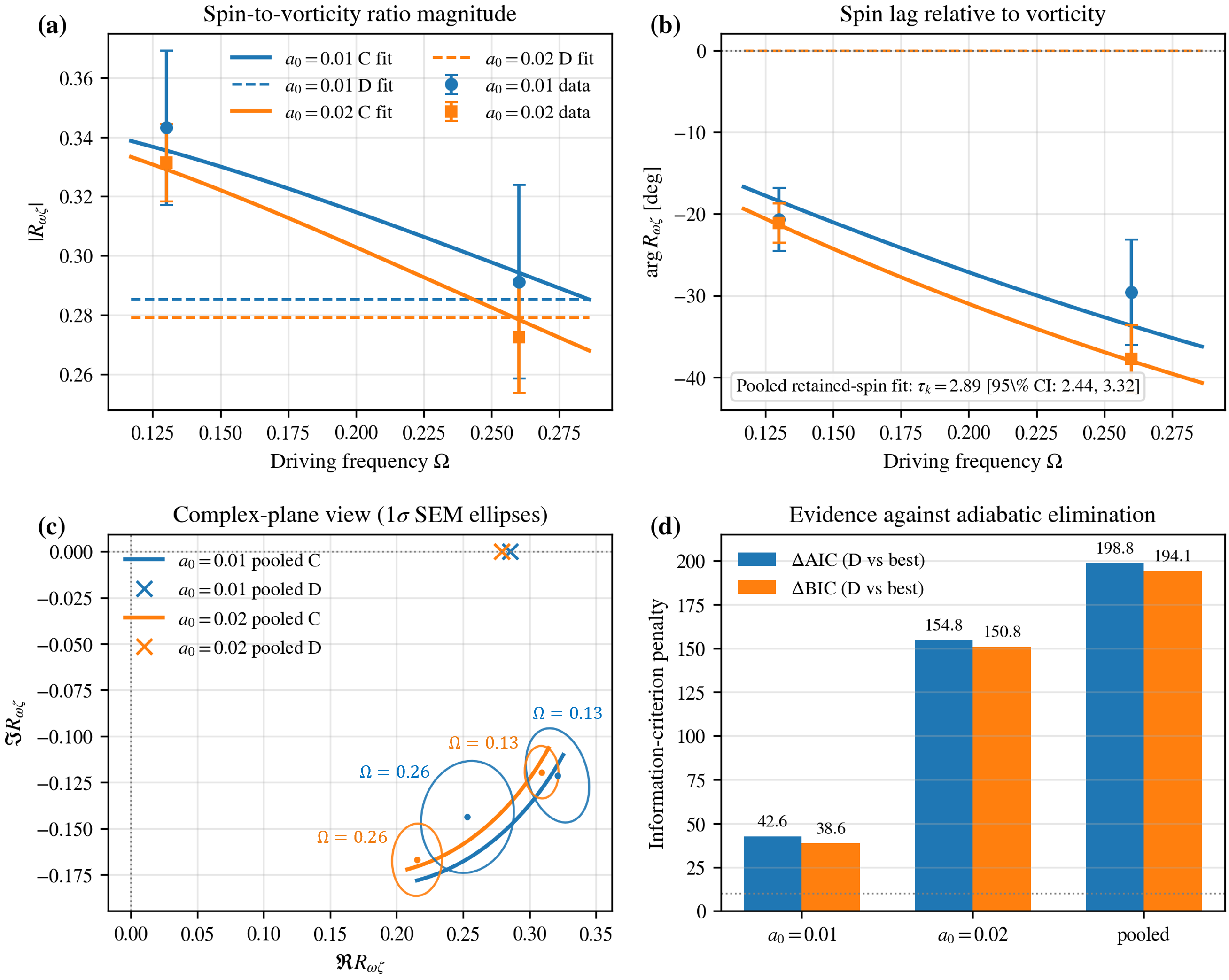}
    \caption{
    Discrimination test using the 99-seed targeted $n=2$ harmonic-forcing EDMD data.
    The complex spin-to-vorticity ratio
    $R_{\omega\zeta}=\hat{\omega}_{z}/\hat{\zeta}_{z}$
    is compared with a retained-spin model (Model C) and an adiabatically eliminated model (Model D).
    (a) Magnitude and (b) phase of $R_{\omega\zeta}$ for $a_{0}=0.01$ and $0.02$.
    The retained-spin fit captures the observed frequency-dependent lag, whereas adiabatic elimination predicts a frequency-independent real ratio.
    (c) Complex-plane representation with $1\sigma$ SEM ellipses.
    (d) Information-criterion penalties against adiabatic elimination:
    $\Delta\mathrm{AIC}/\Delta\mathrm{BIC}=42.6/38.6$ for $a_{0}=0.01$,
    $154.8/150.8$ for $a_{0}=0.02$, and $198.8/194.1$ for the pooled shared-$\tau_k$ fit.
    The pooled retained-spin fit yields $\tau_{k}=2.89$ with bootstrap $95\%$
    confidence interval $[2.44,\,3.32]$.}
    \label{fig:section_VC_spin_ratio_discrimination}
\end{figure*}

A fixed-$k$ discriminator is obtained from the complex spin-to-vorticity ratio
$R_{\omega\zeta}(\Omega)\equiv \avg{\hat{\omega}_z/\hat{\zeta}_z}$.
At fixed $k$, adiabatic elimination predicts a frequency-independent real ratio,
whereas the retained-spin model predicts
$R_{\omega\zeta}=c_0/(1+i\Omega\tau_k)$. For the 99-seed targeted $n=2$
campaign, this test strongly favors retained spin:
$\Delta\mathrm{AIC}/\Delta\mathrm{BIC}=42.6/38.6$ at $a_0=0.01$,
$154.8/150.8$ at $a_0=0.02$, and $198.8/194.1$ in the pooled shared-$\tau_k$
fit. The pooled estimate is $\tau_k=2.89$ with bootstrap 95\% confidence
interval $[2.44,\,3.32]$. Thus even within a single-mode forcing test, the
EDMD data resolve a finite spin-relaxation lag that is incompatible with
instantaneous elimination (Fig.~\ref{fig:section_VC_spin_ratio_discrimination}).

\begin{figure*}[t]
    \centering
    \includegraphics[width=0.94\textwidth]{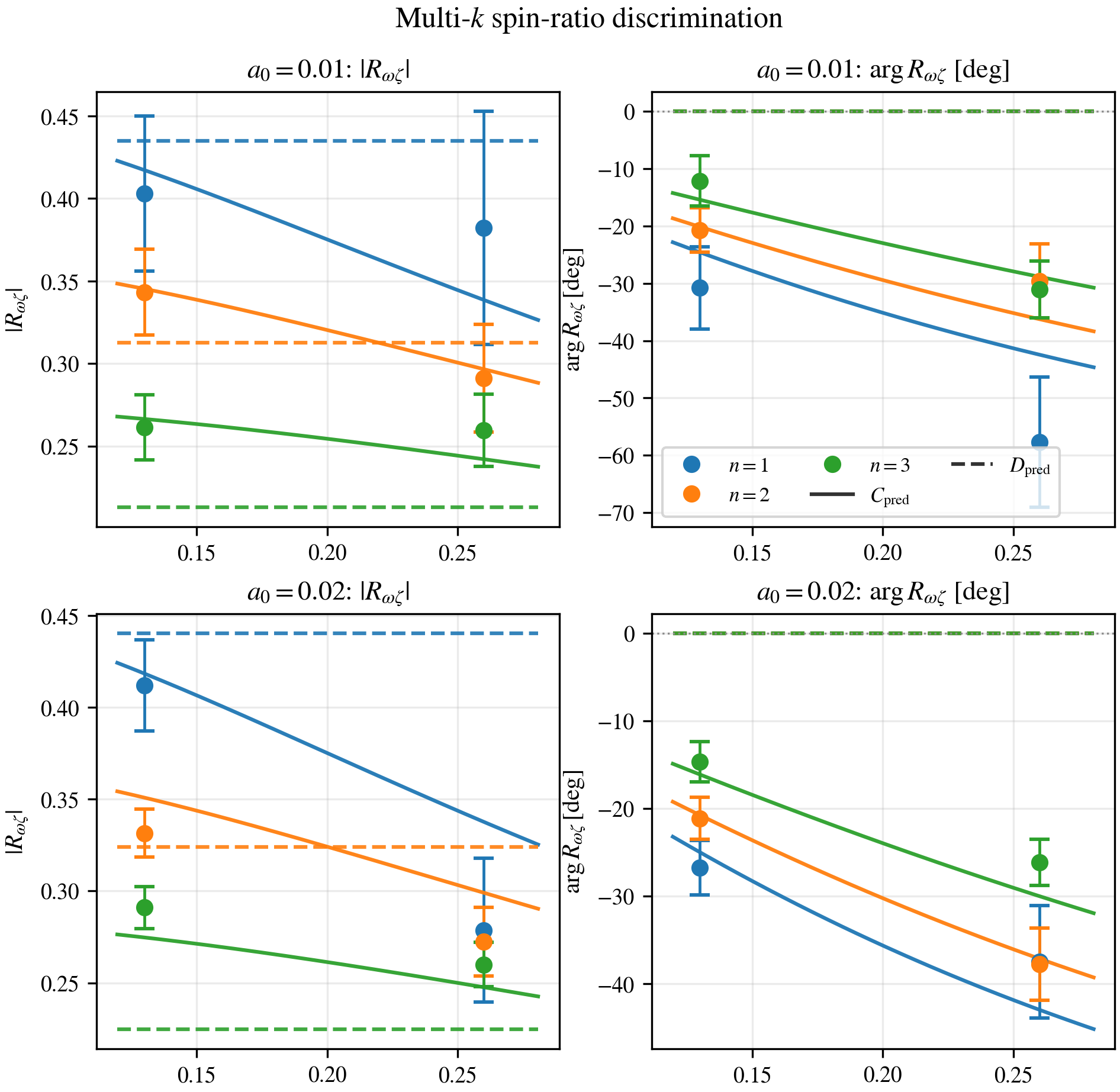}
    \caption{
    Multi-$k$ discrimination test using the extended
    Section~\ref{sec:edmd_harmonic_spin} harmonic-forcing EDMD data at the
    sampled modes $n=1,2,3$ (99 seeds per condition). The plotted quantity is
    the complex spin-to-vorticity ratio
    $R_{\omega\zeta}=\hat{\omega}_{z}/\hat{\zeta}_{z}$.
    The left column shows the magnitude $|R_{\omega\zeta}|$ and the right
    column shows the phase $\arg R_{\omega\zeta}$; the top row corresponds to
    $a_{0}=0.01$ and the bottom row to $a_{0}=0.02$. Symbols with error bars
    are the EDMD ensemble means for the three sampled modes, with uncertainties
    propagated from the real and imaginary SEMs of the complex mean. Solid
    curves ($C_{\mathrm{pred}}$) are the pooled retained-spin prediction,
    whereas dashed lines ($D_{\mathrm{pred}}$) show the corresponding
    adiabatic-elimination prediction, which is frequency independent at fixed
    $k$. Across all three sampled modes, the data exhibit a systematic
    frequency-dependent lag that is captured by the retained-spin model and is
    incompatible with instantaneous elimination. In the pooled fit, the
    physically constrained retained-spin model is strongly favored over
    adiabatic elimination, with
    $\Delta\mathrm{AIC}/\Delta\mathrm{BIC}=604.4/603.2$, and yields
    $\lambda_{M}=2.87$ and $\tau=3.87$ with bootstrap $95\%$ confidence intervals
    $[2.55,\,3.22]$ and $[3.55,\,4.22]$, respectively.}
    \label{fig:section_VC_multik_ratio_discrimination}
\end{figure*}

The stronger statement comes from extending the targeted campaign to modes
$n=1,2,3$. The multi-$k$ ratio $R_{\omega\zeta}(k,\Omega)$ is then well
described by the physically constrained retained-spin form
$R=1/[2(1+\lambda_{M} k^2+i\tau\Omega)]$ and is strongly inconsistent with the
adiabatically eliminated prediction $R=1/[2(1+\lambda_{M} k^2)]$.
The retained-spin model is favored by
$\Delta\mathrm{AIC}/\Delta\mathrm{BIC}=128.6/128.1$ at $a_0=0.01$,
$473.3/472.8$ at $a_0=0.02$, and $604.4/603.2$ in the pooled analysis.
The pooled fit gives $\lambda_{M}=2.87$ and $\tau=3.87$ with bootstrap 95\%
confidence intervals $[2.55,\,3.22]$ and $[3.55,\,4.22]$, respectively.
This is a genuine multi-$k$ discrimination of retained-spin dynamics against
instantaneous elimination, rather than a mere observability statement
(Fig.~\ref{fig:section_VC_multik_ratio_discrimination}).

\begin{figure*}[t]
    \centering
    \includegraphics[width=0.90\textwidth]{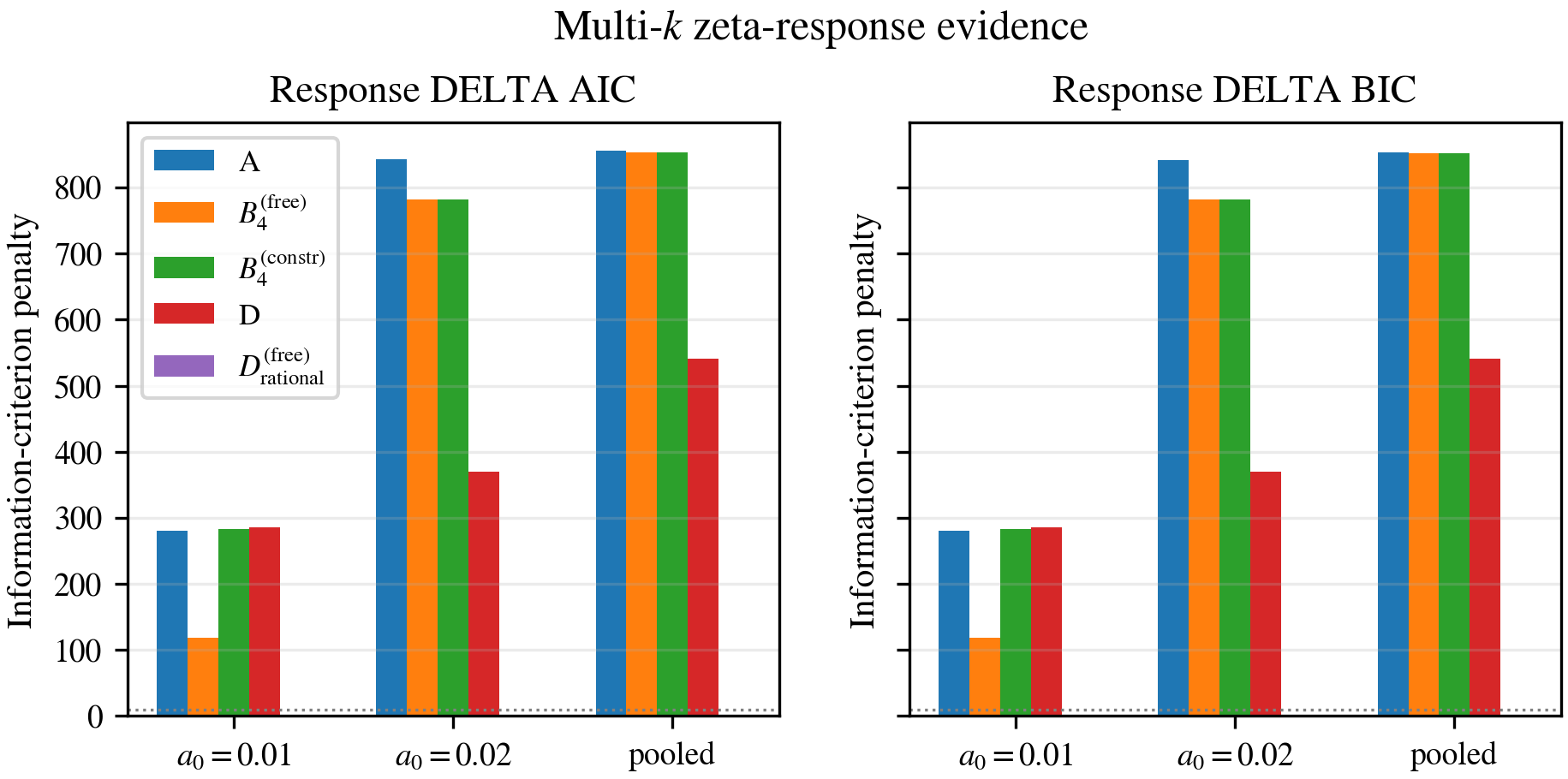}
    \caption{
    Information-criterion evidence from the multi-$k$ vorticity response
    $\avg{\hat{\zeta}_{z}/\hat{f}}$ in the extended
    Section~\ref{sec:edmd_harmonic_spin} campaign.
    Bars show $\Delta$AIC and $\Delta$BIC relative to the flexible one-pole
    reference $D^{(\mathrm{free})}_{\mathrm{rational}}$ for the physical one-field closures
    A, $B^{(\mathrm{free})}_{4}$, $B^{(\mathrm{constr})}_{4}$, and D.
    The pure $k^2$ closure A is strongly disfavored in every dataset.
    The weaker-drive dataset $a_0=0.01$ still prefers the polynomial surrogate
    $B^{(\mathrm{free})}_{4}$ over the rational model D, but the stronger-drive and
    pooled datasets reverse that ordering and strongly favor D over
    $B^{(\mathrm{free})}_{4}$.
    The multi-$k$ vorticity channel therefore carries kernel-shape information
    beyond a pure $k^2$ response and, once statistics are strengthened, becomes
    informative about rational-versus-polynomial one-field structure.}
    \label{fig:section_VC_multik_response_evidence}
\end{figure*}

The same multi-$k$ dataset also clarifies what the one-field vorticity response
can and cannot tell us. The measured $\avg{\hat{\zeta}_z/\hat{f}}$ strongly
rejects a pure $k^2$ closure (Model~A): relative to the flexible one-pole
reference $D^{(\mathrm{free})}_{\mathrm{rational}}$, the penalties are
$\Delta\mathrm{AIC}/\Delta\mathrm{BIC}=281.1/280.1$ at $a_0=0.01$,
$842.2/841.2$ at $a_0=0.02$, and $855.2/852.8$ in the pooled fit.
Between the two nontrivial one-field closures, the weaker-drive dataset
$a_0=0.01$ does not yet prefer the rational kernel D over the polynomial
surrogate $B^{(\mathrm{free})}_{4}$, but the stronger-drive and pooled datasets do:
relative to $B^{(\mathrm{free})}_{4}$, Model~D gains
$412.7/412.2$ information-criterion units at $a_0=0.02$ and
$312.1/310.9$ in the pooled comparison
(Fig.~\ref{fig:section_VC_multik_response_evidence}).
Thus the spin ratio is the cleanest C--D discriminator, while the multi-$k$
vorticity response supplies the one-field kernel-shape information needed for
A/B/D comparisons.
\HLTXT_ORANGE{The polynomial competitors in the multi-$k$ EDMD
vorticity-response analysis are stable fitted $k^4$ one-field families, not
the fixed matched $B^{(6)}$ benchmark used to illustrate the finite-polynomial
instability mechanism in Sec.~\ref{sec:analysis}.}

Taken together, Figs.~\ref{fig:map_CD}--\ref{fig:map_CB} and the EDMD results of
Sec.~V suggest a practical response-based inference strategy in the transverse
sector. First, use fixed-$k$ spin-sensitive forcing to test for a finite
spin lag through $R_{\omega\zeta}$; this is the cleanest retained-spin-versus-
elimination discriminator. Second, if only one pole is resolved in the
macroscopic channel, use multi-$k$ vorticity response to discriminate rational
from polynomial one-field kernels. Third, reserve full coefficient inversion
for broader parameter sweeps and companion Chapman--Enskog transport
calculations. A statistically robust multi-$k$/multi-$\Omega$ framework that
combines both vorticity and spin channels remains an important next step, but
the present EDMD benchmarks show that such an inference program is now feasible
in principle.

\section{Conclusion}\label{sec:conclusion}
This paper develops a response-based framework for distinguishing three
mechanistically different routes to high-curvature transverse behavior:
explicit retained-spin dynamics, adiabatic elimination of a fast spin variable,
and finite polynomial higher-gradient closure. The main contribution is not the
mere presence of a Navier--Stokes-order micropolar closure, which has clear
antecedents, but the demonstration that these routes lead to different pole
structures, different $k$-kernels, and therefore different measurable response
signatures.

\HLTXT_BLUE{The conclusions of the present paper are restricted to this linear
transverse-response setting. They should not be read as a claim of full
nonlinear model validation; extending the same discrimination strategy to
nonlinear dynamics remains an important next step.}

The retained-spin closure used here is derived separately in the companion
paper~\cite{Tsuzuki2026CompanionRetainedSpin}. The present manuscript uses that
closure as a response-theoretic starting point and shows why eliminating an
internal variable generates a rational kernel whose low-$k$ expansion mimics
Burnett-type terms but whose full finite-$k$ behavior is not captured by any
finite polynomial truncation. This distinction becomes operational in the
transverse linear response: Model~C has two poles, Model~D captures the slow
branch with a rational one-pole kernel, and polynomial surrogates deform that
response either through stable overdamping [$B^{(4)}$] or near-critical
amplification and instability [$B^{(6)}$].

The reduced single-mode calculations and the many-particle EDMD benchmarks serve
complementary purposes. The former are controlled tests of the response classes
themselves. The latter now provide a restricted but genuine microscopic
discrimination. Free decay isolates a one-pole hydrodynamic sector, broad
harmonic forcing yields highly coherent responses in $u_x$ and $\zeta_z$, and
targeted fixed-$k$ spin-sensitive forcing resolves a finite spin-relaxation lag
that strongly favors retained spin over instantaneous adiabatic elimination.
Extending that analysis to multiple modes sharpens the conclusion further:
the spin-to-vorticity ratio supports a pooled retained-spin fit over
adiabatic elimination by $\Delta\mathrm{AIC}/\Delta\mathrm{BIC}=604.4/603.2$,
while the stronger-drive and pooled multi-$k$ vorticity data reject a pure
$k^2$ closure and favor the rational eliminated-spin kernel over a polynomial
one-field surrogate. Within that scope, the EDMD results do more than establish
observability; they materially strengthen the closure-discrimination program.
What they still do not provide is a full transport-coefficient inversion.

A quantitatively complete theory will ultimately combine the response-based
discrimination developed here with the coefficient-oriented transport program of
Ref.~\cite{Tsuzuki2026CompanionRetainedSpin}. \HLTXT_BLUE{It also remains important to test the present criteria against
wider parameter sweeps, broader direct kinetic simulations, continuum DNS, and
laboratory experiments.} Within those limits, the present work shows
that transverse response is not merely a formal way of classifying closures: it
is a practical diagnostic language for connecting microscopic rotational
physics to macroscopic effective hydrodynamics.

\appendix
\section{Model-selection protocol for the AIC/BIC analyses}
\label{app:model_selection}
\paragraph*{General likelihood, SEM weighting, and plotted uncertainties.}
The information-criterion comparisons in
Figs.~\ref{fig:section_VC_spin_ratio_discrimination}--\ref{fig:section_VC_multik_response_evidence}
are performed directly on the complex ensemble means, not on separately fitted
magnitudes and phases.  For each forcing condition $j$, let $z_{j,s}\in\mathbb{C}$
denote the seed-level complex response and let
\begin{equation}
\bar z_j = \frac{1}{N_j}\sum_{s=1}^{N_j} z_{j,s}
\end{equation}
be the corresponding ensemble mean.  We then define the real two-vector
\begin{equation}
y_j = \begin{pmatrix} \Re\bar z_j \\ \Im\bar z_j \end{pmatrix},
\qquad
\mu_j(\theta) = \begin{pmatrix} \Re z_j^{\mathrm{mod}}(\theta) \\ \Im z_j^{\mathrm{mod}}(\theta) \end{pmatrix},
\end{equation}
where $z_j^{\mathrm{mod}}(\theta)$ is the model prediction for parameter vector
$\theta$.  The $2\times 2$ covariance used in the weighting is the covariance of
the condition mean,
\begin{equation}
\Sigma_j = \frac{1}{N_j}\,\mathrm{Cov}_{\mathrm{seed}}
\begin{pmatrix}
\Re z_{j,s} \\ \Im z_{j,s}
\end{pmatrix}.
\end{equation}
Thus the real and imaginary parts are fitted jointly, allowing for a nonzero
covariance between them.  The Gaussian log-likelihood is
\begin{equation}
-2\ln L(\theta)
= \sum_j \bigl[y_j-\mu_j(\theta)\bigr]^\top \Sigma_j^{-1}
\bigl[y_j-\mu_j(\theta)\bigr] + \mathrm{const},
\label{eq:app_aicbic_likelihood}
\end{equation}
which is equivalent to SEM-weighted least squares on the whitened residuals.
Because the omitted normalization term depends only on the fixed data covariance,
it cancels in all $\Delta\mathrm{AIC}$ and $\Delta\mathrm{BIC}$ comparisons.
We therefore report
\begin{equation}
\mathrm{AIC}=2p-2\ln L,
\qquad
\mathrm{BIC}=p\ln N_{\mathrm{obs}}-2\ln L,
\label{eq:app_aicbic_def}
\end{equation}
with $p$ the number of fitted parameters and $N_{\mathrm{obs}}$ the adopted
scalar observation count.

Throughout this appendix we use the same lock-in convention as in
Sec.~\ref{sec:edmd_benchmarks}, namely
$q(t)=\Re[\tilde q(\Omega)e^{i\Omega t}]$ for the fitted complex amplitudes.
Relative to the $e^{st}$ convention of
Secs.~\ref{sec:translineardiagno} and \ref{sec:analysis} with $s=-i\Omega$,
this reverses the explicit sign of the $i\Omega$ terms but not the physical
content of the response functions.

The plotted magnitude and phase error bars are propagated from the same
$2\times 2$ SEM covariance.  Writing $\bar z_j=u_j+i v_j$, the Jacobians are
\begin{equation}
\nabla |\bar z_j| = \frac{1}{|\bar z_j|}\begin{pmatrix}u_j\\v_j\end{pmatrix},
\qquad
\nabla \arg\bar z_j = \frac{1}{u_j^2+v_j^2}\begin{pmatrix}-v_j\\u_j\end{pmatrix},
\end{equation}
so that
\begin{equation}
\sigma^2_{|z|,j} \approx \nabla |\bar z_j|^\top \Sigma_j\nabla |\bar z_j|,
\qquad
\sigma^2_{\phi,j} \approx \nabla \arg\bar z_j^\top \Sigma_j\nabla \arg\bar z_j.
\end{equation}
This is the same linear error-propagation rule used to draw the error bars in
Figs.~\ref{fig:section_VC_spin_ratio_discrimination} and
\ref{fig:section_VC_multik_ratio_discrimination}.

\paragraph*{Fixed-$k$ spin-ratio test {\rm (Fig.~\ref{fig:section_VC_spin_ratio_discrimination})}.}
For the targeted Section~\ref{sec:edmd_harmonic_spin} campaign at fixed $k$, the model comparison is
carried out on the complex spin-to-vorticity ratio
\begin{equation}
R_{\omega\zeta}(\Omega)=\left\langle \hat\omega_z/\hat\zeta_z\right\rangle.
\end{equation}
The two candidate forms are
\begin{equation}
R_D(\Omega)=c_0,
\qquad
R_C(\Omega)=\frac{c_0}{1+i\Omega\tau_k},
\end{equation}
corresponding respectively to instantaneous adiabatic elimination (Model~D)
and to explicit retained spin (Model~C).  For each amplitude separately, the
parameter counts are $p_D=1$ and $p_C=2$.  In the pooled shared-$\tau_k$
analysis across amplitudes, the amplitudes have separate $c_0$ values but a
common lag time $\tau_k$, so that
\begin{equation}
p_D=N_a,
\qquad
p_C=N_a+1,
\end{equation}
where $N_a$ is the number of forcing amplitudes.

We evaluate the BIC in Fig.~\ref{fig:section_VC_spin_ratio_discrimination} 
using the seed-level scalar count,
\begin{equation}
N_{\mathrm{obs}}^{(16)} = 2 N_\Omega N_{\mathrm{seed}} \label{eq:ScalarCountSingleAmp}
\end{equation}
for a single-amplitude fit and
\begin{equation}
N_{\mathrm{obs}}^{(16)} = 2 N_a N_\Omega N_{\mathrm{seed}} \label{eq:ScalarCountPooledFit}
\end{equation}
for the pooled fit, reflecting the real and imaginary parts of all seed-level complex responses across the fitted conditions.
Here the likelihood itself remains the condition-mean likelihood of
Eq.~\eqref{eq:app_aicbic_likelihood}; only the BIC penalty count is taken at
the seed level.  We use this convention in
Fig.~\ref{fig:section_VC_spin_ratio_discrimination} as a practical choice for
the fixed-$k$ spin-ratio test, where the number of fitted conditions is small
and each condition mean is itself estimated from many seeds.  For the present
nested C--D comparison, replacing Eqs.~(\ref{eq:ScalarCountSingleAmp})--(\ref{eq:ScalarCountPooledFit}) by a condition-mean count
would shift the reported $\Delta\mathrm{BIC}$ values upward by
$(p_C-p_D)\ln N_{\mathrm{seed}}=\ln N_{\mathrm{seed}}$, but would not change
the qualitative ordering.
Because Section~\ref{sec:edmd_harmonic_spin} at fixed $k$ samples only one wavenumber, this analysis is intended only as a retained-spin
versus elimination test; after $k$ is fixed, one-field polynomial surrogates and
a one-field rational kernel collapse to the same one-pole structure up to an
effective real damping constant.  Confidence intervals for the pooled lag are
obtained by nonparametric bootstrap resampling of the seed-level complex ratios
within each $(a_0,\Omega)$ condition followed by refitting the shared-$\tau_k$
model. We performed $300$ such bootstrap resamples.

\paragraph*{Multi-$k$ spin-ratio test {\rm (Fig.~\ref{fig:section_VC_multik_ratio_discrimination})}.}
For the multi-$k$ extension, the primary observable is again the complex ratio
$R_{\omega\zeta}(k,\Omega)=\langle \hat\omega_z/\hat\zeta_z\rangle$, but we fit two nested model families.  The first is a free-$k$ family,
\begin{equation}
R_D^{\mathrm{free}}(k,\Omega)=c_k,
\qquad
R_C^{\mathrm{free}}(k,\Omega)=\frac{c_k}{1+i\Omega\tau_k},
\end{equation}
with parameter counts $p_D=N_k$ and $p_C=2N_k$.  The second is the
physically structured family quoted in the manuscript,
\begin{equation}
R_D^{\mathrm{phys}}(k,\Omega)=\frac{1}{2(1+\lambda_{M} k^2)},
\qquad
R_C^{\mathrm{phys}}(k,\Omega)=\frac{1}{2(1+\lambda_{M} k^2+i\tau\Omega)},
\end{equation}
with parameter counts $p_D=1$ and $p_C=2$.  The values reported in
Fig.~\ref{fig:section_VC_multik_ratio_discrimination} and in the corresponding
main-text discussion are those of the theory-structured pair
$\{R_D^{\mathrm{phys}},R_C^{\mathrm{phys}}\}$.

In the present multi-$k$ analysis, the default BIC convention is based on the
complex condition means,
\begin{equation}
N_{\mathrm{obs}}^{(17)} = 2 N_{\mathrm{cond}},
\qquad
N_{\mathrm{cond}}=N_a N_k N_\Omega,
\end{equation}
which counts the real and imaginary parts of the condition means used in the fit.
Bootstrap intervals for the pooled physical retained-spin fit are
obtained by independently resampling the seed-level complex ratios within each
$(a_0,k,\Omega)$ condition and refitting $(\lambda_{M},\tau)$. 
We performed $500$ bootstrap resamples for Fig.~\ref{fig:section_VC_multik_ratio_discrimination}.

\paragraph*{Multi-$k$ vorticity-response evidence {\rm (Fig.~\ref{fig:section_VC_multik_response_evidence})}.}
For the multi-$k$ vorticity-response comparison we fit the complex response
\begin{equation}
\chi_{\zeta\zeta}(k,\Omega)=\left\langle \hat\zeta_z/\hat f\right\rangle
\end{equation}
to the one-pole families used in Sec.~\ref{sec:translineardiagno}, namely
\begin{align}
\chi_A(k,\Omega)
&= \frac{-ik}{\nu k^2+i\Omega},
\\
\chi_{B^{\mathrm{free}}}(k,\Omega)
&= \frac{-ik}{\nu k^2+B_1k^4+i\Omega},
\\
\chi_{B^{\mathrm{constr}}}(k,\Omega)
&= \chi_{B^{\mathrm{free}}}(k,\Omega)\quad\text{with}\quad B_1\ge 0,
\\
\chi_D(k,\Omega)
&= \frac{-ik}{\nu k^2+\dfrac{\nu_r\mu k^4}{4\nu_r+\mu k^2}+i\Omega},
\\
\chi_{\mathrm{free}}(k,\Omega)
&= \frac{-ik}{K_k+i\Omega}.
\end{align}
In Fig.~\ref{fig:section_VC_multik_response_evidence} these appear as the bars
labeled $A$, $B_{4}^{(\mathrm{free})}$, $B_4^{(\mathrm{constr})}$, $D$, and
$D_{\mathrm{rational}}^{(\mathrm{free})}$, respectively.  The corresponding
parameter counts are
\begin{equation}
p_A=1,
\qquad
p_{B^{\mathrm{free}}}=2,
\qquad
p_{B^{\mathrm{constr}}}=2,
\qquad
p_D=3,
\qquad
p_{\mathrm{free}}=N_k.
\end{equation}
The condition-mean convention,
\begin{equation}
N_{\mathrm{obs}}^{(18)} = 2 N_{\mathrm{cond}},
\end{equation}
is used for BIC.  The displayed $\Delta\mathrm{AIC}$ and
$\Delta\mathrm{BIC}$ values are reported relative to the minimum AIC/BIC within
the response-model family for each dataset; in the present response comparison,
that minimum is attained by the flexible one-pole reference
$D_{\mathrm{rational}}^{(\mathrm{free})}$.

\paragraph*{Remark on BIC conventions.}
We therefore use slightly different observation-count
conventions for BIC: Fig.~\ref{fig:section_VC_spin_ratio_discrimination}
uses a seed-level penalty count while retaining the condition-mean likelihood
of Eq.~\eqref{eq:app_aicbic_likelihood}, whereas
Figs.~\ref{fig:section_VC_multik_ratio_discrimination} and
\ref{fig:section_VC_multik_response_evidence} use condition means by default.
This choice affects only the penalty term $p\ln N_{\mathrm{obs}}$ in
Eq.~\eqref{eq:app_aicbic_def}; it does not affect the likelihood itself,
any of the AIC values, or the qualitative ordering in the present datasets,
for which the reported information-criterion separations are already large.

\begin{acknowledgments}
This study was supported by JSPS KAKENHI (Grant Number 22K14177) and JST PRESTO (Grant Number JPMJPR23O7).
\end{acknowledgments}

\bibliography{main}

\end{document}